\documentclass[11pt,paper]{article}
\usepackage{jheppub}

\usepackage[usenames,dvipsnames,table]{xcolor}
\usepackage{graphicx,amsmath,amssymb,multirow,array,bm,mathrsfs}
\usepackage{epsf,amsfonts}
\usepackage[numbers,sort&compress]{natbib}
\usepackage{dsfont}

\usepackage{slashed}
\usepackage{subcaption}

\newcommand{\be}{\begin{equation}}
\newcommand{\ee}{\end{equation}}
\newcommand{\bea}{\begin{eqnarray}}
\newcommand{\eea}{\end{eqnarray}}
\newcommand{\tr}{{\text{Tr}}}
\newcommand{\half}{{1 \over 2}}

\newcommand{\bT}{{\mathbf T}}

\newcommand{\Tj }{\mathbf T_j(\lambda)}

\newcommand{\TT}{\mathbf T(\lambda)}
\newcommand{\LL}{\mathbf L(\lambda)}

\newcommand{\C}{\mathbb{C}}
\newcommand{\Z}{\mathbb{Z}}

\title{Thermal Correlation Functions of KdV Charges in 2D CFT}

 \preprint{TCDMATH 18-16}
\author[a,\circ]{Alexander Maloney}
\note[$\circ$]{maloney@physics.mcgill.ca}
\author[b,c,\diamond]{Gim Seng Ng}
\note[$\diamond$]{gng@math.tcd.ie}
\author[d,\star]{Simon F. Ross}
\note[$\star$]{s.f.ross@durham.ac.uk}
\author[a,\triangleright]{Ioannis Tsiares}
\note[$\triangleright$]{ioannis.tsiares@mail.mcgill.ca}

\affiliation[a]{Department of Physics, McGill University, Montr\'{e}al, QC, Canada}
\affiliation[b]{School of Mathematics, Trinity College Dublin, Dublin 2, Ireland}
\affiliation[c]{Hamilton Mathematical Institute, Trinity College Dublin, Dublin 2, Ireland}
\affiliation[d]{Centre for Particle Theory, Department of Mathematical Sciences, Durham University, South Road, Durham DH1 3LE, UK}
\abstract{
Two dimensional CFTs have an infinite set of commuting conserved charges, known as the quantum KdV charges, built out of the stress tensor.  We compute the  thermal correlation functions of the these KdV charges on a circle.  We show that these correlation functions are given by quasi-modular differential operators acting on the torus partition function.  We determine their modular transformation properties, give explicit expressions in a number of cases, and give a general form which determines an arbitrary correlation function up to a finite number of functions of the central charge.  We show that these modular differential operators annihilate the characters of the $(2m+1,2)$ family of non-unitary minimal models.  We also show that the distribution of KdV charges becomes sharply peaked at large level.
}

\begin{document}
\maketitle
\section{Introduction and summary of results} 
\label{sec:intro}

Two dimensional conformal field theories (CFTs) have an infinite-dimensional symmetry algebra -- the Virasoro algebra of local conformal transformations -- which powerfully constrains the structure of the theory.  Our goal is to explore the consequences of this symmetry structure for the finite-temperature behaviour of the theory.  We will study two dimensional CFTs at finite temperature on the circle.  Our interest in this paper is obtaining exact results: we will keep the radius of the circle finite, and will not study the high temperature or large $c$ limits, though these will be discussed in a companion paper \cite{paper2}.

Our starting point is the observation that, from the Virasoro algebra, one can define an infinite set of mutually commuting conserved charges \cite{Sasaki:1987mm,Eguchi:1989hs,Bazhanov:1994ft}. In the large $c$ limit these charges generate the KdV hierarchy of differential equations, so they are often referred to as the quantum KdV charges.  We will follow the notation of \cite{Bazhanov:1994ft} where the charges are denoted $I_{2m-1}$ with $m=1,2,\dots$, and the subscript labels the spin of the charge.  The first KdV charge $I_1$ is just $L_0 = \oint T(z) dz$, the zero mode of the holomorphic part of the stress tensor.  The KdV charge $I_{2m-1}$ is an $m^{th}$ order polynomial in the Virasoro generators $L_n$.

Given this set of mutually commuting charges, one can define a Generalized Gibbs Ensemble (GGE) for two-dimensional CFTs where we introduce a chemical potential for each KdV charge:
\footnote{See \cite{Calabrese:2011vdk,Sotiriadis:2014uza,PhysRevLett.115.157201,Vidmar2016,Pozsgay2017} for some recent work on GGEs and \cite{Langen207,Kinoshita06} for recent experimental realisations. This ensemble was studied from a holographic perspective in \cite{deBoer:2016bov}; see also \cite{Perez:2016vqo}. } 
\begin{equation}
\label{GGEZ}
Z[\beta,\mu_3,\mu_5, \ldots] = \mbox{ Tr }[e^{-\beta E + \mu_3 I_3 + \mu_5 I_5 + \ldots}] = \mbox{ Tr }[  e^{\mu_3 I_3 + \mu_5 I_5 + \ldots} q^{L_0 - c/24}], \quad q \equiv e^{-\beta}\,,
\end{equation}
In this equation we have introduced chemical potentials for the left-moving (holomorphic) generators $L_0, I_3, I_5, \dots$.  In general, of course, one should also introduce potentials for the right-moving (anti-holomorphic) generators ${\bar L}_0, {\bar I}_3, {\bar I}_5, \dots$.  However, all of our results involve the use just of the Virasoro algebra so the left- and right-moving sectors completely factorize.  We will therefore just write equations for the left-moving sector explicitly, remembering that the corresponding right-moving results can be obtained by complex conjugation.
The trace in equation (\ref{GGEZ}) can be taken over the entire Hilbert space of the CFT on a circle, or over a particular Virasoro module; we will give results in both cases.
The dependence of the partition function  (\ref{GGEZ}) on the chemical potentials encodes the consequences of Virasoro symmetry for finite-temperature physics in a useful way. The purpose of this paper is to explore the structure of this ensemble, at finite temperature and finite central charge.

The KdV charges are obtained by  integrating a conserved current built from  positive powers of the stress tensor around a spatial circle. The first three charges are the zero modes of the operators $J_2\equiv T$, $J_4 \equiv (TT)$ and $J_6 \equiv (T(TT)) + \frac{(c+2)}{12} (T'T')$, where $T\equiv T_{zz}$ is the left-moving stress tensor, the round brackets denote conformal normal ordering, and the prime is a spatial derivative.\footnote{Here we adopt the usual notation where the spatial circle has period $2\pi$, but later it will be convenient to adopt a convention where the spatial circle has period 1.} The additional term in $J_6$ is required to ensure that the zero modes commute.  There will be similar terms in all the higher spin operators. The zero modes of these operators can be written explicitly in terms of the Virasoro modes of the stress tensor, as
\bea  \label{kdvcharges}
I_1 &\equiv& L_0 - k , \\
I_3 &\equiv& 2 \sum_{n=1}^\infty L_{-n} L_n + L_0^2 - (2k + \frac{1}{6}) L_0 +  k \left(k +\frac{11}{60}\right), \nonumber\\
 I_5 &\equiv& \sum_{n_1+n_2+n_3=0}: L_{n_1} L_{n_2} L_{n_3} :
+\sum_{n=1}^\infty \left( (4k+\frac{11}{6}) n^2 -1 - 6k\right) L_{-n} L_n \nonumber \\  \nonumber
&&
+\frac{3}{2} \sum_{m=1}^\infty L_{1-2m} L_{2m-1} -(3k+\frac{1}{2}) L_0^2 
+\frac{(18k+5)(12k+1)}{72} L_0 - \frac{k(42k+17)(36k+7)}{1512} \,. 
\eea 
We have defined $k=c/24$ for convenience.
With these explicit expressions one can verify (with some work) that the $I_{2m-1}$ are indeed mutually commuting
\be
[I_{2m-1}, I_{2m'-1}]=0\,.
\ee
These KdV charges act within each Virasoro module, so can be simultaneously diagonalized level by level within each Virasoro representation.
In practise, however, this is a highly difficult task: the authors of \cite{Bazhanov:1994ft} show that the problem of the simultaneous diagonalization of $I_{2m-1}$'s can be mapped to a quantum version of the KdV problem.\footnote{This can also be mapped to a Schroedinger problem \cite{Bazhanov:2003ni}.}

At finite central charge, we are not able to calculate the  GGE partition function for finite values of the chemical potentials in general. Instead, we will consider an expansion for infinitesimal chemical potentials $\mu_i$ for the non-trivial (i.e. $m\ge 2$) KdV charges.  The coefficients in this expansion are the thermal expectation values of $n$-point functions of the KdV charges, $\langle I_3 \rangle$, $\langle I_3^2 \rangle$, $\langle I_5 \rangle$, $\langle I_5^2 \rangle$, $\langle I_3 I_5 \rangle$ etc.  Our focus is on these thermal expectation values, so we will use the notation $\langle X \rangle = Tr \left( X q^{L_0-k}\right)$ for the thermal average. We also use $Z$ to denote the thermal partition function $Z[\beta]$ with no further chemical potentials.  We study the calculation of these thermal expectation values and their modular transformation properties. 

One motivation for understanding the GGE is that it is important in the application of the eigenstate thermalization hypothesis (ETH)  \cite{Deutsch1991,PhysRevE.50.888,Rigol2008,Srednicki99,DAlessio2016} to two-dimensional conformal field theories (CFTs), and more broadly in the understanding of chaos in two-dimensional CFTs.\footnote{See \cite{Fitzpatrick:2015zha,deBoer:2016bov,Perez:2016vqo,Lashkari:2016vgj,Dymarsky:2016ntg,Lashkari:2017hwq,Faulkner:2017hll,He:2017txy,Basu:2017kzo,Brehm:2018ipf,Romero-Bermudez:2018dim,Hikida:2018khg} and the references therein for some recent discussions.} Two dimensional CFTs present an especially interesting case as they have an infinite-dimensional symmetry algebra which does not (except in certain special cases, such as the minimal models) trivialize the dynamics.  Thus 2D CFTs can exhibit chaotic dynamics \cite{Roberts:2014ifa}, and hence could be expected to obey the ETH. The implications of our results for ETH will be discussed in a companion paper \cite{paper2}. A related discussion will appear in \cite{Tolya}.

We will obtain exact results for the thermal expectation values of the KdV charges at finite temperature and finite central charge. We will first discuss the general features of these expectation values and see how these features allow us to determine the form of some of these thermal expectation values without any explicit calculation. We will then discuss the explicit calculation using two different methods. The explicit calculation can in principle be carried out for any KdV charge, but in practice the calculation quickly becomes cumbersome.  We will describe certain techniques (based on the structure of Virasoro null states) which give exact results when a naive-brute force computation is not feasible. 

The key features of these thermal expectation values are:
\begin{itemize}
\item The $n$-point function $\langle I_{2m_1-1} \ldots I_{2m_n-1} \rangle$ in a given Virasoro module is a differential operator acting on the character of that module, regarded as a function of the torus modular parameter (i.e. of temperature).  The form of this differential operator depends only on the  central charge. As a consequence, the thermal expectation values in the full CFT are given by the same differential operators acting on the ordinary thermal partition function $Z[\beta]$. This feature expresses the kinematic nature of the KdV charges; the values of KdV charges on descendent states are entirely determined by the structure of the Virasoro algebra.

\item The one-point functions $\langle I_{2m-1} \rangle$ are modular forms of weight $2m$. Thus the differential operator is a {\it modular} differential operator.  It can be written in terms of the modular covariant Serre derivative, with coefficients which are themselves modular forms (and hence linear combinations of the Eisenstein series $E_4$ and  $E_6$).\footnote{A brief review of necessary features of Eisenstein series and Weierstrass functions is given in appendix \ref{weier}.}

\item
The $n$-point functions $\langle I_{2m_1-1} \ldots I_{2m_n-1} \rangle$ are quasi-modular forms of weight $2 \sum_i m_i$ and depth $n-1$.  Thus the differential operator is a {\it quasi-modular} differential operator which is a linear combination of modular differential operators with coefficients containing up to $n-1$ powers of the Eisenstein series $E_2$. 

We establish this using general results of \cite{Dijkgraaf:1996iy} on the relation between line integrals and surface integrals of operators on the torus.  When applied to $n-$point functions this leads to a certain ``anomaly" under the modular $S$-transformation, which leads to correlation functions which are quasi-modular rather than modular.
\item $I_{r}$ vanishes as an operator in the $(s+2,2)$ minimal model when $r$ is a multiple of $s$ \cite{Bazhanov:1994ft,Bazhanov:1996aq}. This means that the differential operators for $\langle I_r  \dots \rangle $, where $\dots$ denotes any combination of KdV charges, will annihilate the characters of this minimal model.  
In particular, the differential operator for $\langle I_{2m-1}\rangle$ is an $m^{th}$ order differential operator which annihilates the $m$ characters of the $(2m+1,2)$ minimal model when the central charge is set to the appropriate value.
This is a remarkable relation, but it only constrains expectation values for a particular value of the central charge, so will not be particularly useful in computations at general central charge.  We will therefore postpone the discussion of its proof to the end of the paper. 
\end{itemize}

Before diving into details, let us illustrate these statements in the simplest case.  For the first non-trivial KdV charge $I_3$, we find
\be \label{I3result} 
\langle I_3 \rangle
= \left[\partial^2 - \frac{E_2}{6} \partial  + \frac{k}{60} E_4 \right] Z = \left[ D_2 D_0 + \frac{k}{60} E_4 \right] Z, 
\ee
where we introduce the notation $\partial = q \partial_q = - \partial_\beta$. In the second form of the equation we have used the  
Serre derivative, defined as
\be
D_r \equiv \partial-\frac{r}{12} E_2
\ee 
where $E_2$ is an Eisenstein series. These derivatives are useful because they map modular forms to modular forms: $D_r$ applied to a modular form of weight $r$ gives a modular form of weight $r+2$.  Thus, since the partition function is modular invariant, the one point function $\langle I_3\rangle$ will transform as a modular form of weight $4$.

Similarly, we find
 \bea  \label{I3e2result} 
   \langle  I_3^2 \rangle
&=& \left[
D_6 D_4 D_2 D_0 +\frac{1}{90} (3 k+5)E_4 D_2 D_0 -\frac{72k+11}{1080} E_6 D_0 +\frac{k (1221 k+500)}{75600} E_4^2
\right]Z
\nonumber\\
&&
+E_2 \left[\frac{2}{3} D_4 D_2 D_0 + \frac{1}{1080} (72 k+11) E_4 D_0 - \frac{1}{756} k (12 k+5)E_6 \right] Z\,,\label{I3sqcov}
   \eea
The statement that $\langle I_3^2\rangle$ is a quasi-modular form of weight $8$ and depth $1$ reflects the fact that the modular differential operators in the square brackets have weight $8$ and $6$, respectively, and that only one power of $E_2$ appears in this expression.
The appearance of $E_2$ in this expression reflects the existence of an ``anomaly" under modular $S$-transformations.  Indeed, the coefficient of $E_2$ in this expression is precisely the thermal one-point function of an operator appearing in OPE of $J_4$ with itself.  This will be described in more detail in section \ref{features}.  \footnote{A similar structure has actually appeared in the literature before in the context of ${\cal W}$-algebras.  The authors of \cite{Iles:2014gra} computed the modular transformation properties of $\mbox{ Tr } W_0^2 q^{L-0-k}$, where $W_0$ is the zero mode of the spin 3 generator $W(z)$ of the ${\cal W}_3$ algebra.  They discovered that this trace transformed like a modular form, except for an anomaly term that is proportional to $E_2$ times the differential operator appearing in equation (\ref{I3result}).  In light of the discussion in section \ref{features} this is easy to understand: the operator $J_4$, whose zero mode is $I_3$, appears as the coefficient of the $1/z^2$ term in the $W(z) W(0)$ OPE.} 

The above modular differential operators can be applied to the partition function of a CFT or to an individual Virasoro character, depending on whether one wants to obtain these expectation values in a full CFT or in a particular representation.  

The relationship with minimal models is found by evaluating these differential operators in the $(5,2)$ minimal model, i.e. the Yang-Lee model with $k = -\frac{11}{60}$.  One can indeed check that these modular differential operators annihilate the characters of the Yang-Lee model.  Indeed, at $k = -\frac{11}{60}$ the differential operator \eqref{I3result} is precisely the one associated with the null state in the vacuum representation of the Yang-Lee model.\footnote{We refer to \cite{Gaberdiel:2008pr} for a more detailed discussion of the construction of these modular differential operators. It would be interesting to investigate whether our results are relevant for the approach to the classification of CFTs  based on the classification of modular differential operators, advocated in \cite{Mathur:1988na} and used recently in \cite{Hampapura:2015cea, Gaberdiel:2016zke}.}

In section \ref{features} we will give the general argument that the $n$-point functions are modular derivatives of the partition function, involving up to $n-1$ powers of $E_2$. 
This reduces the problem of determining a particular thermal expectation value to fixing a finite number of undetermined coefficients. 

In section \ref{qexp} we show how some of these coefficients can be fixed from the action of the KdV charges at low levels, and by the structure of null states in Virasoro representations. This allows us to determine a number of the thermal expectation values exactly.   The complete list of results so obtained is summarized in appendix \ref{qexpapp}.  
Although the resulting differential operators appear unfamiliar, in fact they can be written in terms of more familiar hypergeometric differential operators, as explained in appendix \ref{hyp}.

Sections \ref{comm} and \ref{rec} describe two other, more straightforward approaches to the computation that can be used to verify the results of section \ref{qexp}. In these sections we carry out the explicit calculations of the thermal expectation values directly from the definition.  This allows us to explicitly confirm the expected structure, and also provides an approach which can in principle be pushed to arbitrary order, although in practice the calculations become laborious and far less efficient than the methods described in section \ref{qexp}. We carry out explicit calculations in two ways: 
\begin{itemize}
\item We can write the KdV charge in terms of Virasoro modes, and cycle the modes around the commutator to relate the thermal expectation value to derivatives of the ordinary thermal partition function of the CFT. We will illustrate this in detail for the computation of $\langle I_3 \rangle$ in section \ref{comm}. This approach is the most straightforward; however, dealing with all the terms rapidly becomes unwieldy as we move to higher powers or charges of higher degree.  Some computational details are relegated to appendix \ref{brute}.
\item We can use Zhu's recursion relations \cite{zhu}, as extended in \cite{Gaberdiel:2012yb} (see also \cite{Gaberdiel:2008pr}). These recursion relations give us expressions for the thermal $n$-point functions of the stress tensor, which we can then appropriately integrate to get thermal expectation values of the KdV charges. The recursion relations are  based on the same kind of cycling of operators around the trace, but they provide a powerful general encoding of these results, which make some of the modular properties of the result more manifest. Using this technique allows us to push the calculation to higher order. We discuss the calculation in section \ref{rec}, though we relegate some of the details to appendices \ref{recursion}, \ref{zero}. 
\end{itemize}

In section \ref{relint}, we discuss the relation to the minimal models, showing that the KdV charges $I_r$ vanish as an operator in the $(s+2,2)$ minimal model whenever $r$ is a multiple of $s$. This makes explicit certain properties of the KdV charges which were asserted in \cite{Bazhanov:1994ft,Bazhanov:1996aq}. The relation fixes the coefficients in the  $n$-point functions for the particular values of the central charge corresponding to the minimal models. It is particularly useful for the one-point functions, which get related to the  differential operators associated to null states \cite{Gaberdiel:2008pr,Leitner:2017}. 

In the final section, we will discuss an important application of these results: to determine the statistics of KdV charges.  In particular, by explicit evaluation of the differential operators on a Virasoro character we can determine the moments of the KdV charges at arbitrary level.  For example, by applying equations  (\ref{I3result})  and  (\ref{I3e2result}) to the Verma module character we can determine the mean and the variance of the $I_3$ at each level in a generic Virasoro representation.
The result of this is that at high level the distribution of KdV charges will become very sharply peaked around its average value.
We will make a few remarks on this in section \ref{stats}, and expand further on the statistics of KdV charges in \cite{paper2}.

In summary, we have determined the general structure of the thermal expectation values of $n$-point correlation function of the KdV charges, and found explicit expressions for a number of cases. These expressions are exact results at finite temperature and finite central charge. This is a useful step towards understanding the GGE for two-dimensional CFTs; the structure is however too complicated to allow for a simple exponentiation to obtain an explicit formula for the GGE partition function at finite $c$ and finite temperature. 

A key element in our derivation is the determination of the modular transformation properties of the $n$-point functions. In our companion paper \cite{paper2}, we will apply these results to explore the high-temperature behaviour of the GGE at finite central charge and consider the comparison to expectation values of KdV charges in a typical high-energy microstate. 

Many open problems remain.  These include finding more efficient methods for determining the remaining free parameters in the thermal expectation values, and applying similar techniques to theories with extended symmetry algebras. 

\section{Modular differential form of the thermal expectation values}
\label{features}

In this section we explain how the thermal expectation values of KdV charges can be written as modular derivatives of the partition function. This fixes the form of the thermal expectation values up to a finite number of coefficients, which are polynomials in the central charge. 

\subsection{Differential operator form}
\label{diffop}

For ordinary conserved charges, such as a $U(1)$ charge, the partition function with a chemical potential for the charge carries additional information about the theory above and beyond what is available in the thermal partition function $Z(\beta)$. That this is not the case for the KdV charges should not be surprising; the KdV charges evaluated on primary states are simply functions of the $L_0$ eigenvalue of the primary, and the KdV charges evaluated on descendents can in principle be determined by applying the Virasoro commutation relations
\be \label{vircom}
[L_m,L_n]=(m-n)L_{m+n} + 2k (m^3-m)\delta_{m+n}.
\ee
One consequence is that the expectation values of KdV charges in a Virasoro module can be written as derivatives of the character with respect to the modular parameter (and hence thermal expectation values in the full CFT are the same derivative of the partition function). 

To understand this a bit more explicitly, consider the thermal expectation value of a product of KdV charges evaluated in a Virasoro module built on a primary state of dimension $h$:
\begin{equation}
\langle I_{2m_1 -1} \ldots I_{2m_n-1} \rangle_h \equiv \mbox{ Tr}_h [I_{2m_1 -1} \ldots I_{2m_n-1} q^{L_0-k}], 
\end{equation}
Here the trace runs over the Virasoro descendents of a primary state with $L_0 = h$. Using the explicit expressions for the KdV charges, the RHS can be rewritten as a polynomial in the Virasoro modes. Virasoro modes $L_r$ for $r \neq 0$ can be cycled around the trace using the commutation relations. We pick up a factor of $q^{-r}$ when $L_r$ commutes past $q^{L_0}$, and commuting it past the other modes gives terms which are lower order polynomials in the Virasoro modes. Thus, we can iteratively rewrite the expression in terms of functions of $q$ times traces of lower polynomials in the Virasoro modes, leaving us ultimately with an expression involving functions of $q$ times traces of polynomials in $L_0$. The factors of $L_0$ can then be rewritten in terms of derivatives of the character $\chi_h(q)= \mbox{Tr}_h [ q^{L_0-k}]$ with respect to $q$. Since $I_{2m-1}$ involves at most $m$ factors of $L_0$, this expression contains derivatives up to order $M = \sum_i m_i$, and our expectation value is an $M^{th}$ order differential operator
\begin{equation}
\langle I_{2m_1 -1} \ldots I_{2m_n-1} \rangle_h =  (\partial^M + F_1(q) \partial^{M-1} + \ldots) \chi_h(q), 
\end{equation}
where we use the notation $\partial = q \partial_q$. The coefficients are functions of $q$ and $k$ (the dependence on $k$ coming from the central term in the Virasoro commutator \eqref{vircom}) but do not depend on the conformal dimension $h$ of the character: the differential operator is universal, depending only on the choice of KdV charges. We can obtain the thermal expectation value in the full CFT by summing over primaries, giving us an expression where the same differential operator acts on the partition function:
\begin{equation}
\langle I_{2m_1 -1} \ldots I_{2m_n-1} \rangle =  (\partial^M + F_1(q) \partial^{M-1} + \ldots) Z[\beta], 
\end{equation}
where $q = e^{-\beta} = e^{2\pi i \tau}$. 

In section \ref{comm}, we will discuss the explicit evaluation of the thermal expectation values using this procedure, but for now, the key point is that it determines a general form for the expectation values, as a differential operator acting on the character or on the partition function.\footnote{This is quite different from the situation in theories with higher spin symmetries, where the GGE for the higher spin charges has genuinely new information. The expectation values of these higher spin charges were studied in \cite{Kraus:2011ds,Gaberdiel:2012yb,Iles:2013jha}. The modular transformations were studied in \cite{Iles:2014gra}, which obtains results in line with the expectations of  \cite{Dijkgraaf:1996iy}, as noted in the next subsection.}

\subsection{Modular transformation properties} 
\label{dijk}

Since the expression obtained in the previous subsection is a function of the modular parameter, it is natural to ask about its behaviour under modular transformations. 

Let us begin with the expectation value of a single KdV charge, where the answer is simple.  The charge is the contour integral of a conserved chiral current $J_{2m}$ around the spatial circle:
\begin{equation}
\langle I_{2m-1} \rangle = \langle \oint J_{2m} \rangle~.
\end{equation}
If we describe the torus by the standard $z$ coordinate, with $z\sim z+1\sim z+\tau$, then $\oint J = \int_0^1 \frac{dz}{2\pi} J(z)$.
The integration around the spatial circle appears to break modular symmetry, since we have picked a particular cycle of the torus around which to integrate. However, since the current $J_{2m}$ is conserved we can freely translate the spatial integral around the torus.  This means that we can average over the location of the spatial circle to turn the one-dimensional integral over the spatial circle into an integral over the entire torus. The result is that  
\begin{equation}
\label{int1}
\langle I_{2m-1} \rangle = \langle \int J_{2m} \rangle. 
\end{equation}
where, following  \cite{Dijkgraaf:1996iy}, we have introduced the following notation for integrals over the torus:
\be
\int J \equiv \int \frac{d^2z}{2 \pi \tau_2} J(z)
\ee 
The factor of $1/\tau_2$ arises because when we average over the torus, we must divide by the area of the torus.  An important point is that with this normalization the integral $\int \frac{d^2z}{2\pi \tau_2}$ is itself modular invariant.
The final step in the argument is to note that the operator $J_{2m}$ is (up to total derivative terms) a quasiprimary with dimension $2m$.  Thus under a modular transformation its one point function $\langle J_{2m}(z) \rangle$ transforms like a modular form of weight $2m$.
The result of equation (\ref{int1}) is therefore that $\langle I_{2m-1} \rangle$ transforms like a modular form of weight $2m$. 

We saw in the previous section that the expectation value is a combination of derivatives of the partition function. It is now natural to rewrite these derivatives in terms of the Serre modular derivative. To simplify the notation, we will denote the Serre derivative of a modular form $A$ of weight $r$ by $D A = D_r A$, where $D$ acts as a derivation on products of modular forms as $D (AB) = (DA) B + A (DB)$. 

The statement that $\langle I_{2m-1} \rangle$ is a modular form of weight $2m$ implies that we can write it as a combination of the derivatives $D^k Z$ where $k =0, \ldots m$, with coefficients which are modular forms of weight $2m-2k$. We can then write  these modular forms in terms of the Eisenstein series $E_4$ and $E_6$ to obtain a general expression for the one point function:
\begin{equation}\label{genform}
\langle I_{2m-1} \rangle = D^m Z + c_1 E_4 D^{m-2} Z + c_2 E_6 D^{m-3} Z + c_3 E_4^2 D^{m-4} Z + \ldots, 
\end{equation}
where the coefficients $c_i$ are functions only of the central charge $k$ of the CFT. We will see that they are polynomials in $k$ of increasing order.\footnote{Note that for sufficiently high weight, there is more than one modular form of a given weight; for example $E_4^3$ and $E_6^2$ are both modular forms of weight $12$. So at higher orders there will generically be more than one term in the sum at a given order in derivatives.} The modular properties thus suffice to determine the $q$-dependence in the differential operator, determining the operator up to a finite number of coefficients. 

For the higher-point functions, there is a subtlety which implies that the differential operators are not exactly modular covariant. For example, consider the two-point function 
\begin{equation}
\langle I_{2m-1} I_{2n-1} \rangle = \langle \oint J_{2m} \oint J_{2n} \rangle. 
\end{equation}
We wish to convert these contour integrals into integrals over the torus by using the fact that the currents $J_{2n}$ and $J_{2m}$ are conserved. But averaging  the two operators over the torus will give a singular contribution where the operators coincide, which is absent in the expression with a pair of contour integrals.  This contribution was analyzed carefully by Dijkgraaf \cite{Dijkgraaf:1996iy}, who showed that 
\begin{equation} \label{modquad}
\langle \oint J_{2m} \oint J_{2n} \rangle = \langle \int J_{2m} \int J_{2n} \rangle - \frac{1}{2 \tau_2} \langle \int [J_{2m} J_{2n}]_2 \rangle, 
\end{equation}
where $[J_{2m} J_{2n}]_2$ denotes the coefficient of $(z-w)^{-2}$ in the OPE $J_{2m}(z) J_{2n}(w)$.\footnote{We use square brackets here to avoid confusion with the round bracket notation for conformal normal ordering.} Again, rewriting the contour integrals in terms of torus integrals allows us to understand the modular transformation properties. The correlation functions on the RHS are modular forms of weight $2m+2n$ and $2m+2n-2$ respectively, but the explicit factor of $\tau_2$ in the second term makes it transform inhomogeneously under a modular transformation. The result is that the two-point function $\langle I_{2m-1} I_{2n-1} \rangle$ is a quasi-modular form. 

When we express this correlation function in terms of a differential operator, this inhomogeneous term is reproduced by allowing a single factor of the Eisenstein series $E_2$ to appear, as $E_2$ has an inhomogeneous transformation under modular transformations which reproduces the transformation of \eqref{modquad}. Hence the expectation value is a combination of a modular form of weight $2m+2n$, and $E_2$ times a modular form of weight $2m+2n-2$.  The general form for the differential operator is
\begin{equation}
\langle I_{2m-1} I_{2n-1} \rangle = \left(D^{m+n} Z + c_1 E_4 D^{m+n-2} Z + \ldots\right) + E_2 \left(D^{m+n-1} Z + d_1 E_4 D^{m+n-3} Z + \ldots\right), 
\end{equation}
where again the coefficients $c_i$, $d_i$ are functions only of $k$.  The first term in parentheses is a modular form of weight $2m+2n$.  The second term in parentheses is a modular form of weight $2m+2n-2$, and is the one point function of the operator $[J_{2m} J_{2n}]_2$ on the torus.

For higher-point functions of the KdV charges, more coincident singularities will appear when we convert contour integrals to torus integrals. The approach of \cite{Dijkgraaf:1996iy} can be applied to express these in terms of inhomogeneous terms involving the second-order OPE singularity of the coinciding operators. The result is that the $n$-point correlation function of KdV charges $\langle I_{2m_1 -1} \ldots I_{2m_n-1} \rangle$ is a quasi-modular form of weight $2 \sum_i m_i$ and depth $n-1$, meaning that the differential operator will contain terms with up to $n-1$ powers of $E_2$, with each term having total weight $2 \sum_i m_i$. 

These modular transformations will be explored in more detail in specific cases in \cite{paper2}, where we use them to relate the high-temperature behaviour of the correlation functions to their low-temperature limit. Here the key point is that they strongly constrain the form of the differential operator appearing in the KdV $n$-point functions.

\section{Determining thermal expectation values: $q$ expansion and null states}
\label{qexp}

In the previous section we derived a general form for the thermal expectation value of a product of KdV charges in terms of a differential operator which is determined up to a finite number of constants. In this section, we discuss to what extent we can use data from low levels in a Virasoro module, and the structure of Virasoro representations, to determine these coefficients. We will discuss only a few representative computations.  A complete list of results obtained using this method is given in Appendix \ref{qexpapp}.

\subsection{One-point functions} 
\label{onept}

We consider first the one-point functions $\langle I_{2m-1} \rangle_h$. The character is $\chi_h(q) = q^{h-k}(1+ q + \ldots)$ for $h \neq 0$; the action of the differential operator on this character will then have a $q$-expansion 
\begin{equation}
\langle I_{2m-1} \rangle_h = q^{h-k} [I_{2m-1}^{L=0}(h,k) + I_{2m-1}^{L=1}(h,k) q + \ldots], 
\end{equation}
where $I_{2m-1}^{L=0}$ is the value of the KdV charge on the primary, and $I_{2m-1}^{L=1}$ is its value on the unique state at level one.\footnote{Note that the KdV charges map each level of the Virasoro representation to itself. That is, the matrix elements of the operator on the states of a Virasoro module in the usual basis of descendent states are block diagonal. In particular, at levels zero and one, where there is a single state, this state is necessarily an eigenstate of the KdV charges, and it makes sense to talk about the value of the KdV charge at levels zero and one.}

It is easy to determine the value of $I_{2m-1}^{L=0}$ from the basic expression for the KdV charges  \eqref{kdvcharges}; it is simply a function of the $L_0$ eigenvalue of the primary, $h$. It will be a polynomial of order $m$ in $h$, as the KdV charge $I_{2m-1}$ involves up to $m$ powers of $L_0$.
We can then compare this to the result that one would obtain if one acts on the character $\chi_h(q)$ with a general differential operator of the form (\ref{genform}), and use this to fix the undetermined coefficients in the differential operator.
When this differential operator acts on the character, the $m$ powers of $h$ come from the $m$ derivatives with respect to $q$ acting on the $q^{h-k}$ term in the character. Thus, knowledge of $I_{2m-1}^{L=0}$ will fix one coefficient at each order in derivatives. This will fix the first few terms in the differential operator uniquely.  For sufficiently small values of $m$ this is enough to determine the differential operator uniquely.

The values of KdV charges $I_3$ up to $I_{15}$ on primary states were tabulated in appendix B of \cite{Bazhanov:1996aq}. We can use this to obtain, for example, 
\begin{equation}
\langle I_3 \rangle_h = [ D^2 + \frac{k}{60} E_4] \chi_h,
\end{equation}
\begin{equation} \label{I5cov}
\langle I_5 \rangle_h = [D^3  + \frac{(6k+1)}{72} E_4 D  - \frac{k(12 k+7)}{756} E_6] \chi_h.
\end{equation}

One can check that the values in minimal models vanish as advertised.  In the Yang-Lee $(5,2)$ minimal model, $k= - \frac{11}{60}$, and there are two primaries: the vacuum with $h=0$ and a scalar of weight $h = -\frac{1}{5}$. The vacuum representation has a null vector at level 4. In \cite{Gaberdiel:2008pr}, this null state was shown to imply a differential equation for the characters of this theory, 
\begin{equation}
[D^2 - \frac{11}{3600} E_4] \chi_{0, -\frac{1}{5}} = 0. 
\end{equation}
This is precisely our differential operator $\langle I_3 \rangle$ for $k = -\frac{11}{60}$, so we see $\langle I_3 \rangle_h=0$ in the Yang-Lee model. Similarly the $(7,2)$ minimal model has $k = -\frac{17}{42}$. The differential operator in \eqref{I5cov} for this value of $k$ reduces to the differential operator which annihilates the characters of the $(7,2)$ minimal model \cite{Leitner:2017}.  

This method allows us to determine the differential operators for $\langle I_{3} \rangle$ up to  $\langle I_{9} \rangle$. However, for $\langle I_{11}\rangle$ and higher, we are confronted with another problem.  The terms in the differential operator are modular forms, and for weight $w\ge 12$ there is no longer a unique modular form of weight $w$.  Thus knowledge of $I_{2m-1}^{L=0}$  alone is no longer sufficient to determine the modular differential operator uniquely.  Fortunately, we can use the structure of null states in Virasoro representations to obtain further constraints.  For example, we know that when $h=0$ the representation has a null state at level $1$.  Thus when applied to the vacuum character $\chi_0 = q^{-k} (1 + q^2 + \ldots)$ our differential operator must not give a term proportional to $q^{-k+1}$.  This provides an extra constraint, which allows us to determine the differential operators up to $\langle I_{13} \rangle$. One could make further progress, for example, by calculating $I_{2m-1}^{L=1}$ at general $h$ using the Virasoro commutation relations, but we have not pursued this. 

A complete tabulation of results is given in appendix \ref{qexpapp}. 

\subsection{Two-point functions}

We can also use constraints from the low orders in the $q$ expansion to fix parameters in our expression for $\langle I_{2m -1} I_{2n-1} \rangle$. For a generic primary, we have the $q$ expansion
\begin{equation}
\langle I_{2m-1} I_{2n-1}  \rangle_h = q^{h-k} [ I_{2m-1}^{L=0} I_{2n-1}^{L=0} +  q I_{2m-1}^{L=1} I_{2n-1}^{L=1}  + q^2 \sum_{i=1}^2 I_{2m-1}^{i,L=2} I_{2n-1}^{i, L=2}+ \ldots ], 
\end{equation}
Thus, we can use the known values of the KdV charges on the primaries, and the values of the KdV charges on the level 1 states, which can be obtained from the differential operators for the one-point functions we worked out in the previous subsection, to constrain the form of the differential operator for the products. Note that we cannot yet determine the individual eigenvalues at level 2, as the preceding calculation only determines the sum $\sum_i  I_{2m-1}^{i,L=2}$. 

This gives us enough data to determine the simplest of the quadratic expressions: 
  \bea
   \langle  I_3^2 \rangle
&=& \left[
D^4 +\frac{1}{90} (3 k+5)E_4 D^2 -\frac{72k+11}{1080} E_6 D +\frac{k (1221 k+500)}{75600} E_4^2
\right]Z
\nonumber\\
&&
+E_2 \left[\frac{2}{3} D^3 + \frac{1}{1080} (72 k+11) E_4 D - \frac{1}{756} k (12 k+5)E_6 \right] Z\,,\label{I3sqcov}
   \eea
and
    \bea
   \langle  I_3 I_5 \rangle
&=& D^5 Z 
+\left(  \frac{k}{10}+\frac{13}{72}\right)   E_4  D^3 Z  \nonumber\\
&&
-\left(\frac{k^2}{63} + \frac{7k}{27}+\frac{19}{216}\right)E_6 D^2 Z 
+\left(\frac{647 k^2}{5040}+\frac{3047 k}{30240}+\frac{55}{4536}\right)E_4^2 D Z \nonumber\\
&&
-\left(\frac{169 k^3}{3780} + \frac{29 k^2}{810}+\frac{37 k}{4320}\right) E_4 E_6  Z \nonumber\\
&& + E_2 \left[   D^4 Z 
+ \left(\frac{5  k}{18}+\frac{5}{54} \right) E_4 D^2 Z 
-\left(\frac{8 k^2}{63}+\frac{19  k}{189}+\frac{55 }{4536}
\right) E_6 DZ 
\right.
\nonumber\\
&&\left. \qquad \qquad +\left(\frac{2k^3}{45} + \frac{77 k^2}{2160}+\frac{37 k}{4320}\right)E_4^2 Z \right]. \label{I3I5}
\eea
More complicated differential operators, such as $\langle I_5^2 \rangle$, are not completely fixed by the constraints at levels 0 and 1 so more constraints are necessary. In this case, we can now use the fact that when
\begin{equation}
\label{null2}
h =  k - 1/24 + 1/4 (3 \sqrt{1/24 - k} - \sqrt{25/24 - k} )^2,
\end{equation}
there is a null state at level 2.  Thus for this value of $h$ there is just one state at level $2$, and $(I_5^2)^{L=2} = (I_5^{L=2})^2$. Determining $I_5^{L=2}$ from the previously obtained differential operator, we have one more constraint on the coefficients in $\langle I_5^2 \rangle$, which allows us to determine 
    \bea \label{I5sq}
   \langle  I_5^2 \rangle
&=&D^6 Z 
+\left(  \frac{k}{6}+\frac{19}{36}\right)   E_4 D^4Z  \\
&&
-\left(\frac{2 k^2}{63} + \frac{23k}{27}+\frac{31}{72}\right)E_6 D^3 Z
+\left(\frac{647 k^2}{1008}+\frac{709 k}{1008}+\frac{821}{5184}\right)E_4^2  D^2 Z \nonumber\\
&&
-\left(\frac{169 k^3}{378} + \frac{425 k^2}{756}+\frac{919 k}{3888}+\frac{31}{1296}\right) E_4 E_6 D  Z \nonumber\\
&&
+\left[E_4^3 \left(\frac{4k^4}{33} + \frac{1187 k^3}{8316}+\frac{3005 k^2}{49896}+\frac{871 k}{85536}\right)+E_6^2 \left(\frac{3539 k^4}{43659} + \frac{3539 k^3}{37422}+\frac{2593 k^2}{64152}+\frac{49 k}{7128}\right)\right] Z \nonumber\\
&&+ E_2 \left[\frac{3}{2}   D^5 Z 
+ \left(\frac{11  k}{12}+\frac{17}{36} \right) E_4 D^3 Z
-\left(\frac{40 k^2}{63}+\frac{265  k}{378}+\frac{205 }{1296}
\right) E_6 D^2 Z 
\right.
\nonumber\\
&&\left. \qquad \qquad +\left(\frac{4k^3}{9} + \frac{485 k^2}{864}+\frac{613 k}{2592}+\frac{745}{31104}\right)E_4^2 DZ 
-\left(\frac{20k^4}{99} + \frac{73 k^3}{308}+\frac{1115 k^2}{11088}+\frac{1459 k}{85536}\right) E_4 E_6
Z \right] .\nonumber
\eea
In appendix \ref{qexpapp}, we show that these expressions vanish on the minimal models as expected. 

\subsection{Higher point functions}

We can proceed in this way for correlation functions of higher power of the KdV charges.
For example, knowledge of $\langle I_{2m-1} \rangle_h$ and $\langle I_{2m-1}^2 \rangle_h$ gives us enough data to determine the two eigenvalues $I_{2m-1}^{i, L=2}$ individually. We can then use the constraints from levels 0, 1, and 2 to constrain expressions for three-point functions. In addition, the generic character has three states at level 3, except when
\begin{equation}
h = k - 1/24 + 
  1/4 (4 \sqrt{1/24 - k} -2 \sqrt{25/24 - k} )^2, 
\end{equation}
where there is a null state at level 3 and only two states are present.  Thus we can determine the $I_{2m-1}^{i, L=3}$ for this value of $h$. 

This gives us enough data to determine 
    \bea \label{I3c}
   \langle  I_3^3 \rangle
&=& D^6Z
+\left(  \frac{k}{20}+\frac{1}{18}\right)   E_4 D^4 Z  \\
&&
-\left(\frac{k}{5}+\frac{233}{1080}\right)E_6 D^3 Z
+\left(\frac{407 k^2}{8400}+\frac{229 k}{840}+\frac{115}{1296}\right)E_4^2  D^2 Z \nonumber\\
&&
-\left(\frac{17 k^2}{100}+\frac{2411 k}{21600}+\frac{11}{864}\right) E_4 E_6 D Z \nonumber\\
&&
+\left[E_4^3 \left(\frac{19069 k^3}{504000}+\frac{71 k^2}{3024}+\frac{25 k}{5184}\right)+E_6^2 \left(\frac{4 k^3}{135}+\frac{103 k^2}{5400}+\frac{67 k}{16200}\right)\right] Z \nonumber\\
&&+ E_2 \left[2  D^5
+ \left(\frac{7  k}{30}+\frac{131}{360} \right) E_4 D^3
-\left(\frac{k^2}{21}+\frac{739  k}{1260}+\frac{49 }{270}
\right) E_6 D^2
\right.
\nonumber\\
&&\left. \qquad \qquad +\left(\frac{101 k^2}{300}+\frac{4811 k}{21600}+\frac{11}{432}\right)E_4^2 D
-\left(\frac{169 k^3}{1260}+\frac{6409 k^2}{75600}+\frac{43 k}{2400}\right) E_4 E_6
\right] Z\nonumber\\
&&+ E_2^2  \left[
D^4
+\left(\frac{19 k}{60}+\frac{23}{240} \right) E_4 D^2
-\left( \frac{k^2}{6}+\frac{k}{9}+\frac{11}{864}\right) E_6 D
+\left(\frac{k^3}{15}+\frac{19 k^2}{450}+\frac{43 k}{4800}\right)E_4^2
\right]Z\,. \nonumber
\eea

In appendix \ref{qexpapp}, we also calculate $\langle I_3^2 I_5 \rangle$, and show that these expressions vanish on the minimal models as expected. 

We can proceed in this manner to obtain expectation values of higher powers of the KdV charges, although this procedure becomes more difficult.  One major obstruction is that the number of states at level $n$ grows rapidly with $n$.  For example, at level 4 we have 5 states, so even if we consider the value of $h$ for which there is a null state at level 4, we still cannot determine the eigenvalues of $I_3$ at level 4 from knowledge of $\langle I_3\rangle$, $\langle I_3^2\rangle$ and $\langle I_3^3\rangle$ alone.  Nevertheless it is still possible to determine $\langle I_3^4\rangle$ by using constraints from minimal models which have a more complicated structure of null states.  This is described, and the explicit result for $\langle I_3^4\rangle$ is given, in appendix \ref{qexpapp}. 

It would be interesting to implement this strategy more systematically, and see whether it is possible to obtain the differential operators for all KdV charges recursively in this manner.  If this is not possible, then more direct (but less efficient) computations may be necessary.  We describe these more direct methods in the next two sections.

\section{Explicit evaluation by commutators}
\label{comm}

The methods described in the previous section are by far the most efficient, but can be checked using a more straightforward evaluation of the thermal expectation of KdV charges which does not rely on the detailed modular structure of the differential operator.  
In this and the following section we will describe two such methods for evaluating the KdV charges.  

In this section we will simply consider the evaluation of the KdV charges \eqref{kdvcharges} as polynomials in the Virasoro modes.  As discussed in section \ref{diffop}, using the Virasoro commutation relations \eqref{vircom}, we can simplify these polynomial expressions and rewrite them in terms of lower polynomials. The key idea is that in the terms which involve non-zero modes, we can move a generator $L_r$ past the $q^{L_0-k}$ in the trace, picking up a factor of $q^{-r}$ from the commutator $[L_r, L_0] =  r L_r$. Then we can move it back past the other generators in the trace to recover the original expression with a factor of $q^{-r}$, plus additional terms coming from the right hand side of the commutator. We can then solve for the original expression in terms of the terms coming from the right hand side, which involve fewer generators. 

Explicitly, for the simplest case of $\langle I_3 \rangle$, 
\begin{eqnarray}
\langle I_3 \rangle &=& \langle 2 \sum_{n=1}^\infty L_{-n} L_n + L_0^2 -\frac{c+2}{12} L_0 +  \frac{c}{24}\left(\frac{c}{24}+\frac{11}{60}\right) \rangle \\
&=&  2\langle \sum_{n=1}^\infty L_{-n} L_n \rangle + \langle L_0^2 \rangle  -(2k+ \frac{1}{6})  \langle L_0 \rangle + k \left(k +\frac{11}{60}\right) Z[q].
\end{eqnarray} 
The factors of $L_0$ can be obtained as derivatives of the partition function, 
\begin{equation}
\langle L_0 -k\rangle = \mbox{ Tr }[ (L_0-k) q^{L_0-k}] = q \frac{d}{dq}   \mbox{ Tr }[q^{L_0-k}] = \partial Z[q].
\end{equation}
Thus
\begin{equation}
\langle I_3 \rangle = 2\langle \sum_{n=1}^\infty L_{-n} L_n \rangle + \partial^2  Z - \frac{1}{6}  \partial  Z   +\frac{k}{60} Z.
\end{equation}
For the first term, we apply the method described above, moving the $L_{-n}$ around the trace:
\begin{equation}
\langle L_{-n} L_n \rangle  = q^n \langle L_n L_{-n} \rangle = q^n \left( \langle L_{-n} L_n \rangle + 2n \langle L_0 \rangle + 2k (n^3-n) Z \right),  
\end{equation} 
so
\begin{equation} \label{lnln}
\langle L_{-n} L_n \rangle  = \frac{q^n}{1-q^n} \left( 2n \partial Z + 2k n^3 Z \right).
\end{equation}
Thus, we can write
\begin{equation}
\langle I_3 \rangle =  4 \sigma_1  \partial Z + 4k \sigma_3 Z  + \partial^2  Z - \frac{1}{6}  \partial  Z   +\frac{k}{60} Z, 
\end{equation}
where  for any odd positive integer $s$, we define
\be
\sigma_{s} \equiv \sum_{n\ge 1} n^{s} \frac{q^n}{1-q^n}\,.
\ee 
These sums are related to  the Eisenstein series through 
\be \label{sigmaE}
E_{2k}
 =
  1-\frac{4k}{B_{2k}}
 \sigma_{2k-1}
 =
 1-\frac{4k}{B_{2k}}
\sum_{n=1}^\infty \frac{n^{2k-1}}{q^{-n}-1}
\ee where $B_{2k}$ are the Bernoulli numbers.
In particular,
\bea
\sigma_1 &=& \frac{1}{24}\left(1-E_2\right) \nonumber\\
\sigma_3 &=& -\frac{1}{240}\left(1-E_4\right)
\eea 
So we can finally write 
\be \label{I3direct}
\langle I_3 \rangle
= \left[\partial^2 - \frac{E_2}{6} \partial  + \frac{k}{60} E_4 \right] Z
=\left[ D^2 + \frac{k}{60} E_4 \right] Z\,,
\ee
reproducing the result obtained in section \ref{onept}. 

We could continue to explore further examples by applying this commutation method. The calculation of $\langle I_3^2 \rangle$ using this method is described in appendix \ref{brute}. However, the calculation quickly becomes lengthy, and the appearance of the Eisenstein series appears somewhat miraculous from this point of view. In the next section we describe an alternative approach which organizes the calculation more cleanly. 
 
\section{Evaluation using Zhu's recursion relations}
\label{rec}

A second approach to the evaluation of thermal expectation values of KdV charges is to compute
an $n$-point function of the stress tensor on the torus, and then perform the necessary contour integrals to extract the desired KdV charges. Expressions for the thermal $n$-point functions of the stress tensor can be obtained by using the recursion relations of Zhu \cite{zhu}, extended in \cite{Gaberdiel:2012yb} (the extension in \cite{Gaberdiel:2012yb} allows us to consider correlation functions also involving zero modes, which arise in the recursion starting from a pure operator correlation function). These recursion relations are reviewed in appendix \ref{recursion}. The basic idea is the same as in the previous calculation: we can obtain an expression for an $n$-point function in terms of $n-1$ point functions by taking one operator in the correlator, expanding it in modes, and commuting each of the modes around the correlator. The method of \cite{zhu} uses the vertex operator algebra formalism to organize this calculation in a convenient way, which makes some of the modular properties more manifest. 

\subsection{One-point functions}

The calculation of one-point functions in this method is fairly straightforward. For example, $I_3$ is the zero mode of the operator $(TT)(u)$, where the round brackets denote conformal normal ordering, and $u$ is a coordinate on the cylinder with compact real part. Given an expression for the thermal two-point function of $T$, $\langle T(u_1) T(u_2) \rangle$, we could implement the conformal normal-ordering by a contour integral, 
\begin{equation} \label{noint}
\langle (TT)(u_1) \rangle = \frac{1}{2 \pi i} \oint_{C} \frac{du_2}{u_2- u_1} \langle T(u_1) T(u_2) \rangle, 
\end{equation}
where $C$ is a contour encircling $u_1$, and then take the zero mode in $u_1$ by a further integral, 
\begin{equation}
\langle I_3 \rangle = \frac{1}{2 \pi i} \int_0^1 du_1 \oint_{C} \frac{du_2}{u_2- u_1} \langle T(u_1) T(u_2) \rangle, 
\end{equation}
where we adopt a convention that $u$ has period 1 in the real direction, $u \sim u+1$, to match the relation between plane and cylinder coordinates usually used in the vertex operator algebra formalism. 

Zhu's recursion relations provide an efficient method for computing the thermal $n$-point functions of the stress tensor and its derivatives. For the case of $I_3$, the relevant torus amplitude in the vertex operator algebra notation is $F( (\tilde \omega, z_1), (\tilde \omega, z_2) ; \tau)$, where we write $q$ as $q = e^{2\pi i\tau}$,  $\tilde \omega = \omega - k \Omega$ is the state corresponding to the stress tensor on the cylinder, and $z_i = e^{2\pi iu_i}$ is the corresponding planar coordinate. Here $\omega$ is the state corresponding to the stress tensor operator on the plane, and $\Omega$ is the global ground state, dual to the identity operator. Note that although we are interested in calculations on the cylinder, it is conventional in the vertex operator formalism to express correlators as functions of the coordinates on the plane. 

The calculation of this correlator is discussed in appendix \ref{recursion}; the result is 
\begin{eqnarray} \label{twopt}
F( (\tilde \omega, z_1), (\tilde \omega, z_2) ; \tau) &=& \partial^2 Z  + \mathcal  P_{2} \left( \frac{z_2}{z_1},q \right) \frac{2}{(2\pi i)^2} \partial Z +  \mathcal P_{4} \left( \frac{z_2}{z_1},q \right) \frac{12k}{(2\pi i)^4}  Z,
\end{eqnarray}
where $\mathcal P_{m}(x ,q)$ are the alternative Weierstrass functions introduced in \cite{zhu}, which are defined in appendix \ref{weier}. We will subsequently usually omit the $q$ argument for compactness of notation. This correlator is a function of $\frac{z_2}{z_1} = e^{2 \pi i (u_2-u_1)}$, as we would expect on general grounds; it just depends on the separation of the two operators on the cylinder. We see that it is a sum of derivatives of the partition function, whose coefficients are Weierstrass functions of ratios of the positions of the insertion points. This pattern continues for higher $n$-point functions-- the coefficients involve  functions of the different ratios $\frac{z_j}{z_i}$ for $i < j$.  The leading derivative term in \eqref{twopt} arises from taking the zero mode in each of the stress tensor operators, and rewriting the thermal expectation value of the zero mode in terms of derivatives, as in the previous section. Any stress tensor $n$-point function will similarly always start with a term with $n$ derivatives of the partition function. 

For $\langle I_3 \rangle$, the integral \eqref{noint} is just extracting the zero mode in the Laurent expansion in  $u_2-u_1$. This Laurent expansion is obtained by rewriting $\mathcal P_m(e^{2\pi i (u_2-u_1)},q)$ in terms of the Weierstrass function $\wp_m(u_2-u_1,\tau)$ as in \eqref{ptowp}. Using the expressions in appendix \ref{weier}, we find that 
\begin{equation}
\mathcal P_2(e^{2\pi i (u_2-u_1)},q) = \frac{1}{(u_2-u_1)^2}+  \frac{(2\pi)^2}{12} E_2 + \mathcal O(u_2-u_1), 
\end{equation}
\begin{equation}
\mathcal P_4(e^{2\pi i (u_2-u_1)},q)  =\frac{1}{(u_2-u_1)^4} +  \frac{(2\pi)^4}{720} E_4+ \mathcal O(u_2-u_1).
\end{equation}
Thus the result is
\begin{equation}
\langle I_3 \rangle = \partial^2  Z - \frac{E_2}{6} \partial Z  + \frac{k}{60} E_4 Z,
\end{equation}
in agreement with the previous results. In this calculation, it is a little clearer how the Eisenstein series appear, as the zero modes of the Weierstrass functions under the conformal normal ordering.

$I_5$ is the zero mode of the operator $J_6= (T (TT)) + (k+\frac{1}{6}) (2\pi)^2 (T' T')$, where again round brackets denote conformal normal ordering, and the derivatives are with respect to the cylinder coordinate $u$.\footnote{This expression is consistent with the one in \cite{Bazhanov:1994ft}, and we have checked explicitly that it leads to the correct expression for $I_5$, which commutes with $I_3$. The factor of $(2\pi)^2$ arises from our convention for the $u$ coordinate, which has period 1, contrary to the convention in \cite{Bazhanov:1994ft}, where it has period $2\pi$.} We can proceed by obtaining expressions for $\langle T(z_1) T(z_2) T(z_3) \rangle$ and $\langle T'(z_1) T'(z_2) \rangle$ from the recursion relation. We then normal order, letting $z_3 \to z_2$ and then $z_2 \to z_1$ in the first term, and $z_2 \to z_1$ in the second term. The result is the one-point function of the operator $\langle J_6(z_1) \rangle$, which is already independent of $z_1$ as expected. 

This gives a result
\begin{equation} \label{I5result}
\langle I_5 \rangle = \partial^3 Z - \frac{1}{2} E_2 \partial^2 Z + \left[ \left( \frac{k}{12} + \frac{1}{36} \right) E_4 + \frac{1}{24} E_2^2 \right] \partial Z - k \left( \frac{k}{63}  + \frac{1}{108} \right) E_6 Z. 
\end{equation}
Writing this in terms of modular covariant derivatives reproduces the result in \eqref{I5result}.  

Given an expression for the operator corresponding to $I_{2m-1}$, it is easy to extend this calculation to find an expression for $\langle I_{2m-1} \rangle$. Since there is some confusion in the literature as to the precise form of these operators, we have instead used the calculation backwards; for small $m$, we can take a generic form for the operator with arbitrary coefficients and evaluate the one-point function, and then fix the coefficients in the expression for the operator by requiring agreement with the expressions in section \ref{onept}. This calculation is described in appendix \ref{ops}. 

\subsection{Higher-point functions}

For higher-point functions, the calculation of the correlation functions of KdV charges from the  correlation functions of the stress tensors and their derivatives becomes more complicated. There are two issues; the conformal normal ordering to obtain a correlation function of the currents $J_{2m}$ from  the expression for the stress tensors becomes more involved, and we need to take  zero modes of the expression for the current correlation function with respect to the angular coordinates on the spatial circles to obtain the correlation functions of KdV charges.  The conformal normal ordering is relatively straightforward, while taking the zero modes is more troublesome. These issues are described in detail in appendix \ref{zero}, we will just illustrate the issues here. 

Let us consider $\langle I_3^2 \rangle$. Applying the recursion methods, we obtain a thermal four-point function  of the stress tensor,
\begin{equation}
F_4 = \partial^4 Z + C_3 \partial^3 Z + (C_2^1 k + C_2) \partial^2 Z + (C_1^1 k + C_1) \partial Z + (C_0^1 k + C_0) k Z,  
\end{equation} 
where the coefficients are functions of the ratios $\frac{z_j}{z_i}$ with $i <j$. These are now products of Weierstrass functions, and the related functions $g^i_k$ defined in appendix \ref{weier}; for example,
\begin{equation}
C_0^1 = \frac{9}{16 \pi^8} \left[ \mathcal P_4 \left( \frac{z_2}{z_1}, q \right)  \mathcal P_4 \left( \frac{z_4}{z_3}, q \right) + \mathcal P_4 \left( \frac{z_3}{z_1}, q \right) \mathcal P_4 \left( \frac{z_4}{z_2}, q \right) + \mathcal P_4 \left( \frac{z_4}{z_1}, q \right) \mathcal P_4 \left( \frac{z_3}{z_2}, q \right) \right]  .
 \end{equation}
In each term, a given $z_i$ appears in the denominator in the argument of at most one Weierstrass function. 

To calculate $\langle I_3^2 \rangle$ from this, we first need to conformal normal order the stress tensors in pairs, by taking $z_4 \to z_3$, $z_2 \to z_1$. For the  Weierstrass functions of $\frac{z_4}{z_3}$ or $\frac{z_2}{z_1}$, we need to make a Laurent expansion using the expression in terms of $\wp_m(u_4-u_3,\tau)$ or $\wp_m(u_2-u_1,\tau)$ in appendix \ref{weier} . For the Weierstrass functions of other arguments, we need to make a Taylor expansion, with coefficients which are functions of $\frac{z_3}{z_1}$. Subleading terms in the Taylor expansion may contribute if they multiply the singular terms in the Laurent expansion of functions of  $\frac{z_4}{z_3}$ or $\frac{z_2}{z_1}$. The conformal normal ordering is described in more detail in appendix \ref{zero}.

This normal ordering will give us an expression for $\langle (TT)(z_3) (TT)(z_1) \rangle$, again as a sum of $\partial^4 Z \ldots Z$, with coefficients involving functions of $\frac{z_3}{z_1}$. To evaluate $\langle I_3^2 \rangle$, we now want to integrate over the spatial circle, that is, we want to integrate over the phase of $z_3$ and $z_1$, which picks out the zero mode in the Laurent expansion of these functions in terms of $\frac{z_3}{z_1}$. For $\langle I_3^2 \rangle$, the functions are products of Weierstrass functions, and we can show that this zero mode can be rewritten in terms of derivatives of Eisenstein series, see appendix \ref{zero}. The result reproduces \eqref{I3sqcov}. Similarly, for $\langle I_3 I_5 \rangle$, we can obtain a result which reproduces \eqref{I3I5}. 

In evaluating $\langle I_5^2 \rangle$, the most difficult term is the six-point function of the stress tensor. Conformal normal ordering gives us a two-point function $\langle (T(TT))(z_1) (T(TT))(z_4) \rangle$. This is a function of the ratio $\frac{z_4}{z_1}$, which will be a sum of $\partial^6 Z \ldots Z$. There are three arguments in the original six-point function which become $z_1$ in this expression, so terms can involve products of up to three Weierstrass functions of $\frac{z_4}{z_1}$. The terms with products of two functions we can deal with analytically, but for the terms with three functions, the zero mode in the Laurent expansion involves a double sum, which we can't analytically simplify in terms of Eisenstein series.

Similarly, in $\langle I_3^3 \rangle$, conformal normal ordering the six-point function of the stress tensor, we obtain an expression for $\langle (TT)(z_1) (TT)(z_3) (TT)(z_5) \rangle$. Now we need to take the zero modes in the arguments $z_1, z_3, z_5$. In any given term, there are at most two Weierstrass functions where the denominator in the argument is $z_1$, and at most two where the denominator in the argument is $z_3$, so we need to deal with products of up to four Weierstrass functions. For the terms with up to three Weierstrass functions, the zero mode can be rewritten analytically in terms of derivatives of Eisenstein series. But when we have a product of four, we again encounter terms where the zero mode is a double sum. We have not found any explicit conversion of these double sums into expressions in terms of Eisenstein series. 

We are able to confirm that the results of the recursion relation agree with the previous results, by  expanding the double sums we encounter in a $q$ expansion and checking order by order that they agree with a specific combination of Eisenstein series. However, these issues with the zero modes make it difficult to extend this recursion relation approach to obtain higher order results in an algorithmic way.

\section{KdV charges for minimal models}
\label{relint}

In this section, we demonstrate that for the minimal models $(2n+1,2)$ for $n \geq 1$, the KdV charges $I_m$ with $m$ divisible by $2n+1$ vanish as operators. These non-unitary minimal models have central charges
\be
c=1-\frac{3 (2 (2n+1)^2}{2n+3}
\ee 
and have $n+1$ Virasoro primary operators with conformal dimensions
\be
h_s=-\frac{(s-1) (2 n-s+2)}{2( 2n+3)} \quad  1 \le s \le n+1.
\ee 

We have already seen evidence for this vanishing in the results for the thermal expectation values in the previous sections. The vanishing of $I_{2n+1}$ in the minimal model $(2n+3,2)$ can be understood using a relation to the null states of the theory  (following the work of \cite{FREUND1989243,DiFrancesco:1991anf}), which we will describe first. We will then give a general argument using integrability techniques from \cite{Bazhanov:1994ft} that all the KdV charges with spin divisible by $2n+1$ indeed vanish in the $(2n+3,2)$ model.

\subsection{Relation to null states}

This section reviews the discussions in \cite{FREUND1989243,DiFrancesco:1991anf}, which elucidate the relations between the vanishing of KdV charges and the null states in the minimal models. We first review the  arguments in the case of Yang-Lee model in detail, and then summarize the basic arguments for the general $(2n+3,2)$ model. The detailed arguments  can be found in \cite{FREUND1989243,DiFrancesco:1991anf} and references therein.

For the $(5,2)$ (i.e. Yang-Lee) theory, the vacuum has a level $4$ null state, corresponding to the operator
\be
\tilde{J}_4\equiv (T T)-\frac{3}{10} T''.
\ee 
The KdV charge $I_3$ is $I_3 = \oint  J_4$ with $J_4= (T T)$. Since the difference between $\tilde{J}_4$ and $J_4$ is a total derivative, the KdV charge can also be written as a zero mode of $\tilde J_4$, 
\be
I_3 = \oint  \tilde{J}_4.
\ee 
But $\tilde{J}_4$ should be set identically to zero as an operator statement since it corresponds to a null state. Hence $I_3=0$ as an operator statement in the Yang-Lee model.

For the general $(2n+3,2)$ theory, a similar argument can be constructed. The vacuum has null states at level $1,  2n+2$ and $ 2n+5$. We will call the operators corresponding to the null states at $2n+2$ and $2n+5$ $\tilde{J}_{2n+2}$ and $\tilde{J}_{2n+5}$ respectively. We would like to write $I_{2n+1}$ in terms of the zero mode of $\tilde{J}_{2n+2}$, finding the difference between $J_{2n+2}$ (whose zero mode is $I_{2n+1}$)  and  $\tilde{J}_{2n+2}$ as above. However, the explicit expression for the current $J_{2n+2}$ is complicated, so we proceed instead by calculating the commutator of $\tilde{J}_{2n+2}$ with $I_3$. Roughly speaking, one finds that $[\oint \tilde{J}_{2n+2} , I_3]\sim L_{-3} \tilde{J}_{2n+2}$. Realizing that $\tilde{J}_{2n+2}$ itself has a null state at level 3, which is precisely $\tilde{J}_{2n+5}\sim L_{-3} \tilde{J}_{2n+2}$, we see that the RHS of this commutator vanishes. Thus, the zero mode of $\tilde{J}_{2n+2}$ will indeed commute with $I_3$ in the minimal model, so it can be identified with  $I_{2n+1}$. Hence  $I_{2n+1} = 0$ in the  $(2n+3,2)$ minimal model. In principle, it seems that this argument should be generalisable to show that all KdV charges with spin divisible by $2n+1$ vanishes, but we did not pursue this further. 

\subsection{Relation to integrability}

We will now prove the general statement using the connection to integrability. There is a long-standing connection between the study of CFT in terms of the KdV charges and minimal models. In the work of Bazhanov-Lukyanov-Zamolodchikov (BLZ) \cite{Bazhanov:1994ft}, an approach to CFT based on quantum versions of the monodromy matrices of KdV theory was introduced.\footnote{See \cite{Negro:2016yuu} for a recent review lecture on the BLZ method.}  Although it is impossible to review the full details here, we will try to motivate the various objects which play a role in our story, while referring the interested readers to the vast literature of integrability for more details.\footnote{We refer the readers to \cite{Faddeev:1996iy,Beisert:2010jr,Beisert,Torrielli:2016ufi,Negro:2016yuu} and the references therein for a sampler of beginner's guides to integrability.}
 
A central object that we will need is the transfer matrix/operator $\Tj$. This arises from the consideration of the Lax matrices, $L$ and $M$. The Lax matrices are linear operators acting on some auxiliary internal space $V$ with some fixed but arbitrary dimension (not necessarily related to the dimensions of phase space). In a classical system with finitely many degrees of freedom (DOF), the entries of the Lax matrices are some functions of phase space while the defining property of the Lax pair is that the equation of motion of the system is equivalent to the Lax equation $dL/dt=[M,L]$. From this, one can show that all moments of $L$, i.e. $\tr_V L^k$ (with the trace being a matrix trace), are constants of motions and that they commute among themselves.\footnote{Actually, we meant that the moments are in {\it involution} among themselves, which means they Poisson commute. This is {\it not} always guaranteed. However, the statement of involutions is equivalent to the existence of a special matrix called the classical $r$-matrix, which satisfies the (classical) Yang-Baxter equation.} For a system with finitely many DOFs, we will only have finitely many conserved charges, and so some of these moments are redundant. 
Therefore, finding all eigenvalues of $L$ provides all conserved charges. There are two generalizations of this formalism. First of all, one has to extend this construction to the case of classical field theory, i.e. infinitely many DOFs.  To do so, in the case of finite DOFs, one constructs a generating polynomial $\sum x^k \tr L^k$ which captures all conserved charges. In the case of field theory, one would need infinitely many conserved charges. So instead of a generating polynomial, it is convenient to promote the generator of conserved charges to a generating {\it function} (of some parameter called the spectral parameter $\lambda$). 
The Lax matrices $L(\lambda)$ and $M(\lambda)$, still satisfying the Lax equation, now have entries which are functions of $\lambda$, and the moments of $L(\lambda)$ provide conserved charges.  We then expand near $\lambda\rightarrow\infty$, and the coefficients in the series expansion provide the (infinitely many) conserved charges. Finally, in a quantum field theory, we shall promote the Lax matrix to a  Lax ``matrix-operator''. What we meant by that  is that the Lax ``matrix-operator'' $\LL$ is first and foremost a matrix in the sense of linear operator acting on internal space $V$, just as in the classical case. It is also a linear operator acting on the Hilbert space. We shall, following the traditions in the field, simply call this the Lax operator. The transfer matrix (or the quantum monodromy matrices) is then given by $\tr_V \LL$, which still remains a quantum operator. In a CFT$_2$, we take  $V$ as the spin-$j$ (where $j$ is a positive half integer) representation of $SL(2,R)$. The transfer matrix is then defined as the trace of the Lax operator, $\Tj\equiv \tr_ j \LL$  and by definition  $\mathbf T_0(\lambda) = \mathbf I$  \cite{Bazhanov:1994ft,Bazhanov:1996aq}.

The non-trivial result of \cite{Bazhanov:1994ft} is that the $\Tj$ are not all independent, but instead they satisfy ``fusion relations'': 
\bea
\label{fusion}
\bT_j(q^{1/2} \lambda) \bT_j( q^{-1/2} \lambda) &=& 1 + \bT_{j-\frac{1}{2}}(\lambda) \bT_{j+\frac{1}{2}} (\lambda)
\eea where
\bea
 q&\equiv& e^{\pi i \beta^2} \quad, \quad\beta\equiv\sqrt{\frac{1-c}{24}}-\sqrt{\frac{25-c}{24}},
\eea
so $\bT_j(\lambda)$ for $j > 1/2$ are determined in terms of $ \bT(\lambda)\equiv \bT_{\frac{1}{2}}(\lambda) $. 
This basic operator also acts as a generating function for the KdV charges, in the sense that there is an asymptotic expansion as $\lambda\to +\infty$
%
\begin{equation}
\log \TT  \approx m \lambda^{1+\xi} - \sum_{k=1}^\infty C_k I_{2k-1} \lambda^{(1+\xi)(1-2k)},
\end{equation}
with 
\begin{equation}
m = 2\sqrt{\pi} \frac{\Gamma(\frac{1}{2} - \frac{\xi}{2})}{\Gamma(1-\frac{\xi}{2})} \Gamma\left( \frac{1}{1+\xi} \right)^{1+\xi}, 
\end{equation}
\begin{equation}
C_k = \frac{1+\xi}{k!} \sqrt{\pi} \left( \frac{\xi}{1+\xi} \right)^k \Gamma\left( \frac{1}{1+\xi} \right)^{(1+\xi)(1-2k)}  \frac{ \Gamma((1+\xi)(k-\frac{1}{2})) }{\Gamma(1+(k-\frac{1}{2}) \xi)}.
\end{equation}

For the $(2n+3,2)$ minimal models,  there are further relations:
 \be
 \mathbf T_{n + \frac{1}{2} - j}(\lambda) = \mathbf T_{j}(\lambda) \quad j=0,\half,1,\frac{3}{2},\ldots ,n. 
 \ee
  In particular, this implies that $\mathbf T_{n + \frac{1}{2}}(\lambda) = \mathbf I$, and it leads to a truncation of the fusion relations to finitely many relations. For the $(2n+3,2)$ minimal models, we have
\begin{equation}
\beta^2 = \frac{2}{2n+3}, \quad q =  e^{\frac{2\pi i}{2n+3}}, \quad \xi = \frac{2}{2n+1}. 
\end{equation} As a notation, we shall denote $t_j(\lambda)$ as the eigenvalues of $\Tj$ in the Fock space of the free field representation. 

The fusion relations are \bea
t_j(q^\half \lambda) t_j(q^{-\half} \lambda) &=& 1+ t_{j+\half}(\lambda)t_{j-\half}(\lambda)  \nonumber\\
t_0( \lambda) &=& t_{n+\half}(\lambda)=1\nonumber\\ 
t_{n+\half-j}(\lambda)&=&t_j(\lambda) \quad, \quad j=0,\half,1,\frac{3}{2},\ldots ,n. 
\eea while the asymptotic expansion of $t(\lambda)\equiv t_\half(\lambda)$ near $\lambda\rightarrow \infty$ is
\be
\log t
\approx m x
-\sum_{k=1}^\infty \frac{C_k}{
x^{2k-1}} I_{2k-1} \,.
\ee  where we have defined $x\equiv \lambda^{\frac{2n+3}{2n+1}}$.
We wish to show that  the KdV charges $I_{2k-1}$ vanish as operators whenever $2k-1$ is divisible by $2n+1$. We shall prove this for $n=1$ and $n=2$ before proceeding to the case of general $n$. 

\subsubsection{Example 1: Yang-Lee minimal model}
For $n=1$,  the only transfer matrices are $t_0,t_{\half},t_1$ and $t_{\frac{3}{2}}$.  The fusion relations then imply $t_0=t_{\frac{3}{2}}=1$ while $t_{1}=t_{\half}= t$. The only remaining constraint is  then
\bea
t(e^{\frac{2\pi i}{3}}x) t(e^{-\frac{2\pi i}{3}} x) &=& 1+ t(x)  \,.
\eea 
Now it is useful to introduce
\be
f(x)\equiv \log t(x)-m x, 
\ee such that the asymptotic expansion is
\be
f(x) \approx   
-\sum_{k=1}^\infty \frac{C_k}{
x^{2k-1}} I_{2k-1}.
\ee 
In terms of $f(x)$, the constraint is 
\be
f(e^{\frac{\pi i}{3}}x)
+f(e^{-\frac{\pi i}{3}}x)-f(x)
=
\log\left[ 1+ \frac{1}{t(x)}\right]
\ee Expanding near $x\rightarrow \infty$, the LHS is 
\be
f(e^{\frac{\pi i}{3}}x)
+f(e^{-\frac{\pi i}{3}}x)-f(x)\approx
-3\sum_{k\ge 1, 3|2k-1}^\infty \frac{C_k}{
x^{2k-1}} I_{2k-1} 
\ee where we need to sum over $k$ such that $2k-1$ is divisible by $3$. The RHS will vanish exponentially since  $t\approx e^{m x -O(1/x)}$. More explicitly, since the argument of the log is close to unity, we can use the taylor expansion of log to write
\be
\log (1+t^{-1})=\sum_{n=1}^\infty (-1)^{n+1} \frac{1}{n!} t^{-n} 
\approx e^{-m x}.
\ee
Thus we conclude that $I_{2k-1}$ vanishes if $3$ divides $2k-1$.

Note that in this example, we take $x\rightarrow \infty$ along real $x$. However, from \cite{Bazhanov:1996aq}, we know $t$ could have zeroes when $\arg x=\pm \pi /3$. The careful way to do the calculation is to factor out the zeroes and apply the fusion relations and asymptotic expressions on the remaining object. The readers are referred to \cite{Bazhanov:1996aq} for a more careful discussion.

\subsubsection{Example 2: $(7,2)$ model}

For $n=2$,  the fusion relations are
\bea
t_\half(e^{\frac{\pi i}{5} } x) t_\half(e^{-\frac{\pi i}{5} }x ) &=& 1+ t_{1}(x) \nonumber\\
t_1(e^{\frac{\pi i}{5} } x) t_1(e^{-\frac{\pi i}{5} } x) &=& 1+ t_{1}(x)t_{\half}(x)  \,.
\eea From \cite{Bazhanov:1996aq}, we also know the asymptotics of $t_j$. We define
\be
f(x)\equiv \log t_\half(x)-m x  \quad 
f_1(x) \equiv  \log t_1 - m \varphi x 
\ee where $\varphi=\exp \left(\frac{i \pi }{5}\right)+\exp \left(-\frac{1}{5} (i \pi )\right)=(1+\sqrt{5})/2$ is the golden mean, so the relations become
\bea
f(e^{\frac{\pi i}{5} } x)
+f(e^{-\frac{\pi i}{5} }x )
-\log\left[1+ \frac{1}{t_{1}(x)}\right]
 &=& f_1(x) \nonumber\\
f_1(e^{\frac{\pi i}{5} } x)+
f_1(e^{-\frac{\pi i}{5} } x) 
-f_{1}(x)-f(x)&=& \log\left[1+ \frac{1}{t_{1}(x)t_{\half}(x)}\right] \,.
  \nonumber\\
\eea  We eliminate $f_1$ using the first equation, so the second equation becomes
\bea
&&f(e^{2\frac{\pi i}{5} } x)
+f(e^{-2\frac{\pi i}{5} }x )
-f(e^{\frac{\pi i}{5} } x)
-f(e^{-\frac{\pi i}{5} }x )
+f(x)\nonumber\\
&=& \log\left[1+ \frac{1}{t_{1}(x)t_{\half}(x)}\right]-\log\left[1+ \frac{1}{t_{1}(x)}\right] 
+\log\left[1+ \frac{1}{t_{1}(e^{-\frac{\pi i}{5} }x)}\right]
+\log\left[1+ \frac{1}{t_{1}(e^{\frac{\pi i}{5} } x)}\right]
 \,.
  \nonumber\\
\eea Expanding near infinity, the LHS nicely becomes
\bea
&&f(e^{2\frac{\pi i}{5} } x)
+f(e^{-2\frac{\pi i}{5} }x )
-f(e^{\frac{\pi i}{5} } x)
-f(e^{-\frac{\pi i}{5} }x )
+f(x)\nonumber\\
&\approx & 5\sum_{k\ge 1, 5|2k-1}^\infty \frac{C_k}{
x^{2k-1}} I_{2k-1} \,.
\eea The RHS can be expanded and is vanishing since $t_1(x)$ and $t_\half(x)$ are exponentially vanishing.  Since
\be
t_1(x) \approx  e^{m \varphi x}
\ee even with the phases, for real $x$ the real part of the exponent is given by $\text{Re}[\varphi e^{\pm i\pi  /5}]=1+\cos \left(\frac{2 \pi }{5}\right)>0$. Thus we conclude that $I_{2k-1}$ vanishes if $5$ divides $2k-1$.

\subsubsection{General $n$}

 
Generalizing to arbitrary $n$ is now straightforward.  Writing the fusion relations in terms of the $f_j$ and eliminating all but $f_\half =f$, we have
\be
-f(x)+\sum_{s=1}^n (-1)^{n+1}\left[ f( Q^s x)+f(Q^{-s}x)\right]=\sum \log(\ldots)
\ee where $Q=e^{i\pi /(2n+1)}$. Expanding at infinity, the LHS becomes
\be
(-1)^n (2n+1) \sum_{k\ge1, 2n+1 |2k-1} \frac{C_k}{x^{2k-1}} I_{2k-1}
\ee while the RHS will be exponentially suppressed. Thus one concludes that $I_{2k-1}$ vanishes when $2n+1$ divides  $2k-1$.

As noted in the $n=1$ calculation,  we take $x\rightarrow \infty$ along real $x$ for the asymptotic expansion while the fusion relations require a  wedge around the real axis. From \cite{Bazhanov:1996aq}, we know that this fine up to the zeroes of $t$. The readers are referred to \cite{Bazhanov:1996aq} for a more careful discussion about how to factor out the zeroes of $t$ and apply the arguments to the remainder.

\section{Application: statistics of KdV charges}
\label{stats}

We have shown that the correlation functions of the KdV charges in a particular Virasoro module can be written as modular differential operators acting on the character. In this section we discuss an application of this result.  So far we have worked in canonical ensemble, where we turn on a potential $q=e^{-\beta}$ conjugate to energy.  However, for a Verma module (the generic Virasoro representation with $c>1$) the modular derivatives take a particularly simple form.  This allows us to extract results in the micro-canonical ensemble, and in particular allows us to write exact formulas for the moments of the KdV charges as a function of level.   The statistics of $I_1=L_0-k$ are trivial, since $I_1$ has the definite value $I_1 = h+n-k$ at level $n$.  The statistical distribution of the higher KdV charges, however, is non-trivial. 
 We illustrate this below by computing the mean and the variance of $I_3$ as a function of level.   

The character of a generic Virasoro representation is
\begin{equation}
\chi_{h}=\frac{q^{h-k}}{\prod_{n=1}^{\infty}(1-q^n)}\,.
\end{equation}
We can then compute
\begin{equation}
\partial \frac{1}{\prod_{n=1}^{\infty}(1-q^n)} =  \frac{1}{\prod_{n=1}^{\infty}(1-q^n)} \sum_{n=1}^\infty \frac{ n q^n}{1-q^n} = \frac{1}{\prod_{n=1}^{\infty}(1-q^n)}  \frac{1}{24} (1-E_2), 
\end{equation}
so
\begin{equation} \label{expp}
\partial \chi_{h}= \left[ h-k + \frac{1}{24} - \frac{E_2}{24} \right] \chi_h = \left[\tilde h -  \frac{E_2}{24} \right] \chi_h,
\end{equation}
where we have introduced the shifted level $\tilde h = h-k+\frac{1}{24}$ to simplify subsequent formulae. 

For example, we have  
\begin{equation} \label{I3exp}
\langle I_3 \rangle_h = \left[ \partial^2 - \frac{E_2}{6} \partial + \frac{k}{60} E_4 \right] \chi_{h} = \left[ \tilde h^2 - \frac{1}{4} \tilde h E_2 + \frac{E_2^2}{192} + \left( \frac{k}{60} + \frac{1}{288} \right) E_4 \right] \chi_h. 
\end{equation}

These results are for the generic character; we can obtain similar results for characters with null states by taking differences. For example the vacuum character $\chi_{vac} = \chi_{h=0} - \chi_{h=1}$, so 
\begin{equation}
\partial \chi_{vac}= \left[ -k + \frac{1}{24} - \frac{E_2}{24} \right] \chi_{vac} - \chi_1  = \left[-k +\frac{1}{24}  -  \frac{E_2}{24}  - \frac{q}{1-q} \right] \chi_{vac},
\end{equation}
which gives for example
\begin{equation}
\langle I_3 \rangle_{vac} = \left[ \left( - k + \frac{1}{24} \right)^2  - \frac{1}{4} \left( - k + \frac{1}{24} \right) E_2 + \frac{E_2^2}{192} + \left( \frac{k}{60} + \frac{1}{288} \right) E_4 + \frac{q}{1-q} \left( 2k + \frac{E_2}{4} - \frac{13}{12} \right) \right] \chi_{vac}. 
\end{equation}

If we were to study the KdV charges in a full CFT we would sum the above expressions over all representations appearing in the theory. In this case it is generally more useful to keep the expression for the KdV correlation function as a differential operator, as this does not include any dependence on the conformal dimension $h$ of the character, so the thermal correlation function is given by the same differential operator acting on the partition function. 

However, if our interest is in an individual character, the above replacement is a significant simplification. 
It can be used to study the statistics of the KdV charges on the $P(n)$ states at level $n$ in the Virasoro representation In section \ref{qexp}, we used knowledge of the value of the KdV charges at low level to fix the form of the differential operator in the one-point functions. We can also use this relation in the inverse sense: given the differential operator, we can act with it on the character and expand the result in a $q$-expansion to obtain an expression for the average value of the KdV charge at each level. This approach was already used implicitly in section \ref{qexp}, where we used the differential operator form of the one-point functions to determine the values of the KdV charges at level one, which was then used as input in the calculation of two-point functions. 
But given \eqref{expp}, we can now proceed more systematically. 

For example, expanding \eqref{I3exp} in a $q$-expansion gives
\begin{equation} \label{I3n}
\langle I_3 \rangle_n = \left[ \tilde h^2 + 6 \left( n- \frac{1}{24} \right) \tilde h  + \frac{(48k+25)}{5} \left(n-\frac{1}{24} \right)^2 \right] P(n) - \frac{4}{5}(6k+5)  \sum_{l=1}^n l \sigma_1(l) P(n-l),  
\end{equation}
where $P(n)$ is the number of partitions of $n$ and the divisor functions  $\sigma_i(n) = \sum_{d | n} d^i$ appear because they are the expansion coefficients of the Eisenstein series expansions: $E_2 = 1- 24 \sum_{n=1}^\infty \sigma_1(n) q^n$ and $E_4 = 1 + 240 \sum_{n=1}^\infty \sigma_3(n) q^n$.  In writing this expression we have used some properties of the divisor functions. 

This expression gives the sum of $I_3$ over all the states at level $n$, that is the trace of the matrix elements between states in this level. To compute the average value ${\overline I_3}(n)$ of the KdV charge at level $n$ we must divide by the number of states, to get:
\begin{equation} \label{I3n}
{\overline I_3}(n) \equiv \frac{\langle I_3 \rangle_n}{P(n)} =  \left[ \tilde h^2 + 6 \left( n- \frac{1}{24} \right) \tilde h  + \frac{(48k+25)}{5} \left(n-\frac{1}{24} \right)^2 \right] - \frac{4}{5}(6k+5)  \sum_{l=1}^n l \sigma_1(l)  \frac{P(n-l)}{P(n)}.
\end{equation}

Applying similar analysis to the other one-point functions gives explicit expressions for the average value of the KdV charges on the states at a given level in the Virasoro representation. Similarly studying the correlation functions will give higher moments and correlations in the distribution of KdV charges on the descendant states at a given level; this is interesting information characterizing the action of the KdV charges within a given Virasoro representation. 

For example, 
\begin{equation}
\begin{aligned} \label{I3sqexp}
\langle I_3^2 \rangle_h &= \left[ \tilde h^4 - \frac{1}{2} \tilde h^3 E_2  - \frac{5}{96} \tilde h^2 E_2^2 + \frac{ 24k+95}{720} \tilde h^2 E_4 + \frac{25}{1152} \tilde h E_2^3 - \frac{(7+24k)}{360} \tilde h E_6+ \frac{168k-19}{2880} \tilde h E_2 E_4 \right.  \\
&\left. -\frac{5}{12288}E_2^4-\frac{19+120k}{46080}E_2^2E_4+\frac{(19536 k (12 k+5)+12425)}{14515200}E_4^2+\frac{(7-24 k (120 k+29))}{181440}E_2 E_6
\right] \chi_h,
\end{aligned}
\end{equation}
which gives, after some manipulation, 
\begin{equation}
\begin{aligned}
\langle I_3^2 \rangle_n &= \left[ \tilde h^2 + 6 \left( n- \frac{1}{24} \right) \tilde h  + \frac{(48k+25)}{5} \left(n-\frac{1}{24} \right)^2 \right]^2 P(n)  +\sum_{l=1}^n \bigg\{ - \frac{8}{5} \tilde h^2 (6k+5) l \sigma_1 \\ & -   \tilde h\left[2 (6k-7) l\sigma_3(l)+\frac{12 }{5}(18k+25)l^2\sigma_1(l)\right] + \left( \frac{52}{5} k^2 +\frac{197}{60} k + \frac{329}{360} \right)   l\sigma_5(l) \\ & - \left( \frac{216}{5} k^2 + \frac{99}{2}k + \frac{333}{20} \right)  l^2\sigma_3(l)- \left( \frac{144}{5} k^2 + 30k +5 \right) l^3\sigma_1(l) \bigg\}P(n-l). 
\end{aligned}
\end{equation}
%
This allows us to compute the variance $\Delta I_3(n)$ of the KdV charge as a function of level:
\begin{equation}
\begin{aligned}
\Big(\Delta I_3(n)\Big)^2 &\equiv \frac{\langle I_3^2 \rangle_n}{P(n)} - \frac{\langle I_3 \rangle_n^2}{P(n)^2}  \\
&= \sum_{l=1}^n \bigg\{   \tilde h\left[-2 (6k-7) l\sigma_3(l)-\frac{12 }{5}(18k+25)l^2\sigma_1(l) + \frac{48}{5} (6k+5) \left( n - \frac{1}{24} \right) l \sigma_1(l) \right] \\ &+ \left( \frac{52}{5} k^2 +\frac{197}{60} k + \frac{329}{360} \right)   l\sigma_5(l)   - \left( \frac{216}{5} k^2 + \frac{99}{2}k + \frac{333}{20} \right)  l^2\sigma_3(l)- \left( \frac{144}{5} k^2 + 30k +5 \right) l^3\sigma_1(l)  \\&+\frac{8}{25}(6k+5)(48k+25) \left( n - \frac{1}{24} \right)^2 l \sigma_1(l)  \bigg\} \frac{P(n-l)}{P(n)}  - \frac{16}{25} (6k+5)^2 \left( \sum_{l=1}^n l \sigma_1(l) \frac{P(n-l)}{P(n)} \right)^2.
\end{aligned}
\end{equation}
Again, identities of the divisor function have lead to significant cancellations.  

We will just make a two brief comments about the interpretation of these results.  The first is that the mean value of ${\overline I_3}(n)$ increases quadratically in both $h$ and $n$ when either of these are taken to be large.  This is not a surprise, since $I_3$ is quadratic in the stress tensor.  Then second (and less obvious) statement is that the normalized variance $\Delta I_3(n) / {\overline I_3}(n)$ vanishes at large $n$.  This means that as we increase the level, the distribution of KdV charges becomes very sharply peaked around its average value ${\overline I_3}(n)$.  This is illustrated in Figure 1 for typical values of $h$ and $k$.  We will study the statistics of KdV charges at high level in more detail in \cite{paper2}.

 \begin{figure}
 \centering
	\begin{subfigure}[b]{.475\textwidth}\centering
	 \includegraphics[width = \textwidth]{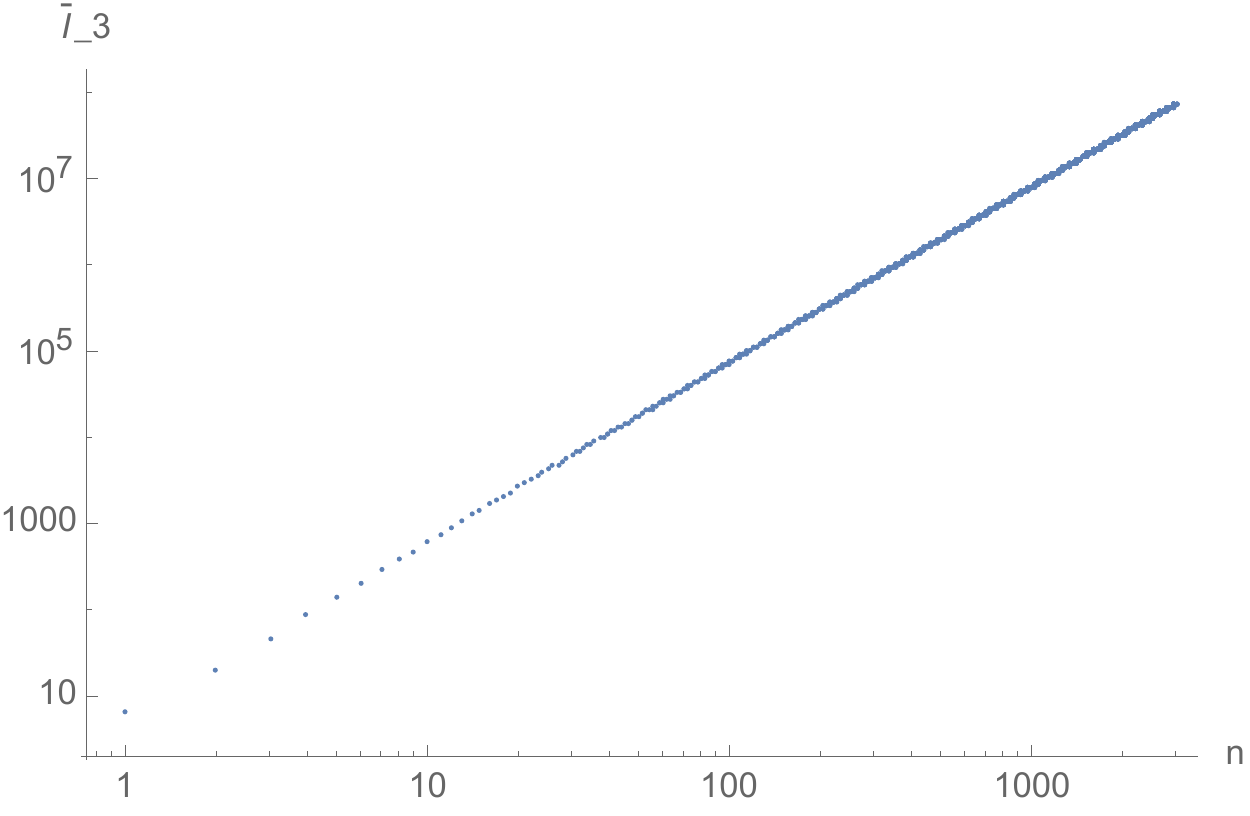}
	\end{subfigure}
	\begin{subfigure}[b]{.475\textwidth}\centering
	 \includegraphics[width = \textwidth]{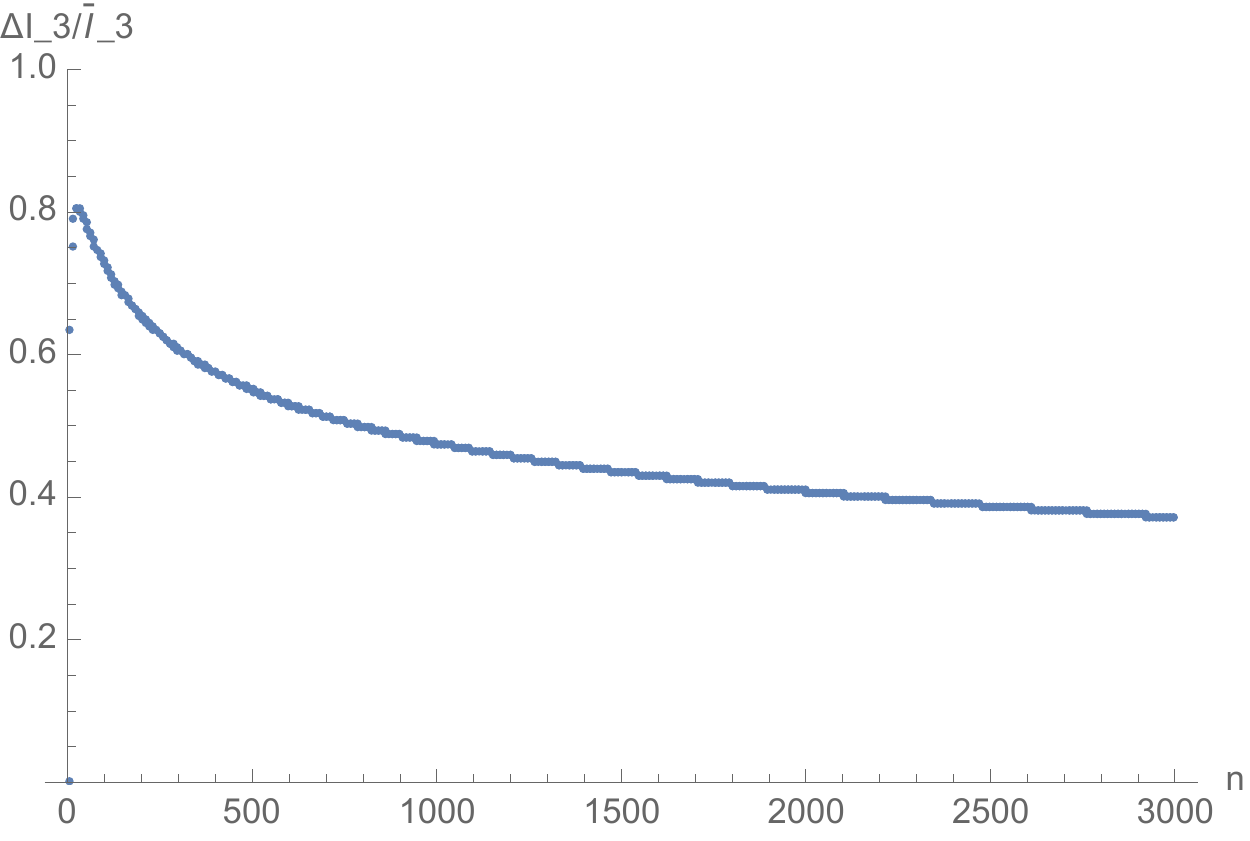}
	\end{subfigure}
\caption{\label{fig1}The mean value ${\overline I_3}(n)$ and the normalized variance $\Delta I_3(n)/ {\overline I_3}(n)$ of the KdV charge $I_3$ as a function of level $n$.  These data are plotted for the Verma module with dimension $h=1$ and central charge $c=6$.}
 \end{figure}




\section*{Acknowledgements}
We thank Vladimir Bazhanov, Anatoly Dymarsky, Jan de Boer, Jeff Harvey, Nima Lashkari, Jan Manschot and Sunil Mukhi for discussions. G. N. is supported by Simons Foundation Grant to HMI under the program ``Targeted Grants to Institutes''. SFR is supported by STFC under consolidated grants ST/L000407/1 and ST/P000371/1. I.T. is supported by the Alexander S. Onassis Foundation under the contract F ZM 086-1.
A.M. acknowledges the support of the Natural Sciences and Engineering Research Council of Canada (NSERC), funding reference number SAPIN/00032-2015.
This work was supported in part by a grant from the Simons Foundation (385602, A.M.).
  
\appendix

\section{Eisenstein series \& Weierstrass functions}
\label{weier}

In our discussion we use the Eisenstein series $E_{2k}(\tau)$, whose expansion is 
\begin{equation}
E_{2k}(\tau) = 1 + \frac{2}{\zeta(1-2k)} \sum_{n=1}^\infty \frac{n^{2k-1} q^n}{1-q^n}.  
\end{equation}
It is useful to record the formula
\begin{equation} \label{derE} 
\partial E_{2k}(\tau) = \frac{2}{\zeta(1-2k)} \sum_{n=1}^\infty \frac{n^{2k} q^n}{(1-q^n)^2}.  
\end{equation}
We also recall the Ramanujan identities 
\begin{equation}
\partial E_{2} = \frac{1}{12} ( E_2^2 - E_4), \quad \partial E_4 = \frac{1}{3} (E_2 E_4 - E_6), \quad \partial E_6 = \frac{1}{2} (E_2 E_6 - E_4^2). 
\end{equation}
The Eisenstein series $E_{2k}(\tau)$ for $k \geq 2$ are modular forms of weight $2k$. Products of $E_4$ and $E_6$ provide a basis for the space of modular forms, so all the higher Eisenstein series can be written as polynomials in $E_4, E_6$. $E_2$ is a quasimodular form, transforming as 
\begin{equation}
E_2(- \tau^{-1}) = \tau^2 E_2(\tau) - \frac{6i}{\pi} \tau
\end{equation}
under the modular $S$ transformation. 

In the recursion relations, we use the Weierstrass functions defined in \cite{zhu}, 
\begin{equation}
\mathcal P_k(x,q) = \frac{(2\pi i)^k}{(k-1)!} \sum_{n\neq 0} \frac{n^{k-1} x^n}{(1-q^n)}, 
\end{equation}
defined for $k \geq 1$, which converge for $|q| < |x| <1$. For compactness, we often omit the $q$ argument in $\mathcal P_k$. Derivatives with respect to $x$ can be simplified using the relation 
\begin{equation}
x \partial_x \mathcal P_k (x,q) = \frac{k}{2\pi i} \mathcal P_{k+1} (x,q). 
\end{equation}
If one takes the argument $x = q_z= e^{2\pi i z}$, this becomes simply
\begin{equation} \label{xderP}
\partial_z \mathcal P_k (e^{2\pi i z} ,q) = k \mathcal P_{k+1} (e^{2\pi i z},q). 
\end{equation}
In the recursion of \cite{Gaberdiel:2012yb}, we also encounter functions 
\begin{equation} \label{gi}
g^i_k(x,q) = \frac{(2\pi i)^{k+i}}{(k-1)!} \sum_{n\neq 0} n^{k-i-1} x^n \partial^i \frac{1}{(1-q^n)}, 
\end{equation}
defined for $k \geq 1$, where we use the notation $\partial = q \partial_q$. For $k >i$, these can be rewritten as $q$ derivatives of the $\mathcal P_k$, 
\begin{equation}
g^i_k(x,q) = (2 \pi i)^{2i} \frac{(k-i-1)!}{(k-1)!} \partial^i \mathcal P_{k-i}(x,q).  
\end{equation}

The $\mathcal P_k(x,q)$ are related to the familiar Weierstrass functions $\wp(z, \tau)$ by  \cite{zhu} 
\begin{equation} \label{ptowp} 
\mathcal P_1(q_z, q) = - \wp_1(z,\tau) + G_2(\tau) z - i \pi, 
\end{equation}
where $q_z = e^{2\pi i z}$, $q = e^{2 \pi i \tau}$, and $G_2(\tau)$ is an Eisenstein series, see below. Using the relation \eqref{xderP} and the similar relation $\partial_z \wp_k (z,\tau) = -k \wp_{k+1}(z,\tau)$, this determines the relation for all higher $k$, 
\begin{equation}
\mathcal P_2(q_z, q) = \wp_2(z,\tau) + G_2(\tau), 
\end{equation}
\begin{equation}
\mathcal P_k(q_z, q) = (-1)^k \wp_k(z,\tau). 
\end{equation}
Note that the arguments of $\mathcal P_k$ are exponentials of the arguments of $\wp_k$. 
To evaluate the behaviour of $\mathcal P_k(x,q)$ near $x=1$, we will need the Laurent expansion of $\wp_k(z,\tau)$ for small $z$, 
\begin{equation} \label{lwp} 
\wp_k(z, \tau) = \frac{1}{z^k} + (-1)^k \sum_{n=1}^\infty \left( \begin{array}{c} 2n+1 \\ k-1 \end{array} \right) G_{2n+2}(\tau) z^{2n+2-k}.  
\end{equation}
Note that the combinatorial factor in the sum vanishes for $2n+2-k < 0$, so the only singular term in the expansion is the explicit $\frac{1}{z^k}$. The $G_{2k}(\tau)$ are Eisenstein series, but with a different normalization,  $G_{2k}(\tau) = 2 \zeta(2k) E_{2k}(\tau)$.

\section{Hypergeometric form of differential operators}
\label{hyp}

The differential operators appearing in the thermal expectation values may appear exotic, but -- at least for the one point functions $\langle I_{2m-1}\rangle$ -- they can be put in a more familiar hypergeometric form by a change of variables.

The essential point is the following.  In the above parameterization, we have represented the KdV charges as derivatives with respect to $q=e^{2\pi i \tau}$ where $\tau$ is the standard modular parameter of the torus defined by the identifications $z\sim z+1 \sim z+\tau$.  However, we can alternatively parameterize a torus as an algebraic curve, as the set of solutions to the equation
\be
\label{elliptic}
y^2 = {w (w-1) \over w-\lambda}
\ee
Here we view the torus as a double cover of the Riemann sphere $\C^*$ parameterized by the $w$ coordinate, with branch points at $w=0,\lambda, 1$ and $\infty$.  The two sheets of the cover are the two roots of equation (\ref{elliptic}).   Indeed, any two-fold cover of $\C^*$ which is branched over four points can put in this form, with $\lambda$ being equal to the cross-ration of the four points.  The parameter $\lambda$ now sweeps out the torus moduli space, and is related to the usual torus modular parameter by
\be
\lambda(\tau)= \left(\frac{\theta_2(\tau)}{\theta_3(\tau)}\right)^4
\ee  
This is the usual modular lambda function.

The important point is that the modular $SL(2,\Z)$ symmetries of the torus now act on $\lambda$ in a very simple way, as the anharmonic transformations generated by
\be\label{anharmonic}
\lambda \to 1-\lambda ~~~\&~~~\lambda \to 1/\lambda~.
\ee
This means that, when written in terms of the parameter $\lambda$, all of the modular properties of our differential operators will be replaced by transformation properties with respect to (\ref{anharmonic}). 

It is straightforward to write all of the ingredients of our differential operators in terms of $\lambda$, suing
\bea
q\partial_q&=&\frac{\theta_3^4}{2}\left(1-\lambda\right)\lambda\partial_\lambda\\
E_2&=&\theta_3^4\left(1-2\lambda\right)+24\frac{\theta_3^4}{2}\left(1-\lambda\right)\lambda\left(\frac{\partial_\lambda\theta_3}{\theta_3}\right) \\
E_4&=&\theta_3^8\left(1-\lambda(1-\lambda)\right)\\
E_6&=&\frac{\theta_3^{12}}{2}\left(2\lambda-1\right)\left(\lambda+1\right)\left(\lambda-2\right)~.
\eea
When we compute the expectation value of a single power of a KdV charge all of the factors of $\left(\frac{\partial_\lambda\theta_3}{\theta_3}\right)$ will cancel among each other; this is a consequence of the fact that $\langle I_{2m-1}\rangle$ is a modular form.  For example,
\be
\begin{aligned}
\langle I_1 \rangle &=\frac{\theta_3^4}{2}\bigg[\left(1-\lambda\right)\lambda \ \partial_\lambda\bigg] Z\\
\langle I_3 \rangle &=\frac{\theta_3^8}{4}\left[(1-\lambda)^2\lambda^2\partial^2_{\lambda}-\frac{2}{3}\lambda(1-\lambda)(2\lambda-1)\partial_{\lambda}+\frac{k}{15}\left(1-\lambda(1-\lambda)\right)\right] Z\\
\langle I_5 \rangle&=\frac{\theta_3^{12}}{8}\bigg[(1-\lambda)^3\lambda^3\partial_\lambda^3-2\lambda^2(1-\lambda)^2(2\lambda-1)\partial_\lambda^2\\
&+\frac{1}{18}\lambda(1-\lambda)\big[(6k+1)(1-\lambda(1-\lambda))-40\lambda(1-\lambda)+4\big]\partial_\lambda
-\frac{k(12k+7)}{189}(\lambda-2)(\lambda+1)(2\lambda-1)\bigg] Z
\end{aligned}
\ee
It is then a straightforward exercise to put these in the standard form of a hypergeometric differential equation with regular singular points at $\lambda=0,1,\infty$.  For example, in terms of the variable ${\tilde Z} = \left(\lambda (1-\lambda)\right)^{\left(1+\sqrt{1-{12 k \over 5}}\right)/6}Z$ the differential operator for $\langle I_3\rangle$ is the usual second order hypergeometric differential operator with parameters
\be
a=\frac{1}{2} + \frac{\sqrt{1-{12 k \over 5}}}{6},~~~~~
b=\frac{1}{2} + \frac{\sqrt{1-{12 k \over 5}}}{2},~~~~~
c=1 + \frac{\sqrt{1-{12 k \over 5}}}{3}~.
\ee

For quasi-modular forms like $\langle I_3^2\rangle$ the results are rather more complicated, and cannot be written as hypergeometric differential operators.  For example,
\be
\begin{aligned}
&\langle{I_3^2}\rangle=\frac{\theta_3^{16}}{16}\bigg\{\bigg[(\lambda -1)^4 \lambda ^4 \partial_{\lambda}^4+\frac{16}{3} (\lambda -1)^3 (2 \lambda -1) \lambda ^3 \partial_{\lambda}^3\\
& \ \ \ \ \ \ \ \ \ \ \ \ \ \ \ \ \ \ \ +\frac{2}{45} (\lambda -1)^2 \lambda ^2 \left[3 k ((\lambda -1) \lambda +1)+565 (\lambda -1) \lambda +115\right] \partial_{\lambda}^2\\
& \ \ \ \ \ \ \ \ \ \ \ \ \ \ \ \ \ \ \  +\frac{1}{90} (\lambda -1) (2 \lambda -1) \lambda  \left[k (80 (\lambda -1) \lambda +8)+491 (\lambda -1) \lambda +40\right] \partial_{\lambda}\\
&\ \ \ \ \ \ \ \ \ \ \ \ \ \ \ \ \ \ \ +\frac{k \left[250 \lambda  \left(2 (\lambda -2) \lambda ^2+\lambda +1\right)+k ((\lambda -1) \lambda  (1207 (\lambda -1) \lambda -586)+7)\right]}{1575}\bigg]\\
& \ \ \ \ \ \ \ \  +\left(\frac{\partial_{\lambda}\theta_{3}}{\theta_3}\right)\bigg[16 (\lambda -1)^4 \lambda ^4 \partial_{\lambda}^3+32 (\lambda -1)^3 (2 \lambda -1) \lambda ^3 \partial_{\lambda}^2\\
& \ \ \ \ \ \ \ \ \ \ \ \ \ \ \ \ \ \ \ \ \ \ \ \  +\frac{4(\lambda -1)^2 \lambda ^2}{15}  \left[24 k ((\lambda -1) \lambda +1)+137 (\lambda -1) \lambda +17\right] \partial_{\lambda}\\
& \ \ \ \ \ \ \ \ \ \ \ \ \ \ \ \ \ \ \ \ \ \ \  \  +\frac{8}{63} k (12 k+5) (\lambda -2) (\lambda -1) (\lambda +1) (2 \lambda -1) \lambda\bigg]\bigg\}Z~.
\end{aligned}
\ee

\section{Thermal expectation values in $q$ expansion}
\label{qexpapp}

Here we give the full results on the determination of the differential operators using the data from the $q$-expansion. 

\subsection{One-point functions} 

We can write 
\begin{equation}
\langle I_{2m-1} \rangle_h = q^{h-k} [I_{2m-1}^{L=0}(h,k) + I_{2m-1}^{L=1}(h,k) q + \ldots], 
\end{equation}
where $I_{2m-1}^{L=0}$ is the value of the KdV charge on the primary, and $I_{2m-1}^{L=1}$ is its value on the unique state at level one. Knowledge of $I_{2m-1}^{L=0}$ will fix one coefficient at each order in derivatives in the differential operator determining the one-point function. The values of KdV charges on primary states were tabulated up to $I_{15}$ in appendix B of \cite{Bazhanov:1996aq}. Using that data allows us to fix 
\begin{equation}
\langle I_3 \rangle_h = [ D^2 + \frac{k}{60} E_4] \chi_h,
\end{equation}
\begin{equation} 
\langle I_5 \rangle_h = [D^3  + \frac{(6k+1)}{72} E_4 D  - \frac{k(12 k+7)}{756} E_6] \chi_h,   
\end{equation}
\begin{equation} \label{I7}
\langle I_7 \rangle_h = [D^4  + \frac{(21k+8)}{90} E_4 D^2  - \frac{(288 k^2 +288 k + 37)}{3240} E_6 D + \frac{k(576k^2 + 618 k+175)}{21600} E_4^2] \chi_h,   
\end{equation}
\begin{multline} \label{I9}
\langle I_9 \rangle_h = [D^5 + \frac{(18k+11)}{36} E_4 D^3  - \frac{(48 k^2 + 68 k+ 17)}{168} E_6 D^2 \\ + \frac{(15552 k^3+  26082 k^2 +12843 k +1375  )}{90720} E_4^2 D \\ - \frac{k(3456 k^3 +  5580 k^2 + 2835 k + 539)}{49896} E_4 E_6] \chi_h.   
\end{multline}

For higher $m$, this calculation leaves undetermined coefficients in $\langle I_{2m-1} \rangle$, as there are now combinations which vanish at leading order in $q$. We can obtain some further results by using the fact that for the ground state, $h=0$, there is no state at level 1, $\chi_0 = q^{-k} (1 + q^2 + \ldots)$, so for $h=0$ we must have $I_{2m-1}^{L=1} =0 $. This provides one extra constraint on the parameters, and allows us to determine
\begin{multline} \label{I11}
\langle I_{11} \rangle_h = [D^6  + \frac{(33k + 28)}{36} E_4 D^4  -\frac{(1056 k^2 + 1936 k + 707)}{1512} E_6 D^3 \\ + \frac{11(1728 k^3+3942 k^2+2749 k+525)}{30240} E_4^2 D^2 \\ - \frac{(46080 k^4+107232 k^3+85392 k^2 + 28343 k + 2625)}{90720} E_4 E_6 D \\ + \frac{k(11612160 k^4+25707456 k^3 +19201338 k^2 +6404625 k + 925925)}{70761600} E_4^3\\ + \frac{k (345600 k^4 + 763632 k^3 + 571116 k^2 +192409 k +  28028 )}{3714984} E_6^2] \chi_h   
\end{multline}
and 
\begin{multline} \label{I13} 
\langle I_{13} \rangle_h = [D^7  +\frac{7(78 k+85)}{360}  E_4 D^5  - \frac{(936 k^2+2106 k+1003)}{648} E_6 D^4 \\ + \frac{(7488 k^3+21606 k^2 +19461 k +5198)}{4320} E_4^2 D^3 \\ -  \frac{(5391360 k^4+16387488 k^3+17595864 k^2+8127678 k+1257419)}{2566080}  E_4 E_6 D^2 \\+\frac{( 23224320 k^5+69987456 k^4 + 76420116 k^3 +39962904 k^2 + 10119985 k +829735)}{17107200}  E_4^3 D \\+ \frac{( 20736000 k^5+61871040 k^4 +66871152 k^3 + 34825248 k^2 +  8822779 k +722939)}{26943840} E_6^2 D \\-\frac{k(6635520 k^5+18959616 k^4 +19130472 k^3 +9228102 k^2 + 2329777 k + 275275   )}{5132160}  E_4^2 E_6 ] \chi_h   
\end{multline}
To go further, we would need more data, for example, from evaluating $ I_{2m-1}^{L=1}$ at general $h$.

One can check that the values in minimal models vanish as advertised:
\begin{itemize}
\item In the Yang-Lee $(5,2)$ minimal model, $k= - \frac{11}{60}$, and there are two primaries: the vacuum with $h=0$ and a scalar of weight $h = -\frac{1}{5}$. The vacuum representation has a null vector at level 4. In \cite{Gaberdiel:2008pr}, this null state was shown to imply a differential equation for the characters of this theory, 
\begin{equation}
[D^2 - \frac{11}{3600} E_4] \chi_{0, -\frac{1}{5}} = 0. 
\end{equation}
This is precisely our differential operator $\langle I_3 \rangle$ for $k = -\frac{11}{60}$, so we see $\langle I_3 \rangle_h=0$ in the Yang-Lee model. One can also verify that  $\langle I_9 \rangle_h = 0$; this is most explicitly seen by using the replacement $D^2 \chi_h = \langle I_3 \rangle_h - \frac{k}{60} E_4 \chi_h$ to rewrite 
\begin{multline} \label{I9b}
\langle I_9 \rangle_h = D^3 \langle I_3 \rangle_h +  \frac{(87 k+55)}{180} E_4 D \langle I_3 \rangle_h   -  \frac{(240 k^2 + 326 k+85)}{840} E_6 \langle I_3 \rangle_h \\ + (11+60 k)  \left[\frac{(48 k + 25)(27 k+25)}{453600}  E_4^2 D - k \frac{(120k+101)(48k+25)}{4989600}  E_4 E_6 \right] \chi_h.   
\end{multline}
\item The $(7,2)$ minimal model has $k = -\frac{17}{42}$. The differential operator in \eqref{I5cov} for this value of $k$ reduces to the differential operator which annihilates the characters of the $(7,2)$ minimal model \cite{Leitner:2017}. 
\item Similarly \eqref{I7} reduces to the operator which annihilates the characters of the  $(9,2)$ minimal model for $k= -\frac{23}{36}$, \eqref{I9b} reduces to the operator which annihilates the characters of the  $(11,2)$ minimal model for $k= - \frac{29}{33}$, and  \eqref{I11} reduces to the operator which annihilates the characters of the  $(13,2)$ minimal model for $k= - \frac{175}{156}$. 
\end{itemize}

\subsection{Two-point functions}

As discussed in section \ref{qexp}, the differential operators for two-point functions are determined by writing 
\begin{equation}
\langle I_{2m-1} I_{2n-1}  \rangle_h = q^{h-k} [ I_{2m-1}^{L=0} I_{2n-1}^{L=0} +  q I_{2m-1}^{L=1} I_{2n-1}^{L=1}  + q^2 \sum_{i=1}^2 I_{2m-1}^{i,L=2} I_{2n-1}^{i, L=2}+ \ldots ], 
\end{equation}
and using the known values of the KdV charges on the primaries, and the values of the KdV charges on the level 1 states, which can be obtained from the differential operators for the one-point functions we worked out in the previous subsection, to constrain the form of the differential operator for the products. 

This determines
  \bea
   \langle  I_3^2 \rangle
&=& \left[
D^4 +\frac{1}{90} (3 k+5)E_4 D^2 -\frac{72k+11}{1080} E_6 D +\frac{k (1221 k+500)}{75600} E_4^2
\right]Z
\nonumber\\
&&
+E_2 \left[\frac{2}{3} D^3 + \frac{1}{1080} (72 k+11) E_4 D - \frac{1}{756} k (12 k+5)E_6 \right] Z\,,
   \eea
and
    \bea
   \langle  I_3 I_5 \rangle
&=& D^5 Z 
+\left(  \frac{k}{10}+\frac{13}{72}\right)   E_4  D^3 Z  \nonumber\\
&&
-\left(\frac{k^2}{63} + \frac{7k}{27}+\frac{19}{216}\right)E_6 D^2 Z 
+\left(\frac{647 k^2}{5040}+\frac{3047 k}{30240}+\frac{55}{4536}\right)E_4^2 D Z \nonumber\\
&&
-\left(\frac{169 k^3}{3780} + \frac{29 k^2}{810}+\frac{37 k}{4320}\right) E_4 E_6  Z \nonumber\\
&& + E_2 \left[   D^4 Z 
+ \left(\frac{5  k}{18}+\frac{5}{54} \right) E_4 D^2 Z 
-\left(\frac{8 k^2}{63}+\frac{19  k}{189}+\frac{55 }{4536}
\right) E_6 DZ 
\right.
\nonumber\\
&&\left. \qquad \qquad +\left(\frac{2k^3}{45} + \frac{77 k^2}{2160}+\frac{37 k}{4320}\right)E_4^2 Z \right]. 
\eea

It is useful to rewrite these expressions in terms of derivatives of $\langle I_3 \rangle$, using the replacement $D^2 Z = \langle I_3 \rangle - \frac{k}{60} E_4 Z$.  This simplifies the form of the coefficients somewhat, and makes explicit the vanishing  for the Yang-Lee model. We have 
  \bea
   \langle  I_3^2 \rangle
&=& \left[
D^2 \langle I_3 \rangle + \left( \frac{k}{60} + \frac{1}{18} \right) E_4 \langle I_3 \rangle   -\frac{60k+11}{1080} E_6 DZ  +\frac{k (60 k+11)}{3780} E_4^2 Z 
\right]
\nonumber\\
&&
+ E_2 \left[\frac{2}{3} D \langle I_3 \rangle + \frac{(60 k+11)}{1080} E_4 DZ - \frac{k (60 k+11)}{3780} E_6 Z \right] \,.
   \eea
We see that this is a combination of modular derivatives of $\langle I_3 \rangle$ and terms which vanish when $k = - \frac{11}{60}$. Thus, this is zero in the Yang-Lee model. Similarly we can write
    \bea
   \langle  I_3 I_5 \rangle
&=& D^3 \langle I_3 \rangle  
+\left(  \frac{k}{12}+\frac{13}{72}\right)   E_4  D \langle I_3 \rangle \\
&&
-\left( \frac{k^2}{63} + \frac{131 k}{540} + \frac{19}{216} \right)  E_6 \langle I_3 \rangle
+\frac{(25+48k)(11+60k)}{22680} E_4^2 D Z -\frac{k(25+48k)(11+60k)}{64800}  E_4 E_6  Z \nonumber\\
&& + E_2 \left[   D^2 \langle I_3 \rangle 
+ \left( \frac{47 k}{180} + \frac{5}{54} \right)   E_4 \langle I_3 \rangle 
-  \frac{(25 + 48 k)(11+60k)}{22680} E_6 DZ 
 +\frac{k(25+48k)(11+60k)}{64800} E_4^2 Z \right].\nonumber
\eea

We would also expect $\langle I_3 I_5 \rangle$ to vanish for the $(7,2)$ minimal model, which can be made explicit by rewriting it in terms of derivatives of $\langle I_5 \rangle$, 
    \bea
   \langle  I_3 I_5 \rangle
&=& D^2 \langle I_5 \rangle  
+\left(  \frac{k}{60}+\frac{1}{6}\right)   E_4  \langle I_5 \rangle \\
&&
-\frac{(17+42k)}{216} E_6 D^2 Z 
+\frac{(1+6k)(17+42k)}{2268} E_4^2 D Z -\frac{k(7+32k)(17+42k)}{30240}  E_4 E_6  Z \nonumber\\
&& + E_2 \left[   D \langle I_5 \rangle 
+\frac{(17+42k)}{216}  E_4 D^2 Z 
- \frac{(1+6k)(17+42k)}{2268} E_6 DZ 
 +\frac{k(7+32k)(17+42k)}{30240} E_4^2 Z \right].\nonumber
\eea

For $\langle I_5^2 \rangle$, we have 
    \bea
   \langle  I_5^2 \rangle
&=&D^6 Z 
+\left(  \frac{k}{6}+\frac{19}{36}\right)   E_4 D^4Z  \\
&&
-\left(\frac{2 k^2}{63} + \frac{23k}{27}+\frac{31}{72}\right)E_6 D^3 Z
+\left(\frac{647 k^2}{1008}+\frac{709 k}{1008}+\frac{821}{5184}\right)E_4^2  D^2 Z \nonumber\\
&&
-\left(\frac{169 k^3}{378} + \frac{425 k^2}{756}+\frac{919 k}{3888}+\frac{31}{1296}\right) E_4 E_6 D  Z \nonumber\\
&&
+\left[E_4^3 \left(\frac{4k^4}{33} + \frac{1187 k^3}{8316}+\frac{3005 k^2}{49896}+\frac{871 k}{85536}\right)+E_6^2 \left(\frac{3539 k^4}{43659} + \frac{3539 k^3}{37422}+\frac{2593 k^2}{64152}+\frac{49 k}{7128}\right)\right] Z \nonumber\\
&&+ E_2 \left[\frac{3}{2}   D^5 Z 
+ \left(\frac{11  k}{12}+\frac{17}{36} \right) E_4 D^3 Z
-\left(\frac{40 k^2}{63}+\frac{265  k}{378}+\frac{205 }{1296}
\right) E_6 D^2 Z 
\right.
\nonumber\\
&&\left. \qquad \qquad +\left(\frac{4k^3}{9} + \frac{485 k^2}{864}+\frac{613 k}{2592}+\frac{745}{31104}\right)E_4^2 DZ 
-\left(\frac{20k^4}{99} + \frac{73 k^3}{308}+\frac{1115 k^2}{11088}+\frac{1459 k}{85536}\right) E_4 E_6
Z \right] .\nonumber
\eea

We can also write it in terms of derivatives of $\langle I_5 \rangle$: 
    \bea
   \langle  I_5^2 \rangle
&=&D^3 \langle I_5 \rangle  
+\left(\frac{k}{12}+\frac{37}{72}\right)   E_4 D \langle I_5 \rangle   \\
&&
-\left(\frac{k^2}{63} + \frac{41k}{54}+\frac{5}{12}\right)E_6 \langle I_5 \rangle
+\frac{11(7 + 12k)(17 + 42 k)}{9072}E_4^2  D^2 Z \nonumber\\
&&
-\frac{(91 + 756k + 1152 k^2)(17+42 k)}{108864} E_4 E_6 D  Z \nonumber\\
&&
+k (17+42k) \left[\frac{147+884 k + 1152 k^2}{399168} E_4^3 + \frac{35 + 354 k + 576 k^2}{299 376} E_6^2 \right] Z \nonumber\\
&&+ E_2 \left[\frac{3}{2}   D^2 \langle I_5 \rangle  
+ \left(\frac{19  k}{24}+\frac{65}{144} \right) E_4 \langle I_5 \rangle 
-\frac{11(7 + 12k)(17 + 42 k)}{9072} E_6 D^2 Z 
\right.
\nonumber\\
&&\left. \qquad \qquad +\frac{(91 + 756k + 1152 k^2)(17+42 k)}{108864} E_4^2 DZ 
- \frac{k(17+42 k) (581+4068 k + 5760 k^2)}{1197504}  E_4 E_6
Z \right] .\nonumber
\eea
Thus, we see that it vanishes for the $(7,2)$ minimal model. 

We have also obtained the differential operator for $\langle I_3 I_7 \rangle$ by these methods: 
\begin{multline}
\langle I_3 I_7 \rangle = D^2 \langle I_7 \rangle + \frac{k+20}{60} E_4  \langle I_7 \rangle - \frac{7(36 k+23)}{540} E_6  D^3 Z + \frac{(36 k+23)(3k+1)}{270} E_4^2 D^2 Z \\- \frac{(36 k+23)(1728 k^2+990 k+115)}{194400} E_4 E_6 D Z \\+ \frac{ k(36k+23)(960 k^2+434 k+55)}{356400} E_4^3 Z + \frac{k(36 k+23)(5760 k^2 + 3132 k + 605)}{3207600} E_6^2 Z 
\\ + E_2 [ \frac{4}{3} D  \langle I_7 \rangle + \frac{7(36 k+23)}{540} E_4 D^3 Z - \frac{(36 k+23)(3k+1)}{270} E_6 D^2 Z \\ +  \frac{(36 k+23)(1728 k^2+990 k+115)}{194400} E_4^2 D Z - \frac{k(36k+23)(7200 k^2+3519k+550)}{1603800} E_4 E_6 Z ] . 
\end{multline}
We have written this in a form which makes it explicit that it vanishes on the $(9,2)$ minimal model; it also vanishes on the Yang-Lee model. 

\subsection{Three-point functions}

In section \ref{qexp}, we obtained 
    \bea 
   \langle  I_3^3 \rangle
&=& D^6Z
+\left(  \frac{k}{20}+\frac{1}{18}\right)   E_4 D^4 Z  \\
&&
-\left(\frac{k}{5}+\frac{233}{1080}\right)E_6 D^3 Z
+\left(\frac{407 k^2}{8400}+\frac{229 k}{840}+\frac{115}{1296}\right)E_4^2  D^2 Z \nonumber\\
&&
-\left(\frac{17 k^2}{100}+\frac{2411 k}{21600}+\frac{11}{864}\right) E_4 E_6 D Z \nonumber\\
&&
+\left[E_4^3 \left(\frac{19069 k^3}{504000}+\frac{71 k^2}{3024}+\frac{25 k}{5184}\right)+E_6^2 \left(\frac{4 k^3}{135}+\frac{103 k^2}{5400}+\frac{67 k}{16200}\right)\right] Z \nonumber\\
&&+ E_2 \left[2  D^5
+ \left(\frac{7  k}{30}+\frac{131}{360} \right) E_4 D^3
-\left(\frac{k^2}{21}+\frac{739  k}{1260}+\frac{49 }{270}
\right) E_6 D^2
\right.
\nonumber\\
&&\left. \qquad \qquad +\left(\frac{101 k^2}{300}+\frac{4811 k}{21600}+\frac{11}{432}\right)E_4^2 D
-\left(\frac{169 k^3}{1260}+\frac{6409 k^2}{75600}+\frac{43 k}{2400}\right) E_4 E_6
\right] Z\nonumber\\
&&+ E_2^2  \left[
D^4
+\left(\frac{19 k}{60}+\frac{23}{240} \right) E_4 D^2
-\left( \frac{k^2}{6}+\frac{k}{9}+\frac{11}{864}\right) E_6 D
+\left(\frac{k^3}{15}+\frac{19 k^2}{450}+\frac{43 k}{4800}\right)E_4^2
\right]Z\,. \nonumber
\eea

This expression also vanishes on the Yang-Lee model, which can be made manifest by rewriting the expression in terms of modular covariant derivatives acting on $\langle I_3 \rangle$, 
    \bea
   \langle  I_3^3 \rangle
&=& D^4 \langle I_3 \rangle
+\left(  \frac{k}{30}+\frac{1}{18}\right)   E_4 D^2 \langle I_3 \rangle  -\left(\frac{8k}{45}+\frac{233}{1080}\right)E_6 D \langle I_3 \rangle 
+\left(\frac{1207 k^2}{25200}+\frac{241 k}{945}+\frac{115}{1296}\right)E_4^2  \langle I_3 \rangle \nonumber\\
&&
-\frac{(5+12 k)(11+60 k)}{4320} E_4 E_6 D Z+\left[ \frac{k(1+3k)(11+60k)}{4860} E_4^3+  \frac{k(41+96k)(11+60k)}{194400} E_6^2 \right] Z \nonumber\\
&&+ E_2 \left[ 2 D^3 \langle I_3 \rangle
+ \left( \frac{k}{5} + \frac{131}{360} \right)  E_4 D \langle I_3 \rangle 
-\left(\frac{k^2}{21}+\frac{697  k}{1260}+\frac{49 }{270}
\right) E_6 \langle I_3 \rangle 
\right.
\nonumber\\
&&\left. \qquad \qquad + \frac{(5+12 k)(11 + 60 k)}{2160}E_4^2 DZ
-\frac{k(3+8k)(11+60k)}{3600} E_4 E_6 Z
\right] \\
&&+ E_2^2  \left[
D^2 \langle I_3 \rangle + \left( \frac{3k}{10} + \frac{23}{240} \right) E_4 \langle I_3 \rangle  - \frac{(5+12 k)(11+60 k)}{4320} E_6 DZ + \frac{k(3+8k)(11+60k)}{7200} E_4^2 Z\nonumber
\right]\,. 
\eea

We can similarly obtain 
\begin{multline}
\langle I_3^2 I_5 \rangle =  D^4 \langle I_5\rangle  + \frac{3k+20}{90} E_4 D^2 \langle I_5 \rangle - \frac{492 k+721}{1080} E_6 D \langle I_5 \rangle + \frac{108126 k^2 + 420120 k + 215575 }{453600} E_4^2 \langle I_5 \rangle \\ - (17 + 42 k) \frac{ 96 k^2+  903 k+ 490}{45360}  E_4 E_6 D^2 Z +  (17 + 42 k) \frac{7632 k^2 + 4092 k+  445}{816480} E_4^3  DZ  \\ + (17 + 42 k) \frac{ 1776 k^2 +1084 k + 133}{233280} E_6^2 DZ -  k (17 + 42 k) \frac{ 3007296 k^2 +1667346 k + 219625}{342921600}  E_4^2 E_6 Z 
\\ + E_2 [ \frac{8}{3} D^3 \langle I_5 \rangle  + \frac{528 k+ 1261}{1080} E_4 D \langle I_5 \rangle  - \frac{ 5400 k^2 + 45366 k +22873}{22680}  E_6  \langle I_5 \rangle  \\ + (17 + 42 k) \frac{576 k^2 +11358 k + 5855}{272160} E_4^2 D^2 Z -  (17 + 42 k) \frac{7026 k^2 +3935 k + 449}{204120} E_4 E_6 DZ 
\\+  k (17 + 42 k) \frac{ 86688 k^2 +51363 k  +7775}{8164800}  E_4^3 Z   +  k (17 + 42 k) \frac{ 11952 k^2+ 5940 k +553 }{1714608}  E_6^2 Z ]
    \\+  E_2^2 [\frac{11}{6} D^2 \langle I_5 \rangle + \frac{11 ( 1272 k +661 )}{12960} E_4 \langle I_5 \rangle - (17 + 42 k) \frac{11 (108 k+53)}{54432}  E_6 D^2 Z \\+  (17 + 42 k) \frac{11 (2592 k^2 + 1428 k +161)}{1632960}  E_4^2 DZ - k (17 + 42 k) \frac{17280 k^2 +9588 k  +1243}{1959552} E_4 E_6 Z ]. 
\end{multline}
This has been written in a form that makes it explicit that it vanishes on the $(7,2)$ minimal model; it can also be shown to vanish on the Yang-Lee model. 

\subsection{Further constraints from particular representations}

We can make some further progress by considering the action of the differential operator on the characters in particular minimal models, where the number of states at low levels is reduced.  For example, we can apply the differential operators for $\langle I_3\rangle$, $\langle I_3^2 \rangle$ and $\langle I_3^3 \rangle$ to any Virasoro character, which fixes completely the eigenvalues of $I_3$ of this representation at any level with 3 or fewer states.  We can then take the general expression for $\langle I_3^4 \rangle$ as a quasi-modular differential operator and consider its action on the same characters, and use this to constrain the unknown coefficients in the differential operator.  We start by using the action on the general Verma module character up to level 3, as well as the vacuum character up to level 5 and the character with $h$ at the value (\ref{null2}) (for which the representation has a null state at level 2) up to level 4.  This still leaves several unfixed numerical coefficients, which can be determined by computing the action on the characters of the Ising, Tri-critical Ising, Potts and Tri-critical Potts model.  The result is: %
\begin{equation}
\begin{aligned} \label{I3q}
\langle  I_3^4 \rangle
&= D^8 Z
+\left(\frac{k}{15}-\frac{1}{9}  \right)   E_4 D^6 Z  \\
&
-\left(\frac{2 k}{5}+\frac{131}{180}\right)E_6 D^5 Z
+\left(\frac{407 k^2}{4200}+\frac{986 k}{945}+\frac{137}{162}\right)E_4^2  D^4 Z \nonumber\\
&
-\left(\frac{17 k^2}{25}+\frac{23633 k}{16200}+\frac{625}{972}\right) E_4 E_6 D^3 Z \nonumber\\
&
+\left[ E_4^3 \left(\frac{19069 k^3}{126000}+\frac{59167 k^2}{75600}+\frac{9445 k}{13608}+\frac{95}{648}\right) +E_6^2 \left(\frac{16 k^3}{135}+\frac{1021 k^2}{1350}+\frac{5177 k}{8100}+\frac{151883}{1166400}\right) \right] D^2 Z \nonumber\\
&
-\left(\frac{28369 k^3}{21000}+\frac{5896477 k^2}{4536000}+\frac{24529 k}{54432}+\frac{9823}{233280}\right) E_4^2 E_6 D Z \nonumber\\
&
+\bigg[E_4^4 \left(\frac{k (9 k (3 k (63408283 k+52729000)+52227500)+67865000)}{5715360000}\right) \nonumber\\
&
+E_6^2 E_4 \left(\frac{k (k (3 k (36160 k+30403)+30142)+4375)}{243000}\right)\bigg] Z 
\nonumber\\
&+ E_2 \bigg[4 D^7
+\left(\frac{8 k}{15}+\frac{19}{20}\right) E_4 D^5
-\left(\frac{2 k^2}{21}+\frac{1621 k}{630}+\frac{83}{30}\right) E_6 D^4 \nonumber\\
&
+\left(\frac{8891 k^2}{6300}+\frac{58039 k}{12600}+\frac{1297}{648}\right) E_4^2 D^3 
-\left(\frac{169 k^3}{315}+\frac{87527 k^2}{18900}+\frac{458393 k}{113400}+\frac{53627}{64800}\right)E_4E_6D^2 \nonumber\\
&
+\left[ E_4^3 \left(\frac{47269 k^3}{21000}+\frac{9962077 k^2}{4536000}+\frac{6478 k}{8505}+\frac{11077}{155520}\right) + E_6^2 \left(\frac{562 k^3}{315}+\frac{96091 k^2}{56700}+\frac{401207 k}{680400}+\frac{8041}{145800}\right) \right] D \nonumber\\
&
-\left(\frac{k (k (9 k (11815180 k+9895317)+29439989)+4269125)}{47628000}\right) E_4^2 E_6
\bigg] Z\nonumber\\
&+ E_2^2  \bigg[
\frac{16}{3} D^6
+\left(\frac{8 k}{5}+\frac{1309}{540}\right) E_4 D^4
-\left(\frac{46 k^2}{63}+\frac{4687 k}{945}+\frac{2299}{1080}\right) E_6 D^3 \nonumber\\
&
+\left(\frac{4 k^3}{15}+\frac{4183 k^2}{900}+\frac{2203 k}{540}+\frac{320401}{388800}\right) E_4^2 D^2 
-\left(\frac{2531 k^3}{630}+\frac{220117 k^2}{56700}+\frac{1838113 k}{1360800}+\frac{73733}{583200}\right)E_4E_6D \nonumber\\
&
+ E_4^3 \left(\frac{301 k^4}{225}+\frac{841 k^3}{750}+\frac{238787 k^2}{648000}+\frac{23 k}{432}\right) + E_6^2 \left(\frac{1177 k^4}{1323}+\frac{14786 k^3}{19845}+\frac{237089 k^2}{952560}+\frac{283 k}{7776}\right)
\bigg]Z \nonumber\\
&+ E_2^3  \bigg[
\frac{22}{9} D^5
+\left(\frac{82 k}{45}+\frac{1291}{1620}\right) E_4 D^3
-\left(\frac{14 k^2}{9}+\frac{37 k}{27}+\frac{59}{216}\right) E_6 D^2 \nonumber\\
&
+\left(\frac{4 k^3}{3}+\frac{581 k^2}{450}+\frac{7291 k}{16200}+\frac{98351}{2332800}\right) E_4^2 D 
-\left(\frac{20 k^4}{27}+\frac{503 k^3}{810}+\frac{37 k^2}{180}+\frac{697 k}{23328}\right)E_4E_6
\bigg]Z\,. \nonumber
\end{aligned}
\end{equation}

\section{Recovering operators from one-point functions}
\label{ops}

The KdV charges are the zero-mode of local operators constructed from the stress tensors and their derivatives. For $I_{2m-1}$, the operator $J_{2m}$ is weight $2m$, and has a component with $m$ stress tensors. The components with fewer stress tensors are determined by requiring that $I_{2m-1}$ commutes with the lower order KdV charges.\footnote{The form of the operator is ambiguous, as we can add any total derivative to the operator without changing the zero mode. In the calculation below we make a convenient arbitrary choice of basis for the independent operators with each number of derivatives.} 
This is an involved calculation, and there is a relative lack of explicit expressions in the literature, so we have taken expressions for the operators with arbitrary coefficients and performed a recursion calculation to obtain their one-point functions, and then compared to the expressions we obtained in section \ref{onept} to fix the coefficients. This confirms the form $J_6 = (T(TT))+ ( 2k + \frac{1}{6}) (2 \pi)^2 (T' T')$, and gives us 
\begin{equation} 
J_8 = (T(T(TT))) + \frac{24 k +2}{3} (2 \pi)^2 (T(T'T'))  + \frac{192 k^2 - 68 k - 7}{60} (2 \pi)^4 (T'' T'') , 
\end{equation}
\begin{multline} 
J_{10}  = (T(T(T(TT)))) +  \frac{60 k + 5}{3} (2 \pi)^2 (T(T(T'T'))) \\ + \frac{192 k^2  - 308 k - 27}{12} (2 \pi)^4 (T(T'' T''))  +\frac{11520 k^3 - 37968 k^2 + 15272 k + 1543}{2520} (2\pi)^6 (T''' T''')  ,
\end{multline}
\begin{multline} 
J_{12}  = (T(T(T(T(TT))))) +  \frac{120 k + 10}{3} (2 \pi)^2 (T(T(T(T'T'))))  + \frac{192 k^2 - 548 k - 47}{4} (2 \pi)^4 (T(T(T'' T''))) \\ + \frac{51840 k^3 - 175104 k^2+ 343755 k  + 29971}{2079}(2 \pi)^4 (T'(T'(T'T')))  -\frac{ 20400 k^2+ 1793 k + 34}{630} (2\pi)^6 (T(T''' T''')) \\ - \frac{691200 k^4 - 1166400 k^3 - 2190600 k^2 + 7248292 k +618371}{27720} (2 \pi)^6 (T'' (T'' T''))\\  - \frac{  
 392601600 k^4  - 1331343360 k^3+ 1188784320 k^2  - 348831368 k - 38627459}{34927200} 
(2 \pi)^8 (T'''' T'''') .
\end{multline}
Note that these operators are only defined up to the addition of total derivative terms, as it is really the zero modes $I_{2m-1}$ that we are interested in. 

For $J_8$ and $J_{10}$, these are useful cross-checks on the calculation, as there are fewer free parameters in the expression for the operator than in our expression for $\langle I_7 \rangle_h$ and $\langle I_9 \rangle_h$. In $J_{12}$, we are starting to get different independent operators at the same derivative order, so the expression has the same number of free parameters as $\langle I_{11} \rangle_h$. For $J_{14}$, there are too many independent operators for us to fix all the coefficients uniquely using the expression for $\langle I_{13} \rangle_h$.  

\section{Brute-Force calculation of thermal expectation values}
\label{brute}

In this appendix we describe some further calculations using the straightforward cycling of the Virasoro operators around the trace. 

\subsection{$I_{3}$ with a primary}

We consider the thermal expectation value of a primary field $V=\sum_{m}e^{2 \pi i m u} v_{m} $ of dimension $h$, where $u$ is the coordinate on the cylinder, with an insertion of the KdV charge $I_3$,
\bea
\langle I_{3}V\rangle&=2\sum_{n\geq1}\langle L_{-n}L_{n}V\rangle+\langle I_{1}^2V\rangle-\frac{1}{6}\langle I_{1}V\rangle+\frac{k}{60}\langle V \rangle  \nonumber\\
 \nonumber\\ 
&=2\sum_{n\geq1}\langle L_{-n}L_{n}V\rangle+\left[\partial^2-\frac{1}{6}\partial+\frac{k}{60}\right]\langle V \rangle
\eea
We can calculate the first term by moving $L$'s around the trace,
\bea 
\langle L_{-n}L_{n}V\rangle=\frac{q^n}{1-q^n}\left(\langle[L_{n},L_{-n}]V\rangle+\frac{1}{1-q^n}\big[L_{n},[V,L_{-n}]\big]\right) \nonumber\\
 \nonumber\\
=\frac{2nq^n}{1-q^n}\langle I_{1}V\rangle+2k\frac{n^3q^n}{1-q^n}\langle V\rangle+\frac{q^n}{\left(1-q^n\right)^2}\langle\big[L_{n},[V,L_{-n}]\big]\rangle \label{e2}
\eea 
To calculate the last term in \eqref{e2} we expand $V$ in modes and use the commutation relation
\be
\big[L_{m},v_{n}\big]=\left(\left(h-1\right)m-n\right)v_{m+n}. 
\ee
This gives us
\bea
\frac{q^n}{\left(1-q^n\right)^2}\langle\big[L_{n},[V,L_{-n}]\big]\rangle &=& h(h-1)\frac{n^2q^n}{(1-q^n)^2}\langle V\rangle+\frac{q^n}{(1-q^n)^2}\langle\sum_{m}^{}e^{2\pi imu}m(n-m)v_{m}\rangle \\
&=&h(h-1)\frac{n^2q^n}{(1-q^n)^2}\langle V\rangle+\frac{nq^n}{(1-q^n)^2}(-i)(2\pi)^{-1} \partial_{u}\langle V \rangle+\frac{q^n}{(1-q^n)^2}(2\pi)^{-2} \partial^{2}_{u}\langle V\rangle. \nonumber
\eea
By translation invariance, $\langle V \rangle$ does not depend on the location of $V$ on the torus, therefore the derivative terms vanish in the last expression.

The result is 
\be
\langle I_{3}V\rangle= (\partial^2  - \frac{E_2}{6} \partial  + \frac{k}{60} E_4) \langle V\rangle+\frac{h-h^2}{12}\partial E_{2}\langle V\rangle.
\ee
Note that the derivative operator appearing in the first term is precisely the same one that appears in $\langle I_3 \rangle$ in \eqref{I3result}. It is interesting that for conserved currents, with $h=1$, we would get just this term. The torus one-point function of a chiral primary operator is a modular form of weight $h$, so it is instructive to re-write our result in terms of the Serre derivative $D_{h}=\partial-\frac{h}{12}E_{2}$,
\be
\langle I_{3}V\rangle=D_{h+2}D_{h}\langle V\rangle+ \left[\frac{h^2 - 2h}{144}+\frac{k}{60}\right] E_4 \langle V\rangle +\frac{h}{6}E_{2} D_h \langle V\rangle .
\ee
As with the calculations quadratic in KdV charges, we get a modular form of weight $h+4$ plus $E_2$ times a modular form of weight $h+2$. 

\subsection{$\langle I_{3}^2 \rangle$}

We will describe the brute-force calculation of $\langle I_3^2\rangle$ without the use of the recursion relations. Using the definition \eqref{kdvcharges}, we have 
\bea
\langle I_3^2\rangle &=&4\sum_{n,m\geq1}\langle L_{-n}L_{n}L_{-m}L_{m}\rangle+\left[4\partial^2-\frac{2}{3}\partial+\frac{k}{15}\right]\sum_{n\geq1}\langle L_{-n}L_{n}\rangle\nonumber\\
&&+\left[\partial^4-\frac{1}{3}\partial^3+\left(\frac{k}{30}+\frac{1}{36}\right)\partial^2-\frac{k}{180}\partial+\left(\frac{k}{60}\right)^2\right]Z.
\eea
We write $\mathcal{I}\equiv\sum_{n\geq1} L_{-n}L_{n}$. From \eqref{lnln}, we have
\be
\langle \mathcal{I}\rangle=\left[2\sigma_{1}\partial+2k\sigma_{3}\right]Z,
\ee
and hence
\bea
\langle I_3^2\rangle&=&4\sum_{n,m\geq1}\langle L_{-n}L_{n}L_{-m}L_{m}\rangle\nonumber\\
&&+\bigg[\partial^4+\left(8\sigma_{1}-\frac{1}{3}\right)\partial^3+\left(16\partial\sigma_{1}+8k\sigma_{3}-\frac{4}{3}\sigma_{1}+\frac{k}{30}+\frac{1}{36}\right)\partial^2 \nonumber\\
&&+\left(8\partial^2 \sigma_{1}+16k\partial\sigma_{3}-\frac{4}{3}\partial\sigma_{1}-\frac{4k}{3}\sigma_{3}+\frac{2k}{15}\sigma_{1}-\frac{k}{180}\right)\partial \nonumber\\
&&+\left(8k\partial^2 \sigma_{3}-\frac{4k}{3}\partial\sigma_{3}+\frac{2k^2}{15}\sigma_{3}+\frac{k^2}{3600}\right)\bigg] Z
\eea
We will calculate the term $\sum_{n,m\geq1}\langle L_{-n}L_{n}L_{-m}L_{m}\rangle$ by moving the $L$'s around the trace,
\bea
\sum_{n,m\geq1}\langle L_{-n}L_{n}L_{-m}L_{m}\rangle&=&\sum_{n\geq1}\langle L_{-n}L_{n}\mathcal{I}\rangle\nonumber\\
&=&\sum_{n\geq1}\frac{q^n}{1-q^n}\left(\langle[L_{n},L_{-n}]\mathcal{I}\rangle+\frac{1}{1-q^n}\langle\big[L_{n},[\mathcal{I},L_{-n}]\big]\rangle\right)\nonumber\\
&=&\left[2\sigma_{1}\partial+2k\sigma_{3}\right]\langle\mathcal{I}\rangle+\sum_{n\geq1}\frac{q^n}{(1-q^n)^2}\langle\big[L_{n},[\mathcal{I},L_{-n}]\big]\rangle. \label{4l}
\eea
The nested commutator term for $n,m\geq1$ reads
\bea
\big[L_{n},[\mathcal{I},L_{-n}]\big]&=&\bigg[L_{n},\big[\sum_{m=1}^{\infty}L_{-m}L_{m},L_{-n}\big]\bigg]\\
&=&\sum_{m=1}^{\infty}\bigg\{(m+n)^2L_{n-m}L_{m-n}+(n-m)^2L_{-(m+n)}L_{(m+n)}\bigg\}\nonumber\\
&&+\sum_{m=1}^{\infty}\big\{\left(4n^2-2m^2\right)L_{-m}L_{m}\big\}+8k(n^4-n^2)I_{1}+4k^2(n^6-n^2)\mathds{1}.\nonumber
\eea
For the terms involving $L_{n-m}L_{m-n}$ and $L_{-(m+n)}L_{(m+n)}$ we re-label the indices as  $m-n\equiv l$ and $m+n\equiv l$ correspondingly for the first two summands (for some fixed $n\geq1$), and then split the summation ranges,
\bea
\big[L_{n},[\mathcal{I},L_{-n}]\big]&=&\sum_{l=1-n}^{-1}\big\{(2n+l)^2L_{-l}L_{l}\big\}+4n^2L_{0}^{2}+\sum_{l=1}^{\infty}\big\{(2n+l)^2L_{-l}L_{l}\big\}\\
&&+\sum_{l=1}^{\infty}\big\{(2n-l)^2L_{-l}L_{l}\big\}-\sum_{l=1}^{n-1}\big\{(2n-l)^2L_{-l}L_{l}\big\}-n^2L_{-n}L_{n}\nonumber\\
&&+\sum_{m=1}^{\infty}\big\{\left(4n^2-2m^2\right)L_{-m}L_{m}\big\}+8k(n^4-n^2)I_{1}+4k^2(n^6-n^2)\mathds{1}.\nonumber
\eea
The finite sums involving $L_{-l}L_{l}$, cancel and we are left with
\bea
\big[L_{n},[\mathcal{I},L_{-n}]\big]&=&\sum_{l=1}^{n-1}(2n-l)^2[L_{l},L_{-l}]+4n^2L_{0}^{2}-n^2L_{-n}L_{n} \\
&&+12n^2\sum_{m=1}^{\infty}L_{-m}L_{m}+8k(n^4-n^2)I_{1}+4k^2(n^6-n^2)\mathds{1}\nonumber\\
&=&12n^{2}\mathcal{I}+4n^2I_{1}^{2}+8kn^4I_{1}+4k^2n^6\mathds{1}\nonumber\\
&&+\left(-\frac{5}{6}n^{2}-n^3+\frac{11}{6}n^4\right)I_{1}+k\left(\frac{n^2}{10}+\frac{n^4}{6}-n^5+\frac{11}{15}n^6\right)\mathds{1}-n^{2}L_{-n}L_{n}.\nonumber
\eea
Going back to \eqref{4l} we get
\bea
\sum_{n,m\geq1}\langle L_{-n}L_{n}L_{-m}L_{m}\rangle&=&\bigg[\left(4\partial\sigma_1+4\sigma_{1}^{2}\right)\partial^{2}\\
&&+\left(-\frac{5}{6}\partial\sigma_{1}+\left(8k+\frac{11}{6}\right)\partial\sigma_{3}-\partial^{2}\sigma_{1}+28\sigma_{1}\partial\sigma_{1}+8k\sigma_{1}\sigma_{3} \right)\partial\nonumber\\
&&+\left(24k\sigma_{3}\partial\sigma_{1}+4k\sigma_{1}\partial\sigma_{3}+\frac{k}{10}\partial\sigma_{1}+\frac{k}{6}\partial\sigma_{3}+\frac{11k}{15}\partial\sigma_{5}+4k^2\sigma_{3}^{2}+4k^2\partial\sigma_{5}-k\partial^{2}\sigma_{3}\right)\bigg]Z.\nonumber
\eea
Putting everything together, and using the relation \eqref{sigmaE} between the sigmas and the Eisenstein series, we  recover \eqref{I3sqcov}.

\section{Recursion relations for stress tensors and derivatives} 
\label{recursion}

In our second approach to the calculation of the thermal expectation values of KdV charges, we make use of the thermal $n$-point functions of stress tensors and derivatives of the stress tensor, calculated using the recursion relations of \cite{zhu, Gaberdiel:2012yb}. The basic idea of the recursion relation is the same as in the direct computation: commuting the modes around the trace. We take an $n$-point function 
\begin{equation}
\mathrm{Tr} [ V_1(z_1) \ldots V_n(z_n) q^{L_0-k}], 
\end{equation}
pick one of the operators in the $n$-point function, expand it in modes $V_r$. For each non-zero $r$, we can  commute the mode around the trace. The commutators with the other operators give a series of terms involving $n-1$ point functions, and commuting the mode past $q^{L_0 - k}$ will give the original expression back multiplied by $q^r$. Thus, we can obtain an expression for the non-zero modes $V_r$ in terms of a sum of $n-1$ point functions where one of the operators in the $n-1$ point function has the operator mode $V_r$ acting on it, and we sum over the $n-1$ other operators. The zero mode cannot be treated in this way, so it is left as a separate contribution. Thus, the full $n$ point function can be written as an $n-1$ point function with an insertion of the zero mode $V_0$, and a sum over $r$ of the sum of $n-1$ point functions where one of the operators in the $n-1$ point function has the operator mode $V_r$ acting on it. To continue the recursion, we need also to commute the modes of the operators past any zero modes present in the correlation function.  

In \cite{zhu, Gaberdiel:2012yb}, the calculation is carried out in the vertex operator algebra formalism; to describe this we need to introduce some notation. For a state $a$ we have a vertex operator $V(a, z)$. We will often omit the explicit vertex operator label; so for example we write the modes of the operator $V(a,z)$ as $a_n$, with 
\begin{equation}
V(a,z) = \sum_n \frac{a_n}{z^{n+h_a}}. 
\end{equation}
We also introduce the square bracket modes
\begin{equation}
a[n] = (2 \pi i)^{-n-1} \sum_{j = n+1-h_a}^\infty c(h_a, j+h_a -1,n) a_j,  
\end{equation}
where the expansion coefficients $c$ are defined by 
\begin{equation}
(\log (1+z))^s (1+z)^{h-1} = \sum_{j = s}^\infty c(h,j,s) z^j.   
\end{equation}
We write the torus $n$-point function of the vertex operators associated to states $a^1, \ldots a^n$ as $F( (a^1, z_1), \ldots (a^n, z_z); \tau)$. Note that the coordinates are coordinates on the plane; for the torus, the fundamental region can be taken to be $1 < |z| < |q|$. In the recursion, we will also encounter $n$-point functions with additional insertions of zero modes of some operators. We write $F( b_0^\ell;  (a^1, z_1), \ldots (a^n, z_z); \tau)$ for the $n$-point function with $\ell$ powers of the zero mode of the vertex operator corresponding to the state $b$. 

The key result from \cite{Gaberdiel:2012yb} is the recursion relation 
\begin{multline} \label{recursionrel}
F(b_0^\ell;  (a^1, z_1), \ldots (a^n, z_z); \tau) = F(b_0^\ell a^1_0;  (a^2, z_2), \ldots (a^n, z_n); \tau) \\ + \sum_{i=0}^\ell \sum_{j=2}^n \sum_{m = 0}^\infty \left( \begin{array}{c} \ell \\ i \end{array} \right) g^i_{m+1} \left( \frac{z_j}{z_1} \right) F( b_0^{\ell -i} ; (a^2, z_2), \ldots (d^i[m] a^j, z_j) \ldots (a^n,z_n); \tau),
\end{multline}
where the functions $g^i_k$ are defined in \eqref{gi}, and
\begin{equation}
d^i[m] = (-1)^i ( (b[0])^i a^1)[m]. 
\end{equation}
For the case $\ell=0$, this reduces to the recursion relation of \cite{zhu}. This has the structure sketched at the beginning of this section: we have expanded the operator $V(a^1,z_1)$ in modes $a^1_m$. The first term is the zero mode $a_0$. The second line is a sum of $n-1$ point functions, with a mode $a^1[m]$ acting on one of the operators, summed over modes and over which operator it acts on. Note that the sum over modes $a^1_m$ has been re-organised in terms of the square bracket modes. The functions appearing in this case are simply $g^0_k = \mathcal P_k$. When we generalise to $\ell \neq 0$, dealing with commuting the $a^1_m$ past the zero modes $b_0^\ell$ turns out to be nicely expressed in terms of the sum over $i$ above; see \cite{Gaberdiel:2012yb} for details. 

We want to apply this recursion relation to calculate the $n$-point functions involving just stress tensors and their derivatives. The stress tensor on the cylinder corresponds to the state $\tilde \omega = \omega - k \Omega$, where $\Omega$ is the global vacuum state, and $\omega = L_{-2} \Omega$ is the state corresponding to the stress tensor on the plane. The zero mode $\tilde \omega_0  = L_0 - k$. The square bracket modes form a representation of the Virasoro algebra, up to normalization; the Virasoro modes are given by $L_{[n]} = (2 \pi i)^2 \tilde \omega[n+1]$. Explicitly, we have
\begin{equation}
\tilde \omega[0] = \frac{1}{(2\pi i)^2} L_{[-1]} = \frac{1}{(2\pi i)} ( L_{-1} + L_0), 
\end{equation}
\begin{equation}
\tilde \omega[1] = \frac{1}{(2\pi i)^2} L_{[0]} = \frac{1}{(2\pi i)^2} ( L_{0} + \frac{1}{2} L_1 - \frac{1}{6} L_2 + \ldots ), 
\end{equation}
\begin{equation}
\tilde \omega[2] = \frac{1}{(2\pi i)^2} L_{[1]} = \frac{1}{(2\pi i)^3} ( L_{1}  - \frac{1}{12} L_3 + \ldots ), 
\end{equation}
and so on. Note that the first expression only involves two terms, as there are just two non-zero coefficients $c(2,j,0)$, and the coefficient of the term involving $L_2$ in the RHS of the last expression vanishes. 

In the recursion relation, we will have states obtained by acting with modes of the stress tensor on the stress tensor. The positive modes do not generate any new states, but acting with $\tilde \omega[0]$ does. Since $\tilde \omega[0] \propto L_{[-1]}$, these states correspond to the derivatives of the stress tensor with respect to the cylinder coordinate. They have no zero mode part, $(\tilde \omega[0]^n \tilde \omega)_0 = 0$. Acting with one of the positive stress tensor modes on this state gives
\begin{equation}
\tilde \omega[m] \tilde \omega[0]^n \tilde \omega = c_{mn} \tilde \omega[0]^{n+1-m} \tilde \omega
\end{equation}
for $m \leq n+1$, where
\begin{equation}
c_{mn} = \frac{1}{(2\pi i)^{2m}} (n+m+1) \frac{n!}{(n-m+1)!}.
\end{equation}
When $m= n+2$, $\tilde \omega[n+2] \tilde \omega[0]^n \tilde \omega =0$. When $m=n+3$, 
\begin{equation}
\tilde \omega[n+3] \tilde \omega[0]^n \tilde \omega =  d_n \Omega,  
\end{equation}
where 
\begin{equation}
d_n = \frac{2k}{(2\pi i)^{2n+4}} n! ((n+2)^2-1)(n+2) . 
\end{equation}
Applying recursion will also require the action of modes of the states $\tilde \omega[0]^n \tilde \omega$ on other states of the same form. This can be determined recursively; for $m \leq n+i+1$, 
\begin{equation}
(\tilde \omega[0]^i \tilde \omega)[m] \tilde \omega[0]^n \tilde \omega = c^i_{mn} \tilde \omega[0]^{n+i+1-m} \tilde \omega
\end{equation}
where $c^i_{mn} = c^{i-1}_{mn} - c^{i-1}_{m n+1}$. For $m = n+i+2$, $(\tilde \omega[0]^i \tilde \omega)[n+i+2] \tilde \omega[0]^n \tilde \omega =0$. For $m=n+i+3$, 
\begin{equation}
(\tilde \omega[0]^i \tilde \omega)[n+i+3] \tilde \omega[0]^n \tilde \omega =  d^i_n \Omega,  
\end{equation}
where 
\begin{equation}
d^i_n = \frac{2k}{(2\pi i)^{2n+2i+4}} (n+i)! ((n+i+2)^2-1)(n+i+2) . 
\end{equation}

Since all the states obtained are of the form $\tilde \omega[0]^n \tilde \omega$, or the vacuum state $\Omega$ corresponding to the identity operator, the most general correlation function we need to consider in the recursion is $F(\tilde \omega_0^\ell; (\tilde \omega[0]^{n_1} \tilde \omega, z_1) \ldots (\tilde \omega[0]^{n_n} \tilde \omega, z_n); \tau)$. Applying the recursion formula \eqref{recursionrel} to this case, we have 
\begin{multline} 
F(\tilde \omega_0^\ell; (\tilde \omega[0]^{n_1} \tilde \omega, z_1) \ldots (\tilde \omega[0]^{n_n} \tilde \omega, z_n); \tau) \\  = \delta_{n_1,0} F(\tilde \omega_0^{\ell+1} ; (\tilde \omega[0]^{n_2} \tilde \omega, z_2) \ldots (\tilde \omega[0]^{n_n} \tilde \omega, z_n); \tau)  \\ + \sum_{i=0}^\ell \sum_{j=2}^n \sum_{m = 0}^\infty \left( \begin{array}{c} \ell \\ i \end{array} \right) (-1)^i g^i_{m+1} \left( \frac{z_j}{z_1} \right) F( \tilde \omega _0^{\ell -i} ; (\tilde \omega[0]^{n_2} \tilde \omega, z_2) \ldots ((\tilde \omega[0]^{n_1+i} \tilde \omega)[m] \tilde \omega[0]^{n_j} \tilde \omega,z_j)  \ldots (\tilde \omega[0]^{n_n} \tilde \omega, z_n); \tau) , \\
= \delta_{n_1,0} F(\tilde \omega_0^{\ell+1} ; (\tilde \omega[0]^{n_2} \tilde \omega, z_2) \ldots (\tilde \omega[0]^{n_n} \tilde \omega, z_n); \tau)  \\ +  \sum_{i=0}^\ell \sum_{j=2}^n \sum_{m = 0}^{M} \left( \begin{array}{c} \ell \\ i \end{array} \right) (-1)^i c_{m n_j}^{i+n_1} g^i_{m+1} \left( \frac{z_j}{z_1} \right) F( \tilde \omega _0^{\ell -i} ; (\tilde \omega[0]^{n_2} \tilde \omega, z_2) \ldots ( \tilde \omega[0]^{M-m} \tilde \omega,z_j)  \ldots (\tilde \omega[0]^{n_n} \tilde \omega, z_n); \tau) \\ + \sum_{i=0}^\ell \sum_{j=2}^n \left( \begin{array}{c} \ell \\ i \end{array} \right) (-1)^i d^{i+n_1}_{n_j} g^i_{M+3} \left( \frac{z_j}{z_1} \right)  F( \tilde \omega _0^{\ell -i} ; (\tilde \omega[0]^{n_2} \tilde \omega, z_2) \ldots [\mbox{ no } z_j] \ldots ( \tilde \omega[0]^{\tilde n_j} \tilde \omega,z_j) ; \tau) 
\end{multline}
where the upper limit on the sum on $m$ is $M = n_j + n_1 + i +1$, and in the last term the $j$th argument of the correlator is absent because the state is the vacuum $\Omega$, so the operator is the identity operator. This expresses the $n$-point function of the stress tensor in terms of $n-1$-point and $n-2$ point functions.  This recursion can be straightforwardly implemented in Mathematica. 

The simplest case is the two-point function of the stress tensor. Applying the recursion relation, we get 
\begin{equation}
F( (\tilde \omega, z_1), (\tilde \omega, z_2) ; \tau) = F(\tilde \omega_0, (\tilde \omega, z_2) ; \tau) + \sum_{m=0}^1 c_{m0} \mathcal  P_{m+1} \left( \frac{z_2}{z_1} \right) F( (\tilde \omega[0]^{1-m} \tilde \omega, z_2); \tau) + d_0 \mathcal P_4 \left( \frac{z_2}{z_1} \right) F(\tau). 
\end{equation}
The one-point functions of the operator on the torus get a contribution only from the zero mode of the operator, so this simplifies to (using the fact that $(\tilde \omega[0] \tilde \omega)_0 = 0$)  
\begin{equation}
F( (\tilde \omega, z_1), (\tilde \omega, z_2) ; \tau) = F(\tilde \omega_0^2 ; \tau) + \frac{2}{(2\pi i)^2} \mathcal P_{2} \left( \frac{z_2}{z_1} \right) F( \tilde \omega_0; \tau) +  \frac{12 k}{(2 \pi i)^4} \mathcal P_4 \left( \frac{z_2}{z_1} \right) F(\tau). 
\end{equation}
The $\tilde \omega_0 = L_0 -k$ can be rewritten in terms of $q$-derivatives, to give 
\begin{equation}
F( (\tilde \omega, z_1), (\tilde \omega, z_2) ; \tau) = \partial^2 Z  + \mathcal  P_{2} \left( \frac{z_2}{z_1} \right) \frac{2}{(2\pi i)^2} \partial Z +  \mathcal P_{4} \left( \frac{z_2}{z_1} \right) \frac{12k}{(2\pi i)^4}  Z. 
\end{equation}

For the higher-point functions, if we start with an $n$-point function just of stress tensors, $F( (\tilde \omega, z_1), \ldots , (\tilde \omega, z_n) ; \tau)$, the leading derivative term comes from just taking the zero mode in each of the operators, so it will have $n$ $q$-derivatives, 
\begin{equation}
F( (\tilde \omega, z_1), \ldots , (\tilde \omega, z_n) ; \tau) = \partial^n Z  + \ldots  . 
\end{equation}
If we have a more general correlator, the $\tilde \omega[0]^n \tilde \omega$ do not have zero modes, so the leading derivative order is the sum of the number of stress tensor operators plus half the number of other operators. 

At each stage in the recursion, we eliminate the dependence on one of the arguments $z_i$ in the correlators, generating coefficients which are functions of the ratio $\frac{z_j}{z_i}$ for $j > i$. Thus, the final form is a sum of products of functions $g^i_k \left(  \frac{z_j}{z_i} \right)$ for $j > i$, where in a given term $z_i$ can only appear in the denominator of one of the functions in the product. 

\section{Extracting the KdV correlation functions}
\label{zero} 

The recursion relations give us expressions for correlation functions of the stress tensors and their derivatives. This method can be applied iteratively to efficiently obtain such expressions up to quite high order. However, to obtain the desired correlation functions of KdV charges, we have two further steps: first, to obtain a correlation function of the currents $J_{2m}$, we need to appropriately conformal normal order the stress tensors. Second, we need to integrate each of the currents over the spatial circle, taking zero modes of the expression for the current correlation function with respect to the angular coordinates.  The conformal normal ordering is relatively straightforward, while taking the zero modes is more troublesome. 

\subsection{Conformal normal ordering} 

Having obtained an expression for the relevant $n$-point correlation functions of the stress tensor and its derivatives using the recursion relation, to extract the correlation functions of the operators corresponding to the KdV charges, we have to conformal normal order products of operators. Operators are conformal normal ordered in pairs; if the correlation function involves $V(z_i)$ and $W(z_j)$, we obtain a correlator involving $(VW)(z_i)$ by taking $z_j \to z_i$, and keeping the zero mode term in the Laurent expansion in $u$, where we write $z_j = e^{2 \pi i u} z_i$.

The Laurent expansion of $\mathcal P_k \left( \frac{z_j}{z_i} \right) = \mathcal P_k \left( e^{2\pi i u} \right)$ is obtained by rewriting it in terms of the Weierstrass function $\wp_k(u)$ as in \eqref{ptowp} and using the Laurent expansion for $\wp_k(u)$ given in \eqref{lwp}. This gives 
\begin{equation} \label{lp1} 
\mathcal P_{1}(e^{2\pi i u}) = - \frac{1}{u} - i \pi + \ldots,
 \end{equation}
\begin{equation}
\mathcal P_{2k}(e^{2\pi i u}) = \frac{1}{u^{2k}} +  2 \zeta(2k) E_{2k} + \ldots, \quad k \geq 1
 \end{equation}
\begin{equation}
\mathcal P_{2k+1}(e^{2\pi i u}) = -\frac{1}{u^{2k+1}} + \ldots, \quad k \geq1. 
 \end{equation}
The higher terms do not matter, as a given term in the expression for the correlator can involve at most one of these functions of $\frac{z_j}{z_i}$. Weierstrass functions of other arguments are treated in Taylor series, using the expression \eqref{xderP} for the derivatives. This gives 
\begin{eqnarray} 
\mathcal P_n \left( \frac{z_j}{z_k} \right) &=& \mathcal P_n \left( \frac{z_i}{z_k} e^{2\pi i u} \right) =  \sum_{m=0}^\infty \frac{u^m}{m!} \partial_u^m \mathcal P_n\left( \frac{z_i}{z_k} e^{2 \pi i u}  \right)|_{u=0} \\ &=& \sum_{m=0}^\infty \frac{u^m}{m!} \frac{(k+m-1)!}{(k-1)!} \mathcal P_{n+m} \left( \frac{z_i}{z_k}  \right) = \sum_{m=0}^\infty \left( \begin{array}{c} k+m-1 \\ m \end{array} \right) u^m \mathcal P_{n+m} \left( \frac{z_i}{z_k}  \right) .\nonumber 
\end{eqnarray} 
If the $z_j$ is in the denominator, we simply get a sign difference; 
\begin{equation}
\mathcal P_n \left( \frac{z_k}{z_j} \right)  = \sum_{m=0}^\infty \left( \begin{array}{c} k+m-1 \\ m \end{array} \right) (-u)^m \mathcal P_{n+m} \left( \frac{z_k}{z_i}  \right) .
 \end{equation}
The higher terms in the Taylor series matter only when the Weierstrass function $\mathcal P_n \left( \frac{z_j}{z_k} \right)$ appears multiplying a Weierstrass function $\mathcal P_m  \left( \frac{z_j}{z_i} \right)$; then the singular term in the expansion of the latter can give zero mode contributions to the product from higher terms in the Taylor series. The recursion relation expression for the original correlator is a sum of terms, each of which is a product of functions $g^i_k \left(\frac{z_r}{z_s} \right)$. For each term, we apply the above expansions to write dependence on $z_j$ as a power series in $u$, and we keep the $u^0$ term in the product. 

This calculation is conceptually straightforward, but becomes slow as the number of operators being normal ordered increases. Note there is one also potential conceptual problem; there are functions  $g^i_k$ which can appear in the recursion relation which cannot be expressed as  derivatives of the Weierstrass functions $\mathcal P_k$. If these were to appear with the argument $\frac{z_j}{z_i}$, we would not know how to evaluate the Laurent expansion. This has not arisen in practice in the calculations we have performed, but we do not have a general argument to exclude it.

\subsection{Zero modes of the operators} 

After conformal normal ordering, we will have an $n$-point function of the operators corresponding to some set of KdV charges, with arguments $z_1, \ldots z_n$. To obtain the thermal expectation value of the product of KdV charges, we simply integrate this expression over the real parts of the coordinates $u_i$ on the cylinder. This integration picks out the part independent of $z_i$ in a Laurent expansion of the resulting expression in powers of $z_i$. Evaluating this zero mode is the most non-trivial part of our computation from the recursion relation. 

For $\langle I_3 \rangle$ or $\langle I_5 \rangle$, this final stage of the computation is trivial; after conformal normal ordering, we are considering the one-point function $\langle (TT)(z) \rangle$ or $\langle J_6(z) \rangle$, and this one-point function is already independent of $z$. 

For $\langle I_3^2 \rangle$, we obtain an expression for $\langle (TT)(z_3) (TT)(z_1) \rangle$. There were two arguments $z_2, z_1$ in the original four-point function corresponding to $z_1$ in this correlator, so a given term can involve at most two Weierstrass functions of $\frac{z_3}{z_1}$. Thus, we are interested in zero modes of products. The product is
\begin{equation}
\mathcal P_{m_1}(x) \mathcal P_{m_2}(x) = \frac{(2\pi i)^{m_1 + m_2}}{(m_1-1)! (m_2-1)!} \sum_{n_1 \neq 0} \frac{n_1^{m_1-1}}{(1-q^{n_1})} x^{n_1} \sum_{n_2 \neq 0} \frac{n_2^{m_2-1}}{(1-q^{n_2})} x^{n_2},  
 \end{equation}
so the zero mode comes from terms with $n_2= -n_1$, 
\begin{equation}
\mathcal P_{m_1}(x) \mathcal P_{m_2}(x)|_{x^0} = \frac{(2\pi i)^{m_1 + m_2}}{(m_1-1)! (m_2-1)!} \sum_{n \neq 0} \frac{n^{m_1+m_2-2} (-1)^{m_2} q^n}{(1-q^{n})^2} .
\end{equation}
For $m_1 + m_2$ odd, this vanishes, as the terms with $n <0$ cancel the terms with $n>0$. For $m_1 + m_2$ even, 
\begin{equation}
\mathcal P_{m_1}(x) \mathcal P_{m_2}(x)|_{x^0} = \frac{2(2\pi i)^{m_1 + m_2}}{(m_1-1)! (m_2-1)!} (-1)^{m_2} 2\sum_{n=1}^\infty \frac{n^{m_1+m_2-2} q^n}{(1-q^n)^2}. 
\end{equation}
This can be rewritten in terms of Eisenstein series by recognising this sum as the derivative of an Eisenstein series \eqref{derE}, so the zero mode is, for $m_1 + m_2$ even, $m_1 + m_2 >2$,\footnote{
Note that this expression is invalid when $m_1 + m_2 = 2$; such products do not contribute in any of the examples we have considered, but we again do not have a general argument to exclude them from appearing in further calculations. }
\begin{equation} \label{pair1}
\mathcal P_{m_1}(x) \mathcal P_{m_2}(x)|_{x^0}  =\frac{(2\pi i)^{m_1 + m_2} (-1)^{m_2}}{(m_1-1)! (m_2-1)!} \zeta(3- m_1 -m_2)   \partial E_{m_1 + m_2 -2}. 
 \end{equation}
Applying this relation to the expression for $\langle (TT)(z_1) (TT)(z_3) \rangle$ gives a result for $\langle I_3^2 \rangle$ which agrees with the previous computations.  

For $\langle I_3^3 \rangle$, we obtain an expression for $\langle (TT) (z_1) (TT) (z_3) (TT)(z_5) \rangle$. There are two arguments  in the original six-point function of the stress tensor  corresponding to each of the arguments in this correlator, so a given term in the expression for this correlator can involve a product of up to four Weierstrass functions. In the terms with two Weierstrass functions, in addition to the kind of terms which appeared in $\langle I_3^2 \rangle$, there can also be terms involving derivatives of the Weierstrass functions. The ones that are relevant are, for $m_1+m_2$ even, $m_1+m_2 >2$, 
\begin{equation} \label{pair2}
\partial \mathcal P_{m_1}(x) \mathcal P_{m_2}(x)|_{x^0}  =\frac{ (2\pi i)^{m_1 + m_2} (-1)^{m_2}}{2(m_1-1)! (m_2-1)!}\zeta(3- m_1 -m_2)  \partial^2 E_{m_1 + m_2 -2},
 \end{equation}
\begin{equation} \label{pair3} 
\partial \mathcal P_{m_1}(x) \partial  \mathcal P_{m_2}(x)|_{x^0}  =\frac{(2\pi i)^{m_1 + m_2} (-1)^{m_2}}{6(m_1-1)! (m_2-1)!} ( \zeta(3-m_1-m_2) \partial^3 E_{m_1 + m_2 -2} - \zeta(1-m_1-m_2) \partial E_{m_1+m_2}),
 \end{equation}
and
\begin{equation} \label{pair4} 
\partial^2 \mathcal P_{m_1}(x)  \mathcal P_{m_2}(x)|_{x^0}=\frac{(2\pi i)^{m_1 + m_2} (-1)^{m_2}}{6(m_1-1)! (m_2-1)!} ( 2\zeta(3-m_1-m_2) \partial^3 E_{m_1 + m_2 -2} + \zeta(1-m_1-m_2) \partial E_{m_1+m_2}),
 , \end{equation}
The zero modes of products of three functions which occur in $\langle I_3^3 \rangle$ are, with no derivatives, for $m_1 + m_2 + m_3$ even, 
\begin{equation} \label{triple1}
\mathcal P_{m_1}\left( \frac{z_i}{z_j} \right)  \mathcal P_{m_2} \left( \frac{z_j}{z_k} \right)    \mathcal P_{m_3} \left( \frac{z_k}{z_i} \right)|_{z_i^0 z_j^0 z_k^0}  =\frac{(2\pi i)^{m_1 + m_2+m_3} (-1)^{m_3}}{2(m_1-1)! (m_2-1)! (m_3-1)!} \zeta(5-m_1-m_2-m_3) \partial^2 E_{m_1 + m_2 +m_3 -4} ,
 \end{equation}
and for $m_1 + m_2 + m_3$ odd, 
\begin{equation} \label{triple2} 
\mathcal P_{m_1}\left( \frac{z_i}{z_j} \right)  \mathcal P_{m_2} \left( \frac{z_j}{z_k} \right)    \mathcal P_{m_3} \left( \frac{z_k}{z_i} \right)|_{z_i^0 z_j^0 z_k^0} =\frac{(2\pi i)^{m_1 + m_2+m_3} (-1)^{m_3}}{2(m_1-1)! (m_2-1)! (m_3-1)!} \zeta(4-m_1-m_2-m_3) \partial E_{m_1 + m_2 +m_3 -3} ,
 \end{equation}
and with one derivative, for $m_1+m_2+m_3$ even, 
\begin{multline} \label{triple3} 
\partial \mathcal P_{m_1}\left( \frac{z_i}{z_j} \right)  \mathcal P_{m_2} \left( \frac{z_j}{z_k} \right)    \mathcal P_{m_3} \left( \frac{z_k}{z_i} \right)|_{z_i^0 z_j^0 z_k^0}=  \\ \frac{(2\pi i)^{m_1 + m_2+m_3} (-1)^{m_3}}{2(m_1-1)! (m_2-1)! (m_3-1)!} ( \zeta(5-m_1-m_2-m_3) \partial^3 E_{m_1 + m_2 -4}  - \zeta(3-m_1-m_2-m_3) \partial E_{m_1+m_2-2}).
 \end{multline}

Terms with products of four Weierstrass functions have a zero mode which involves a double sum, and we have been unable to simplify these in terms of Eisenstein series explicitly.  For example, the zero mode of 
\begin{equation} 
\mathcal P_{m_1} \left( \frac{z_3}{z_1} \right) \mathcal P_{m_2} \left( \frac{z_5}{z_3} \right)\mathcal P_{m_3} \left( \frac{z_5}{z_1} \right) \mathcal P_{m_4} \left( \frac{z_3}{z_1} \right) 
 \end{equation}
is obtained by setting $n_1 + n_4 = n_2 = - n_3$, so the zero mode involves a double sum, 
\begin{equation} 
\frac{ (2 \pi i)^{m_1 + m_2 + m_3 + m_4} }{(m_1 -1)! (m_2 - 1)! (m_3-1)! (m_4 - 1)!} \sum_{n_1, n_4} \frac{n_1^{m_1-1} n_4^{m_4-1} (n_1 + n_4)^{m_2-1} (-n_1 - n_4)^{m_3-1}}{(1- q^{n_1})(1-q^{n_4}) (1-q^{n_1 + n_4}) (1-q^{-n_1 -n_4}) },
 \end{equation}
where the sum does not include terms with $n_1 = 0, n_4 = 0$ or $n_1 + n_4 = 0$. We have not been able to simplify this analytically in terms of Eisenstein series. However, we can rewrite this as 
\begin{eqnarray} \nonumber
&& \frac{ (2 \pi i)^{m_1 + m_2 + m_3 + m_4} }{(m_1 -1)! (m_2 - 1)! (m_3-1)! (m_4 - 1)!}\left[ - \sum_{n_1 = 1}^{\infty} \sum_{n_4 = 1}^\infty  \frac{n_1^{p_1} n_4^{p_2}}{(1- q^{n_1})(1-q^{n_4}) } \frac{ (n_1 + n_4)^{p_3} q^{n_1+n_4} (1+q^{n_1 + n_4})}{(1-q^{n_1 + n_4})^2}  \right. \\ &&  + (-1)^{p_1+p_3} \sum_{n_1 = 2}^{\infty} \sum_{n_4 = 1}^{n_1-1} \frac{n_1^{p_1} n_4^{p_2}}{(1- q^{n_1})(1-q^{n_4}) }  \frac{(n_1-n_4)^{p_3}  q^{n_1} (1 + q^{n_1-n_4})}{(1-q^{n_1-n_4})^2} \nonumber 
\\ &&  \left. + (-1)^{p_1} \sum_{n_4 = 2}^{\infty} \sum_{n_1 = 1}^{n_4-1} \frac{n_1^{p_1} n_4^{p_2}}{(1- q^{n_1})(1-q^{n_4}) }  \frac{(n_4-n_1)^{p_3}  q^{n_4} (1 + q^{n_4-n_1})}{(1-q^{n_4-n_1})^2} \right] . 
 \end{eqnarray}
 We can then do a $q$ expansion; comparing to the $q$ expansions of quasimodular forms of the expected order, we were able to show that the particular combination which appears in $\langle I_3^3 \rangle$ can be rewritten in terms of Eisenstein series. 
 
For $\langle I_5^2 \rangle$, after conformal normal ordering we have an expression for $\langle J_6(z_1) J_6(z_4) \rangle$. This can involve products of up to three Weierstrass functions of $\frac{z_4}{z_1}$. The products of pairs of Weierstrass functions which appear are cases we have treated already. The product of three Weierstrass functions has a zero mode which involves a double sum:
\begin{equation} 
\mathcal P_{m_1} (x) \mathcal P_{m_2} (x) \mathcal P_{m_3} (x) = \frac{ (2\pi i)^{m_1 + m_2 + m_3}}{(m_1-1)! (m_2-1)! (m_3-1)!} \sum_{n_1 \neq 0} \sum_{n_2 \neq 0} \sum_{n_3 \neq 0} \frac{ n_1^{m_1-1} n_2^{m_2-1} n_3^{m_3-1} }{(1-q^{n_1})(1-q^{n_2})(1-q^{n_3})} x^{n_1+n_2+n_3}, 
\end{equation}
so the zero mode is the terms with $n_1+n_2+n_3=0$. Again, we were not able to rewrite this in terms of Eisenstein series explicitly, but we were able to reproduce the $q$ expansion of the result from a $q$ expansion of a combination of Eisenstein series. 

\subsection{Tracing the modular weight through the calculation} 

The argument of \cite{Dijkgraaf:1996iy}, reviewed in section \ref{dijk}, implies that the results for the expectation values of the KdV charges are combinations of powers of $E_2$ with coefficients which are modular forms of particular weights. One of the motivations for developing the recursion relation approach was to have a method of calculation which makes this structure manifest. Unfortunately, while the relation to Eisenstein series is clearer in this approach, there are still issues. 

In the recursion relation calculation for an $n$-point function of the stress tensor, in each term $2n = 2 n_\partial + \sum_i m_i$, where $n_\partial$ is the number of $q$-derivatives (acting either on the partition function or a Weierstrass function) and $\sum_i m_i$ is the sum of the indices of the Weierstrass functions appearing. This property is usually preserved under conformal normal ordering, as $\mathcal P_{2k}\left( \frac{z_j}{z_i} \right)$ is replaced by $E_{2k}$, or contributes a singular term which, if it multiplies functions like $ \mathcal P_m\left( \frac{z_j}{z_k} \right)$, will pick out corresponding higher terms in the Taylor expansion. However, there is an exception; the $-i\pi$ term in \eqref{lp1} does not obey this rule. 

In the zero modes of operators, the zero mode mostly has a modular weight corresponding to the sum $ 2 n_\partial + \sum_i m_i$ for the product whose zero mode we're taking, but not always. In the simple products of pairs of Weierstrass functions this rule of thumb is obeyed: the resulting zero mode in \eqref{pair1}  has weight $m_1 + m_2$, and similarly the zero mode in \eqref{pair2} has weight $m_1 + m_2 +2$.  However, in \eqref{pair3} and \eqref{pair4}, we have combinations of terms of weight $m_1 + m_2 +4$ and $m_1 + m_2 + 2$. Similarly, in the product of triples of Weierstrass functions, \eqref{triple1} has weight $m_1 + m_2 + m_3$ as expected, but \eqref{triple2} has weight $m_1 + m_2 + m_3-1$, and \eqref{triple3} is a combination of weight $m_1 + m_2 + m_3 +2$ and $m_1 + m_2 + m_3$. The zero mode in the product of four Weierstrass functions also has contributions of weight $m_1 + m_2 + m_3 + m_4$ and $m_1 + m_2 + m_3 + m_4 -2$. 

Thus, the expression for a KdV charge obtained from the $n$-point function of the stress tensor can have components of different modular weight. The first place these difficulties arise is in the computation of $\langle I_3^3 \rangle$, where there are components of modular weight $12$ and $10$, but the components of modular weight $10$ cancel. This cancellation is required to reproduce the expected structure, but it is unfortunate that the structure is not more manifest.

\bibliographystyle{utphys}
\bibliography{kdvDraft1}

\providecommand{\href}[2]{#2}\begingroup\raggedright\begin{thebibliography}{10}

\bibitem{paper2}
A.~Maloney, S.~G. Ng, S.~F. Ross, and I.~Tsiares, ``{Generalized Gibbs Ensemble
  and the Statistics of KdV Charges in 2D CFT},''
\href{http://arxiv.org/abs/1810.11054}{{\ttfamily arXiv:1810.11054 [hep-th]}}.

\bibitem{Sasaki:1987mm}
R.~Sasaki and I.~Yamanaka, ``{Virasoro Algebra, Vertex Operators, Quantum
  {Sine-Gordon} and Solvable Quantum Field Theories},''
{\em Adv. Stud. Pure Math.} {\bfseries 16} (1988) 271--296.

\bibitem{Eguchi:1989hs}
T.~Eguchi and S.-K. Yang, ``{Deformations of Conformal Field Theories and
  Soliton Equations},''
\href{http://dx.doi.org/10.1016/0370-2693(89)91463-9}{{\em Phys. Lett.}
  {\bfseries B224} (1989) 373--378}.

\bibitem{Bazhanov:1994ft}
V.~V. Bazhanov, S.~L. Lukyanov, and A.~B. Zamolodchikov, ``{Integrable
  structure of conformal field theory, quantum KdV theory and thermodynamic
  Bethe ansatz},'' \href{http://dx.doi.org/10.1007/BF02101898}{{\em Commun.
  Math. Phys.} {\bfseries 177} (1996) 381--398},
\href{http://arxiv.org/abs/hep-th/9412229}{{\ttfamily arXiv:hep-th/9412229
  [hep-th]}}.

\bibitem{Calabrese:2011vdk}
P.~Calabrese, F.~H.~L. Essler, and M.~Fagotti, ``{Quantum Quench in the
  Transverse Field Ising Chain},''
  \href{http://dx.doi.org/10.1103/PhysRevLett.106.227203}{{\em Phys. Rev.
  Lett.} {\bfseries 106} no.~22, (2011) 227203},
\href{http://arxiv.org/abs/1104.0154}{{\ttfamily arXiv:1104.0154
  [cond-mat.str-el]}}.

\bibitem{Sotiriadis:2014uza}
S.~Sotiriadis and P.~Calabrese, ``{Validity of the GGE for quantum quenches
  from interacting to noninteracting models},''
  \href{http://dx.doi.org/10.1088/1742-5468/2014/07/P07024}{{\em J. Stat.
  Mech.} {\bfseries 1407} (2014) P07024},
\href{http://arxiv.org/abs/1403.7431}{{\ttfamily arXiv:1403.7431
  [cond-mat.stat-mech]}}.

\bibitem{PhysRevLett.115.157201}
E.~Ilievski, J.~De~Nardis, B.~Wouters, J.-S. Caux, F.~H.~L. Essler, and
  T.~Prosen, ``Complete generalized gibbs ensembles in an interacting theory,''
  \href{http://dx.doi.org/10.1103/PhysRevLett.115.157201}{{\em Phys. Rev.
  Lett.} {\bfseries 115} (Oct, 2015) 157201}.
  \url{https://link.aps.org/doi/10.1103/PhysRevLett.115.157201}.

\bibitem{Vidmar2016}
L.~Vidmar and M.~Rigol, ``Generalized gibbs ensemble in integrable lattice
  models,'' {\em Journal of Statistical Mechanics: Theory and Experiment}
  {\bfseries 2016} no.~6, (2016) 064007.
  \url{http://stacks.iop.org/1742-5468/2016/i=6/a=064007}.

\bibitem{Pozsgay2017}
B.~Pozsgay, E.~Vernier, and M.~A. Werner, ``On generalized gibbs ensembles with
  an infinite set of conserved charges,'' {\em Journal of Statistical
  Mechanics: Theory and Experiment} {\bfseries 2017} no.~9, (2017) 093103.
  \url{http://stacks.iop.org/1742-5468/2017/i=9/a=093103}.

\bibitem{Langen207}
T.~Langen, S.~Erne, R.~Geiger, B.~Rauer, T.~Schweigler, M.~Kuhnert,
  W.~Rohringer, I.~E. Mazets, T.~Gasenzer, and J.~Schmiedmayer, ``Experimental
  observation of a generalized gibbs ensemble,''
  \href{http://dx.doi.org/10.1126/science.1257026}{{\em Science} {\bfseries
  348} no.~6231, (2015) 207--211}.
  \url{http://science.sciencemag.org/content/348/6231/207}.

\bibitem{Kinoshita06}
T.~Kinoshita, T.~Wenger, and D.~S. Weiss, ``A quantum newton's cradle,'' {\em
  Nature} {\bfseries 440} (04, 2006) 900 EP --.
  \url{http://dx.doi.org/10.1038/nature04693}.

\bibitem{deBoer:2016bov}
J.~de~Boer and D.~Engelhardt, ``{Remarks on thermalization in 2D CFT},''
  \href{http://dx.doi.org/10.1103/PhysRevD.94.126019}{{\em Phys. Rev.}
  {\bfseries D94} no.~12, (2016) 126019},
\href{http://arxiv.org/abs/1604.05327}{{\ttfamily arXiv:1604.05327 [hep-th]}}.

\bibitem{Perez:2016vqo}
A.~Pérez, D.~Tempo, and R.~Troncoso, ``{Boundary conditions for General
  Relativity on AdS$_{3}$ and the KdV hierarchy},''
  \href{http://dx.doi.org/10.1007/JHEP06(2016)103}{{\em JHEP} {\bfseries 06}
  (2016) 103},
\href{http://arxiv.org/abs/1605.04490}{{\ttfamily arXiv:1605.04490 [hep-th]}}.

\bibitem{Bazhanov:2003ni}
V.~V. Bazhanov, S.~L. Lukyanov, and A.~B. Zamolodchikov, ``{Higher level
  eigenvalues of Q operators and Schroedinger equation},''
  \href{http://dx.doi.org/10.4310/ATMP.2003.v7.n4.a4}{{\em Adv. Theor. Math.
  Phys.} {\bfseries 7} no.~4, (2003) 711--725},
\href{http://arxiv.org/abs/hep-th/0307108}{{\ttfamily arXiv:hep-th/0307108
  [hep-th]}}.

\bibitem{Deutsch1991}
J.~M. Deutsch, ``Quantum statistical mechanics in a closed system,''
  \href{http://dx.doi.org/10.1103/PhysRevA.43.2046}{{\em Phys. Rev. A}
  {\bfseries 43} (Feb, 1991) 2046--2049}.

\bibitem{PhysRevE.50.888}
M.~Srednicki, ``Chaos and quantum thermalization,''
  \href{http://dx.doi.org/10.1103/PhysRevE.50.888}{{\em Phys. Rev. E}
  {\bfseries 50} (Aug, 1994) 888--901}.

\bibitem{Rigol2008}
M.~Rigol, V.~Dunjko, and M.~Olshanii, ``Thermalization and its mechanism for
  generic isolated quantum systems,'' {\em Nature} {\bfseries 452} (04, 2008)
  854 EP --. \url{http://dx.doi.org/10.1038/nature06838}.

\bibitem{Srednicki99}
M.~Srednicki, ``The approach to thermal equilibrium in quantized chaotic
  systems,'' {\em Journal of Physics A: Mathematical and General} {\bfseries
  32} no.~7, (1999) 1163. \url{http://stacks.iop.org/0305-4470/32/i=7/a=007}.

\bibitem{DAlessio2016}
L.~{D'Alessio}, Y.~{Kafri}, A.~{Polkovnikov}, and M.~{Rigol}, ``{From quantum
  chaos and eigenstate thermalization to statistical mechanics and
  thermodynamics},''
  \href{http://dx.doi.org/10.1080/00018732.2016.1198134}{{\em Advances in
  Physics} {\bfseries 65} (May, 2016) 239--362},
  \href{http://arxiv.org/abs/1509.06411}{{\ttfamily arXiv:1509.06411
  [cond-mat.stat-mech]}}.

\bibitem{Fitzpatrick:2015zha}
A.~L. Fitzpatrick, J.~Kaplan, and M.~T. Walters, ``{Virasoro Conformal Blocks
  and Thermality from Classical Background Fields},''
  \href{http://dx.doi.org/10.1007/JHEP11(2015)200}{{\em JHEP} {\bfseries 11}
  (2015) 200},
\href{http://arxiv.org/abs/1501.05315}{{\ttfamily arXiv:1501.05315 [hep-th]}}.

\bibitem{Lashkari:2016vgj}
N.~Lashkari, A.~Dymarsky, and H.~Liu, ``{Eigenstate Thermalization Hypothesis
  in Conformal Field Theory},''
  \href{http://dx.doi.org/10.1088/1742-5468/aab020}{{\em J. Stat. Mech.}
  {\bfseries 1803} no.~3, (2018) 033101},
\href{http://arxiv.org/abs/1610.00302}{{\ttfamily arXiv:1610.00302 [hep-th]}}.

\bibitem{Dymarsky:2016ntg}
A.~Dymarsky, N.~Lashkari, and H.~Liu, ``{Subsystem ETH},''
  \href{http://dx.doi.org/10.1103/PhysRevE.97.012140}{{\em Phys. Rev.}
  {\bfseries E97} (2018) 012140},
\href{http://arxiv.org/abs/1611.08764}{{\ttfamily arXiv:1611.08764
  [cond-mat.stat-mech]}}.

\bibitem{Lashkari:2017hwq}
N.~Lashkari, A.~Dymarsky, and H.~Liu, ``{Universality of Quantum Information in
  Chaotic CFTs},'' \href{http://dx.doi.org/10.1007/JHEP03(2018)070}{{\em JHEP}
  {\bfseries 03} (2018) 070},
\href{http://arxiv.org/abs/1710.10458}{{\ttfamily arXiv:1710.10458 [hep-th]}}.

\bibitem{Faulkner:2017hll}
T.~Faulkner and H.~Wang, ``{Probing beyond ETH at large $c$},''
\href{http://arxiv.org/abs/1712.03464}{{\ttfamily arXiv:1712.03464 [hep-th]}}.

\bibitem{He:2017txy}
S.~He, F.-L. Lin, and J.-j. Zhang, ``{Dissimilarities of reduced density
  matrices and eigenstate thermalization hypothesis},''
  \href{http://dx.doi.org/10.1007/JHEP12(2017)073}{{\em JHEP} {\bfseries 12}
  (2017) 073},
\href{http://arxiv.org/abs/1708.05090}{{\ttfamily arXiv:1708.05090 [hep-th]}}.

\bibitem{Basu:2017kzo}
P.~Basu, D.~Das, S.~Datta, and S.~Pal, ``{Thermality of eigenstates in
  conformal field theories},''
  \href{http://dx.doi.org/10.1103/PhysRevE.96.022149}{{\em Phys. Rev.}
  {\bfseries E96} no.~2, (2017) 022149},
\href{http://arxiv.org/abs/1705.03001}{{\ttfamily arXiv:1705.03001 [hep-th]}}.

\bibitem{Brehm:2018ipf}
E.~M. Brehm, D.~Das, and S.~Datta, ``{Probing thermality beyond the
  diagonal},''
\href{http://arxiv.org/abs/1804.07924}{{\ttfamily arXiv:1804.07924 [hep-th]}}.

\bibitem{Romero-Bermudez:2018dim}
A.~Romero-Bermudez, P.~Sabella-Garnier, and K.~Schalm, ``{A Cardy formula for
  off-diagonal three-point coefficients; or, how the geometry behind the
  horizon gets disentangled},''
  \href{http://dx.doi.org/10.1007/JHEP09(2018)005}{{\em JHEP} {\bfseries 09}
  (2018) 005},
\href{http://arxiv.org/abs/1804.08899}{{\ttfamily arXiv:1804.08899 [hep-th]}}.

\bibitem{Hikida:2018khg}
Y.~Hikida, Y.~Kusuki, and T.~Takayanagi, ``{Eigenstate thermalization
  hypothesis and modular invariance of two-dimensional conformal field
  theories},'' \href{http://dx.doi.org/10.1103/PhysRevD.98.026003}{{\em Phys.
  Rev.} {\bfseries D98} no.~2, (2018) 026003},
\href{http://arxiv.org/abs/1804.09658}{{\ttfamily arXiv:1804.09658 [hep-th]}}.

\bibitem{Roberts:2014ifa}
D.~A. Roberts and D.~Stanford, ``{Two-dimensional conformal field theory and
  the butterfly effect},''
  \href{http://dx.doi.org/10.1103/PhysRevLett.115.131603}{{\em Phys. Rev.
  Lett.} {\bfseries 115} no.~13, (2015) 131603},
\href{http://arxiv.org/abs/1412.5123}{{\ttfamily arXiv:1412.5123 [hep-th]}}.

\bibitem{Tolya}
A.~Dymarsky and K.~Pavlenko, ``{Generalized Gibbs Ensemble of 2d CFTs at large
  central charge in the thermodynamic limit},''
\href{http://arxiv.org/abs/1810.11025}{{\ttfamily arXiv:1810.11025 [hep-th]}}.

\bibitem{Dijkgraaf:1996iy}
R.~Dijkgraaf, ``{Chiral deformations of conformal field theories},''
  \href{http://dx.doi.org/10.1016/S0550-3213(97)00153-3}{{\em Nucl. Phys.}
  {\bfseries B493} (1997) 588--612},
\href{http://arxiv.org/abs/hep-th/9609022}{{\ttfamily arXiv:hep-th/9609022
  [hep-th]}}.

\bibitem{Bazhanov:1996aq}
V.~V. Bazhanov, S.~L. Lukyanov, and A.~B. Zamolodchikov, ``{Integrable quantum
  field theories in finite volume: Excited state energies},''
  \href{http://dx.doi.org/10.1016/S0550-3213(97)00022-9}{{\em Nucl. Phys.}
  {\bfseries B489} (1997) 487--531},
\href{http://arxiv.org/abs/hep-th/9607099}{{\ttfamily arXiv:hep-th/9607099
  [hep-th]}}.

\bibitem{Iles:2014gra}
N.~J. Iles and G.~M.~T. Watts, ``{Modular properties of characters of the
  W$_{3}$ algebra},'' \href{http://dx.doi.org/10.1007/JHEP01(2016)089}{{\em
  JHEP} {\bfseries 01} (2016) 089},
\href{http://arxiv.org/abs/1411.4039}{{\ttfamily arXiv:1411.4039 [hep-th]}}.

\bibitem{Gaberdiel:2008pr}
M.~R. Gaberdiel and C.~A. Keller, ``{Modular differential equations and null
  vectors},'' \href{http://dx.doi.org/10.1088/1126-6708/2008/09/079}{{\em JHEP}
  {\bfseries 09} (2008) 079},
\href{http://arxiv.org/abs/0804.0489}{{\ttfamily arXiv:0804.0489 [hep-th]}}.

\bibitem{Mathur:1988na}
S.~D. Mathur, S.~Mukhi, and A.~Sen, ``{On the Classification of Rational
  Conformal Field Theories},''
\href{http://dx.doi.org/10.1016/0370-2693(88)91765-0}{{\em Phys. Lett.}
  {\bfseries B213} (1988) 303--308}.

\bibitem{Hampapura:2015cea}
H.~R. Hampapura and S.~Mukhi, ``{On 2d Conformal Field Theories with Two
  Characters},'' \href{http://dx.doi.org/10.1007/JHEP01(2016)005}{{\em JHEP}
  {\bfseries 01} (2016) 005},
\href{http://arxiv.org/abs/1510.04478}{{\ttfamily arXiv:1510.04478 [hep-th]}}.

\bibitem{Gaberdiel:2016zke}
M.~R. Gaberdiel, H.~R. Hampapura, and S.~Mukhi, ``{Cosets of Meromorphic CFTs
  and Modular Differential Equations},''
  \href{http://dx.doi.org/10.1007/JHEP04(2016)156}{{\em JHEP} {\bfseries 04}
  (2016) 156},
\href{http://arxiv.org/abs/1602.01022}{{\ttfamily arXiv:1602.01022 [hep-th]}}.

\bibitem{zhu}
Y.~Zhu, ``Modular invariance of characters of vertex operator algebras,'' {\em
  Journal of the American Mathematical Society} {\bfseries 9} no.~1, (1996)
  237--302.

\bibitem{Gaberdiel:2012yb}
M.~R. Gaberdiel, T.~Hartman, and K.~Jin, ``{Higher Spin Black Holes from
  CFT},'' \href{http://dx.doi.org/10.1007/JHEP04(2012)103}{{\em JHEP}
  {\bfseries 04} (2012) 103},
\href{http://arxiv.org/abs/1203.0015}{{\ttfamily arXiv:1203.0015 [hep-th]}}.

\bibitem{Leitner:2017}
M.~Leitner, ``An algebraic approach to minimal models in cfts,''
  \href{http://arxiv.org/abs/1705.08294}{{\ttfamily arXiv:1705.08294
  [math-ph]}}.

\bibitem{Kraus:2011ds}
P.~Kraus and E.~Perlmutter, ``{Partition functions of higher spin black holes
  and their CFT duals},'' \href{http://dx.doi.org/10.1007/JHEP11(2011)061}{{\em
  JHEP} {\bfseries 11} (2011) 061},
\href{http://arxiv.org/abs/1108.2567}{{\ttfamily arXiv:1108.2567 [hep-th]}}.

\bibitem{Iles:2013jha}
N.~J. Iles and G.~M.~T. Watts, ``{Characters of the $W_3$ algebra},''
  \href{http://dx.doi.org/10.1007/JHEP02(2014)009}{{\em JHEP} {\bfseries 02}
  (2014) 009},
\href{http://arxiv.org/abs/1307.3771}{{\ttfamily arXiv:1307.3771 [hep-th]}}.

\bibitem{FREUND1989243}
P.~Freund, T.~Klassen, and E.~Melzer, ``S-matrices for perturbations of certain
  conformal field theories,''
  \href{http://dx.doi.org/https://doi.org/10.1016/0370-2693(89)91165-9}{{\em
  Physics Letters B} {\bfseries 229} no.~3, (1989) 243 -- 247}.

\bibitem{DiFrancesco:1991anf}
P.~Di~Francesco and P.~Mathieu, ``{Singular vectors and conservation laws of
  quantum KdV type equations},''
  \href{http://dx.doi.org/10.1016/0370-2693(92)90714-F}{{\em Phys. Lett.}
  {\bfseries B278} (1992) 79--84},
\href{http://arxiv.org/abs/hep-th/9109042}{{\ttfamily arXiv:hep-th/9109042
  [hep-th]}}.

\bibitem{Negro:2016yuu}
S.~Negro, ``{Integrable structures in quantum field theory},''
  \href{http://dx.doi.org/10.1088/1751-8113/49/32/323006}{{\em J. Phys.}
  {\bfseries A49} no.~32, (2016) 323006},
\href{http://arxiv.org/abs/1606.02952}{{\ttfamily arXiv:1606.02952 [math-ph]}}.

\bibitem{Faddeev:1996iy}
L.~D. Faddeev, ``{How algebraic Bethe ansatz works for integrable model},'' in
  {\em {Relativistic gravitation and gravitational radiation. Proceedings,
  School of Physics, Les Houches, France, September 26-October 6, 1995}},
  pp.~pp. 149--219.
\newblock 1996.
\newblock
\href{http://arxiv.org/abs/hep-th/9605187}{{\ttfamily arXiv:hep-th/9605187
  [hep-th]}}.
\newblock

\bibitem{Beisert:2010jr}
N.~Beisert {\em et al.}, ``{Review of AdS/CFT Integrability: An Overview},''
  \href{http://dx.doi.org/10.1007/s11005-011-0529-2}{{\em Lett. Math. Phys.}
  {\bfseries 99} (2012) 3--32},
\href{http://arxiv.org/abs/1012.3982}{{\ttfamily arXiv:1012.3982 [hep-th]}}.

\bibitem{Beisert}
N.~Beisert, ``{Lecture Notes: Integrability in QFT and AdS/CFT},''.
  \url{http://edu.itp.phys.ethz.ch/hs14/14HSInt/IntAdSCFT14Notes.pdf}.

\bibitem{Torrielli:2016ufi}
A.~Torrielli, ``{Lectures on Classical Integrability},''
  \href{http://dx.doi.org/10.1088/1751-8113/49/32/323001}{{\em J. Phys.}
  {\bfseries A49} no.~32, (2016) 323001},
\href{http://arxiv.org/abs/1606.02946}{{\ttfamily arXiv:1606.02946 [hep-th]}}.

\end{thebibliography}\endgroup

\end{document}